\pdfoutput=1
\documentclass{article}
\usepackage{arxiv}
\usepackage{natbib} 
\usepackage{xcolor}
\usepackage{bookmark}
\usepackage{hyperref}
\usepackage[linesnumbered,ruled,vlined]{algorithm2e}
\usepackage[utf8]{inputenc} 
\usepackage[T1]{fontenc}    
\usepackage{commath}
\usepackage{hyperref}       
\usepackage{url}      
\usepackage{graphics} 
\usepackage{mathtools}
\usepackage{booktabs}       
\usepackage{amsfonts}       
\usepackage{nicefrac}       
\usepackage{microtype}      
\usepackage{lipsum}
\usepackage{graphicx}
\graphicspath{ {./images/} }
\usepackage{color}
\usepackage{tikz}
\usetikzlibrary{shapes.geometric}
\usepackage{diagbox}
\usetikzlibrary{shapes,decorations,arrows,calc,arrows.meta,fit,positioning}
\tikzset{
    -Latex,auto,node distance =1 cm and 1cm,semithick,
    state/.style ={ellipse, draw, minimum width = 0.7cm},
    point/.style = {circle, draw, inner sep=0.04cm,fill,node contents={}},
    bidirected/.style={Latex-Latex,dashed},
    el/.style = {inner sep=2pt, align=left, sloped}
    state/.style={circle, draw, minimum size=0.2cm},
            rectangle state/.style={rectangle, draw, minimum size=0.2cm},
    double state/.style={rectangle, draw, double, minimum size=0.2cm, double distance=0.5pt}, 
    double arrow/.style={double, line width=1pt, double distance=2pt}
}
\usepackage{caption}
\usepackage{subcaption}
\usepackage{graphicx}
\usepackage{multirow}
\usepackage{array}
\usepackage[flushleft]{threeparttable}
\usepackage{adjustbox}
\usepackage{float}

\usepackage{url} 
\usepackage{xcolor}
\usepackage{fancyhdr}
\usepackage[linesnumbered,ruled,vlined]{algorithm2e}
\usepackage[utf8]{inputenc} 
\usepackage{commath}
\usepackage{hyperref}       
\usepackage{url}            
\usepackage{mathtools}
\usepackage{booktabs}       
\usepackage{amsfonts}       
\usepackage{nicefrac}       
\usepackage{microtype}      
\usepackage{lipsum}
\usepackage{graphicx}
\graphicspath{ {./images/} }
\usepackage{color}
\usepackage{tikz}
\usepackage{caption}
\usepackage{subcaption}
\usepackage{graphicx}
\usepackage{multirow}
\usepackage{array}
\usepackage[flushleft]{threeparttable}
\usepackage{adjustbox}
\usepackage{float}

\newtheorem{theorem}{Theorem}

\newtheorem{corollary}{Corollary}
\newtheorem{lemma}{Lemma}

\usetikzlibrary{shapes,decorations,arrows,calc,arrows.meta,fit,positioning}
\tikzset{
    -Latex,auto,node distance =1 cm and 1 cm,semithick,
    state/.style ={ellipse, draw, minimum width = 0.7 cm},
    point/.style = {circle, draw, inner sep=0.04cm,fill,node contents={}},
    bidirected/.style={Latex-Latex,dashed},
    el/.style = {inner sep=2pt, align=left, sloped}
}

\usepackage{amssymb}
\usepackage{pifont}

\usepackage{latexsym}
\usepackage{savesym}
\usepackage{mathtools}
\savesymbol{iint}
\usepackage{txfonts}
\restoresymbol{TXF}{iint}
\usepackage{amsfonts}
\usepackage{psfrag}
\usepackage{graphicx}
\usepackage{subcaption}
\usepackage{mathrsfs} 
\usepackage{caption}
\usepackage{subcaption}

\usepackage[english]{babel}
\title{A Double Machine Learning Approach for the Evaluation of COVID-19 Vaccine Effectiveness under the Test-Negative Design: Analysis of Québec Administrative Data}

\author{ Cong Jiang$^{1}$, Denis Talbot$^{2,3}$, Sara Carazo$^{4}$, Mireille E Schnitzer$^{1,5}$ \\
$^{1}$ \normalsize{\textit{Faculty of Pharmacy, Université de Montréal, Québec, Canada}}\\ $^{2}$ \normalsize{\textit{Département de médecine sociale et préventive, Université Laval, Québec, Canada}}\\
    $^{3}$ \normalsize{\textit{Axe santé des populations et pratiques optimales en santé, Québec, Canada}}\\
    $^{4}$ \normalsize{\textit{Institut national de santé publique du Québec, Québec, Canada}}\\ 
    $^{5}$ \normalsize{\textit{British Columbia Centre for Disease Control, Vancouver, Canada}}\\ \texttt{\{cong.jiang, mireille.schnitzer\}@umontreal.ca}
\\}
\date{} 
\begin{document}
\maketitle
\begin{abstract}
The test-negative design (TND), which is routinely used for monitoring seasonal flu vaccine effectiveness (VE), has recently become integral to COVID-19 vaccine surveillance, notably in Québec, Canada. Some studies have addressed the identifiability and estimation of causal parameters under the TND, but efficiency bounds for nonparametric estimators of the target parameter under the unconfoundedness assumption have not yet been investigated. Motivated by the goal of improving adjustment for measured confounders when estimating COVID-19 VE among community-dwelling people aged $\geq 60$ years in Québec, we propose a one-step doubly robust and locally efficient estimator called TNDDR (TND doubly robust), which utilizes cross-fitting (sample splitting) and can incorporate machine learning techniques to estimate the nuisance functions and thus improve control for measured confounders. We derive the efficient influence function (EIF) for the marginal expectation of the outcome under a vaccination intervention, explore the von Mises expansion, and establish the conditions for $\sqrt{n}-$consistency, asymptotic normality and double robustness of TNDDR. The proposed estimator is supported by both theoretical and empirical justifications.

\end{abstract}

\section{Introduction}\label{sec:intro}
Clinical trials first demonstrate the efficacy of a vaccine in controlled settings; however, once vaccines are rolled out, jurisdictions may diverge from the tested protocol, and the vaccine is used on a broader population than in the clinical trials. Consequently, there is a rapid demand for observational analyses to assess the effectiveness of vaccines across heterogeneous populations and over extended periods. The test-negative design (TND) is an observational study design that recruits care-seeking individuals who meet a standard clinical case definition, aiming to estimate the effectiveness of vaccination in guarding against disease outcomes resulting from an infectious disease of interest \citep{jackson2013test, lipsitch2016observational}. It was initially proposed to assess the effectiveness of seasonal influenza vaccines \citep{skowronski2007estimating}, increasingly formalized by Jackson
and Nelson (2013)\cite{jackson2013test}; Foppa et al. (2013)\cite{foppa2013case}; Dean (2019)\cite{dean2019re}; Sullivan et al. (2016)\cite{sullivan2016theoretical} and others, and was employed globally to evaluate vaccine effectiveness (VE) against COVID-19 caused by emerging SARS-CoV-2 variants \citep{ dean2021covid} in addition to VE against seasonal influenza, rotavirus, and dengue fever \citep{ boom2010effectiveness,anders2018awed}. Distinct from the case-control study, the TND typically involves the recruitment of participants with a common symptom presentation who are being tested for the infectious disease in question. The “cases” are defined as participants who test positive for the target infection, and the “controls” are those who test negative, with the latter presumably afflicted by some other, non-target, infectious disease. 

In the province of Qu\'ebec, Canada, the director of public health has entrusted the mandate of monitoring vaccine effectiveness to the \textit{Institut national de santé publique du Qu\'ebec} (INSPQ). The utilization of a TND presents notable advantages in terms of expediency and cost-effectiveness, while also holding promise in addressing unobserved confounding variables such as healthcare-seeking behavior disparities between vaccinated and unvaccinated individuals  \citep{jackson2006evidence, jackson2013test}. In Qu\'ebec, data pertaining to cases and controls have been sourced from administrative and surveillance databases. Through the integration of various provincial databases utilizing public health insurance identification numbers, these robust datasets offer comprehensive insights into administered COVID-19 vaccines, polymerase chain reaction (PCR) tests, hospitalizations (with reasons), some sociodemographic information, diagnosed comorbidities, and deaths. The analyses for COVID-19 VE surveillance are crucial as they serve to inform policy decisions, such as the implementation of delayed second vaccine doses in Qu\'ebec comparatively with other jurisdictions \citep{skowronski2022two}. 

While TND is convenient and has recently become integral to COVID-19 vaccine surveillance, it is prone to several potential biases, including confounding and collider bias \citep{sullivan2016theoretical, ortiz2024potential, li2023double}. The identification and estimation of statistical and causal estimands related to VE under the TND have been investigated in past work. Up until recently, logistic regression was the only statistical method proposed to adjust for confounders to estimate VE under this design \citep{mesidor2023test, chung2021effectiveness, bernal2021effectiveness}. Under strong modeling assumptions and conditional independence between vaccination and the condition of having some other infection leading to inclusion criteria, the logistic regression produces estimates of a \textit{conditional risk ratio} of the association between vaccination and the probability of medically-attended symptomatic infection \citep{jackson2013test, schnitzer2022estimands}. Then, the ``vaccine effectiveness'' estimand is typically expressed as 1 minus the conditional risk ratio estimand. However, this approach is susceptible to bias when there is effect modification of the VE according to a confounder, which cannot be remedied by the specification of the logistic regression \citep{schnitzer2022estimands}. While a proposed inverse probability of treatment weighting (IPW) estimator for the \textit{marginal risk ratio} is valid under effect modification, an important limitation of this estimator is that it requires the correct specification of a parametric model for the probability of vaccination \citep{schnitzer2022estimands}. Both of the above techniques depend on conditional exchangeability in order to equate their respective risk ratio to a \textit{causal risk ratio}. Another identification strategy, employed to bypass unmeasured confounding and collider pathways, involved making alternative assumptions regarding the presence of negative control exposure and outcome variables \citep{lipsitch2010negative, shi2020multiply}. Through these assumptions, Li et al. (2023)\cite{li2023double} aimed to identify and estimate the causal risk ratio of infection with SARS-CoV-2. Nevertheless, valid negative control exposures or outcomes may not always be available. This is the case in the databases used in Québec for assessing COVID-19 VE, where a convincing negative control exposure is not available.

Despite the considerable progress in VE estimation under the TND, no previous work derived efficiency bounds for semiparametric estimators of the marginal risk ratio (or VE). The field of semiparametric theory provides a valuable framework for benchmarking efficiency and devising efficient estimators under limited statistical assumptions \citep{robins1994estimation, robins1992recovery, tsiatis2006semiparametric, kennedy2017semiparametric, coston2022role}. In particular, empirical process theory offers potent tools to better comprehend the asymptotic behavior of semiparametric estimators,  and to seamlessly integrate machine learning and other modern methods into causal inference research \citep{kennedy2016semiparametric}. Utilizing data-adaptive modeling (e.g., statistical or machine learning) in this context offers the advantage of improving the flexibility and accuracy of the nuisance estimator, and thus contributes to a more efficient estimation of the target parameters, effectively addressing certain parametric modeling limitations.


As a motivating example, we consider the estimation of VE against hospitalization due to a COVID-19 infection during the Omicron BA.5 variant period using the administrative data from Qu\'ebec, Canada. More specifically, we aimed to estimate the relative VE of having received a booster mRNA dose within 6 months compared to a latest dose administered 6 or more months ago among adults aged 60 years or older. While rich information on potential confounders is available in the database, bias due to incorrect modeling remains possible when using parametric methods, such as a logistic regression or inverse probability weighting. The semiparametric estimator we propose minimizes the risk of bias due to model misspecification through estimating the probability of treatment and outcomes conditional on measured confounders data-adaptively using machine learning methods. As such, this estimator provides more robust evidence to adequately guide public policies.

The paper is structured as follows: Section \ref{sec:back} introduces the TND's basic structure and defines the target parameter of VE, along with presenting a new identifiability formula. Section \ref{sec:meth} derives the efficient influence function and proposes the double machine learning TNDDR (TND Doubly Robust) algorithm for VE estimation under the TND, and establishes its theoretical large-sample properties. Section \ref{sec:sim} investigates the proposed estimator's double robustness, $\sqrt{n}$-consistency, and confidence interval coverage through simulation studies. In Section \ref{sec:real}, we address the motivating research questions of the study, while Section \ref{sec:conc} summarizes the conclusions and offers a discussion of the findings.

\section{Background and Problem Setting}\label{sec:back}
In this section, we set up the structure of the setting we will be addressing, defining notation for the data, distributions, and statistical quantities under a hypothetical simple random sample and the TND applied to the entire population. The target population of inference is thus the entire population. Then, we define our target parameter and describe the previously proposed IPW estimator. Further, we provide an alternative identifiability formula for the target parameter and present a related outcome regression-based estimator. This estimator uses \textit{debiasing weights}, tailored to the TND setting, which can be solely contingent upon outcome modeling. 

\begin{figure}[ht]
    \begin{subfigure}[b]{0.50\textwidth}
        \centering
        \begin{tikzpicture}
            \tikzset{
                state/.style={circle, draw, minimum size=0.8cm, inner sep=0},
                double state/.style={circle, draw, double, minimum size=0.8cm, inner sep=0, line width=1pt},
                double arrow/.style={double, line width=0.5pt, double distance=1pt}
            }
            \node[state] (V) {$V$};
            \node[state] (I2) [right=of V] {$I_2$};
            \node[state] (W) [right=of I2] {$W_2$};
            \node[state] (H) [white, right=of W] {$H_2$};
            \node[rectangle state] (C) [below left=of V, yshift=-1.7cm, xshift=-.2cm] {$\boldsymbol{C}$};
            \node[state] (I1) [right=of C, xshift=2.0cm] {$I_1$};
            \node[state] (W1) [right=of I1, xshift=-.1cm] {$W_{1}$};
            \node[state] (H1) [white, right=of W1, xshift=.0cm] {$H_{1}$};
            \node[rectangle state] (Hosp) [below=of H,  yshift=.1cm] {$H$}; 
            \node[rectangle state] (Sym) [below=of W,  yshift=.1cm] {$W$}; 
            \node[state] (U) [above left=of V, yshift=.1cm, xshift=-.4cm] {$\boldsymbol{U}$};

            \path (V) edge (I2);
            \path (I2) edge (W);
            \path (I1) edge (W1);
            \path (V) edge[bend left=45] (W);
            \path[dashed] (U) edge (I2);
            \path[dashed] (U) edge[bend right=30] (I1);
            \path[dashed] (U) edge[bend right=30] (W1);
            \path[dashed] (U) edge[bend right=50] (Hosp);
            \path[dashed] (U) edge (W);
        \draw[double arrow] (W) -- (Sym);
        \draw[double arrow] (W1) -- (Sym);
        \path (Sym) edge (Hosp);
        \end{tikzpicture}
        \caption{\normalsize\  TND DAG with selection}
        \label{fig:subfig4}
    \end{subfigure}
    \begin{subfigure}[b]{0.50\textwidth}
        \centering
        \begin{tikzpicture}
            \tikzset{
                state/.style={circle, draw, minimum size=0.8cm, inner sep=0},
                double state/.style={circle, draw, double, minimum size=0.8cm, inner sep=0, line width=1pt},
                double arrow/.style={double, line width=0.5pt, double distance=1pt}
            }
            \node[state] (V) {$V$};
            \node[state] (I2) [right=of V] {$I_2$};
            \node[state] (W) [right=of I2] {$W_2$};
            \node[state] (H) [white, right=of W] {$H_2$};
            \node[rectangle state] (C) [below left=of V, yshift=-1.7cm, xshift=-.2cm] {$\boldsymbol{C}$};
            \node[state] (I1) [right=of C, xshift=2.0cm] {$I_1$};
            \node[state] (W1) [right=of I1, xshift=-.1cm] {$W_{1}$};
            \node[state] (H1) [white, right=of W1, xshift=.0cm] {$H_{1}$};
            \node[state] (Hosp) [below=of H,  yshift=.2cm] {$H$}; 
            \node[state] (Sym) [below=of W,  yshift=.2cm] {$W$}; 
            \node[state] (U) [above left=of V, yshift=.1cm, xshift=-.4cm] {$\boldsymbol{U}$};
            \node[state] (Y) [above=of H,  yshift=.18cm] {$Y$}; 
            \path (V) edge (I2);
            \path (I2) edge (W);
            \path (I1) edge (W1);
            \path (V) edge[bend left=45] (W);
            \path[dashed] (U) edge (I2);
            \path[dashed] (U) edge[bend right=30] (I1);
            \path[dashed] (U) edge[bend right=30] (W1);
            \path[dashed] (U) edge[bend right=53] (Hosp);
            \path[dashed] (U) edge (W);
        \draw[double arrow] (Hosp) -- (Y);
        \draw[double arrow] (I2) -- (Y);
        \draw[double arrow] (W) -- (Sym);
        \draw[double arrow] (W1) -- (Sym);
        \path (Sym) edge (Hosp);
        \end{tikzpicture}
        \caption{\normalsize\  TND DAG demonstrating compound outcome $Y$}
        \label{fig:subfig2}
    \end{subfigure}
    \caption{Directed acyclic graphs (DAGs) of a TND study that recruits patients hospitalized for infectious disease symptoms. In the context of COVID-19, $V$ denotes the vaccination status against COVID-19, $\boldsymbol{C}$ denotes measured confounders, $I_2$ and $I_1$ represent the infection statuses of SARS-CoV-2 infection and other infections, respectively. $W$ indicates the presence of COVID-19-like symptoms, where $W = \max\{W_1, W_2\},$ and the subscripts denote whether the symptoms are due to SARS-CoV-2 infection ($W_2$) or other infections ($W_1$). $W_1$ and $W_2$, which are latent, thus deterministically determine $W$, which is represented by double-lined arrows. $H$ denotes hospitalization for symptoms. Latent variable $\boldsymbol{U}$ denotes any (possibly unmeasured) common causes of infection, symptoms, and hospitalization that do not also affect vaccination. The right-hand DAG (b) also illustrates a structure allowing for control exchangeability, i.e., \( \{I_1= H=1\} \perp V \mid \boldsymbol{C}\) which must hold in the unselected population.  Illustrating the no-unmeasured confounding assumption, DAG (b) includes the compound outcome \(Y\), a deterministic function of SARS-CoV-2 infection and hospitalization (i.e., $Y = \mathbb{I}(I_2=1, H=1)$), indicated by double-lined arrows.} 
    \label{TNDdag}
\end{figure}

\subsection{Notation and target parameter of vaccine effectiveness}
We consider a TND study that samples people hospitalized for COVID-19-like illness. Note that our structure could also represent a TND study that samples people tested for symptoms compatible with COVID-19. The directed acyclic graphs (DAGs) depicted in Figure \ref{TNDdag} give the key variables and structure that we will consider. Let $V$ (categorical or binary) denote the status of vaccination against COVID-19, $\boldsymbol{C}$ be measured confounders, and $\boldsymbol{U}$ be unmeasured variables, assumed to not affect $V$. Confounders $\boldsymbol{C}$  in our application include age groups, gender, multimorbidity and epidemiological observation timeframe, and the variables $\boldsymbol{U}$ could include unknown COVID-19 and influenza susceptibility. Let $I_2$ indicate infection by SARS-CoV-2, and let $I_1 = 1$ represent the presence of another infection (i.e., not SARS-CoV-2) that can potentially cause COVID-19-like symptoms. The variable  $W$ indicates the presence of COVID-19-like symptoms, which are recorded in the Québec administrative data. By assumption, symptoms $W$ can only develop if an infection is present, i.e., $I_1 =1 \cup I_2 =1$. Define latent variables $W_2$ and $W_1$ as indicators denoting whether the symptoms are due to SARS-CoV-2 infection ($W_2$) or other infections ($W_1$), respectively; thus, the presence of symptoms is represented as $W = \max\{W_1, W_2\}$. Let $H$ denote hospitalization for symptoms. The complete data from the underlying population are defined as $\boldsymbol{Z}^{C} = (\boldsymbol{C}, V, I_1, I_2, W, H)$. Suppose that $\boldsymbol{Z}^{C}$ under simple random sampling has (unknown) probability distribution $\mathbb{P}$, which is assumed to lie within some {model} (i.e., set of distributions) $\mathcal{P}$. 

Our focus is on the TND study that samples those with some infection ($I_1 =1 \cup I_2 =1$), with symptoms ($W=1$), who are then hospitalized due to these symptoms ($H = 1$). As shown in DAG (a) in Figure \ref{TNDdag}, the boxes around $W$ and $H$ represent how the study sampling method conditions on these variables. The box around $C$ represents statistical adjustment for the measured confounders. Using indicator function $\mathbb{I}$, we define $S = \mathbb{I}({I_1 =1} \cup {I_2 =1},W=1, H=1)$ to represent the presence of the inclusion criteria for the TND, and define $q_0 = \mathbb{P}(S=1) = \mathbb{P}({I_1 =1} \cup {I_2 =1}, W=1, H=1)$ as the marginal probability of meeting the inclusion criteria for the study. 
It is important to note that, by design, the set of infected and symptomatic individuals with $W=1\cap H=1$ is equivalent to the set where $H=1$. Consequently, eligibility for TND selection ($S$) can be defined solely by hospitalization ($H$), i.e., $S=H$. In the TND, we only observe a sample of size $n$ from a subset of the complete data, whose data structure is represented by  $\boldsymbol{Z} = (\boldsymbol{C}, V, I_1, I_2)\mathbb{I}(H=1)$. The observed data, $\boldsymbol{Z}$, under TND sampling, are distributed according to $\mathbb{P}_{TND}(\boldsymbol{Z} = \boldsymbol{z}) := \mathbb{P}(\boldsymbol{Z} = \boldsymbol{z} \mid H=1)$, where $\boldsymbol{z}$ represents a realization of the observed data. Specifically, the relevant compound or combined outcome of interest is defined as $Y = \mathbb{I}(I_2=1, H=1)$, {i.e., hospitalized (or medically-attended) symptomatic SARS-CoV-2 infection}. This outcome involves three consecutive steps: infection with SARS-CoV-2, development of symptoms, and subsequent hospitalization due to these symptoms. When $Y = 1$, it indicates the cases within the TND sample, while the condition $\{Y = 0, H=1\}$ corresponds to the controls present in the sample. 
Note that this compound outcome definition is consistent with the implied outcome of using traditional logistic regression models under a TND using similar sampling to what we describe, as discussed in Schnitzer (2022)\cite{schnitzer2022estimands}.

Our objective is to directly estimate a marginal risk ratio (mRR) under distribution $\mathbb{P}$, contrasting some vaccination statuses $v$ v.s. $v_0$, where $v, v_0 \in \mathcal{V},$ and $\mathcal{V}$ denotes exposure values of interest. In our motivating example, $v$ indicates ``a booster mRNA vaccine administered within 6 months'' and $v_0$ indicates ``a latest dose of mRNA vaccine administered over 6 months ago''. Define $p_{\boldsymbol{C}} (\boldsymbol{c}):= \mathbb{P}(\boldsymbol{C} = \boldsymbol{c})$ as the marginal probability density function of the covariates $\boldsymbol{C}\in\boldsymbol{\mathscr{C}}$, under simple random sampling. Also, define 
$\mathbb{E}_{\mathbb{P}}$ as an expectation  taken with respect to the  distribution of $\boldsymbol{C}$. As such, $\mathbb{E}_{\mathbb{P}}\left[\mathbb{P}(Y=1 \mid V=v,  \boldsymbol{C}  = \boldsymbol{c} ) \right] = \int_{ \boldsymbol{\mathscr{C}} } \mathbb{P}(Y =1 \mid V=v,  \boldsymbol{C}  = \boldsymbol{c} )    p_{ \boldsymbol{C}  }(\boldsymbol{c}  ) d\boldsymbol{c}$.  The marginal risk ratio can then be defined as 
\begin{align} \label{target}
    \psi_{mRR}  := \frac{\mathbb{E}_{\mathbb{P}}\left[\mathbb{P}(Y=1 \mid V=v,  \boldsymbol{C}  = \boldsymbol{c} ) \right] }{\mathbb{E}_{\mathbb{P}}\left[\mathbb{P}(Y=1 \mid V=v_0,  \boldsymbol{C}  = \boldsymbol{c} ) \right]} .
\end{align}
 Then, the “vaccine effectiveness” estimand is given as $\textit{VE} = 1 - \psi_{mRR}$.

Causal inference is possible under further assumptions, including the existence of a well-defined potential outcome $Y^*(v)$, which is the outcome that would be observed if we set an individual's vaccination status to $V=v$. Specifically,  under causal assumptions of no interference, consistency, positivity and conditional exchangeability for the compound outcome $Y$, we have that $\psi_{mRR}={\mathbb{P}\left(Y^*(v)=1\right) }/{ \mathbb{P}\left(Y^*(v_0)=1\right) }$ \cite{schnitzer2022estimands}. We also note that these same identifiability assumptions are required if using logistic regression to estimate causal effects. However, in our methods development, we formally target the non-causal estimand defined in Equation (\ref{target}). 

\subsection{Identification and inverse probability  weighting approach for estimating VE}    
To estimate the marginal risk ratio (i.e., $1 - VE$) using TND-sampled data, we define $p_{\boldsymbol{C}} (\boldsymbol{c} \mid H=1) := \mathbb{P}(\boldsymbol{C} = \boldsymbol{c} \mid H=1)$ as the marginal probability density function of the covariates $\boldsymbol{C}$ under TND sampling. This distribution, denoted by $\mathbb{P}_{TND}$, is assumed to fall within some model $\mathcal{P}_{TND}$. We also define $\mathbb{E}_{TND}$ as the expectation with respect to the distribution of $\boldsymbol{C}$ under TND sampling, such that $\mathbb{E}_{TND}\left( \boldsymbol{Z}\right) = \int_{ \boldsymbol{\mathscr{C}} } \boldsymbol{Z}  p_{ \boldsymbol{C}  }(\boldsymbol{c}  \mid H=1) d\boldsymbol{c},$ for any variable $\boldsymbol{Z}$. Schnitzer (2022)\cite{schnitzer2022estimands} introduced an inverse probability weighting procedure for estimating VE, leveraging an identifiability result that $$
    \mathbb{E}_\mathbb{P}\{\mathbb{P}(Y=1\mid V=v, \boldsymbol{C}) \} = \mathbb{E}_{TND}\left\{Y \mathbb{I}(V=v) q_0 \middle/ \mathbb{P}(V=v\mid \boldsymbol{C})\right\} = \mathbb{E}_{TND}\left\{Y \mathbb{I}(V=v) q_0 \middle/ \mathbb{P}_{TND}(V=v\mid Y=0, \boldsymbol{C})\right\}.$$
The second equality is based on an equivalence between the propensity score based under simple random sampling and the propensity score conditional on the control data of the TND sample, i.e., $\mathbb{P}(V=v\mid \boldsymbol{C})=\mathbb{P}_{TND}(V=v\mid Y = 0, \boldsymbol{C})$. This equivalence holds under the assumption that being infected with another infection and being hospitalized for symptoms is independent of vaccination conditional on covariates, i.e., $\{I_1= H=1\} \perp V \mid  \boldsymbol{C}$.  We will refer to this assumption as ``control exchangeability'' since it means that controls are exchangeable across exposure groups. Under our DAG, control exchangeability holds with the additional assumption of ``no overlap'' between the infection of interest and other infections. This leads to a control group $\{Y=0,H=1\}=\{I_1= H=1\}$ that is representative of the source population in terms of the distribution of vaccination. The typical TND statement related to this assumption is that the vaccine does not have an impact on non-target infections \citep{feng2017assessment}, which is necessary but not sufficient for control exchangeability. 
When infection overlap is possible (i.e., $\mathbb{P}(I_1=I_2=1) \neq 0$), the control exchangeability assumption may still hold approximately in some instances. 

Under control exchangeability, the related estimator for comparing vaccination statuses $v$ versus $v_0$ using inverse probability weighting can be expressed as
\begin{align*}
   \hat{\psi}^{ipw}_{mRR} =  \left.\frac{1}{n}\sum_{k=1}^n \frac{Y_k\mathbb{I}(V_k=v)}{\hat{\mathbb{P}}_{TND,n}(V=v\mid Y=0, \boldsymbol{C}=\boldsymbol{c}_k)} \middle/ \frac{1}{n}\sum_{k=1}^n\frac{Y_k\mathbb{I}(V_k=v_0)}{\hat{\mathbb{P}}_{TND,n}(V=v_0 \mid Y=0, \boldsymbol{C}=\boldsymbol{c}_k)}\right., 
\end{align*}
where $\hat{\mathbb{P}}_{TND,n}(V=v\mid Y=0, \boldsymbol{C}=\boldsymbol{c})$ denotes the propensity score, which is derived by (I) modeling the conditional probability of vaccine status using only the control group, and (II) utilizing this model to estimate the probabilities for the entire sample\cite{schnitzer2022estimands}.

\subsection{Alternative identifiability formula and VE estimation with weighted outcome regression}
Building on these preliminary results, we derive an alternative formula for identifying the marginal risk ratio.  First, we define the debiasing weight as follows:
\begin{equation} \label{wet}
\omega_v(\boldsymbol{c}) := \frac{\mathbb{P}(H=1 \mid V=v ,  \boldsymbol{C}  = \boldsymbol{c}  )}{\mathbb{P}(H=1 \mid  \boldsymbol{C}  = \boldsymbol{c} )},
\end{equation}
which is the ratio of conditional selection probabilities.
Then, we have the following identifiability theorem:
\begin{theorem}\label{ProP1}
(Alternative identifiability formula) The marginal risk ratio $\psi_{mRR}$ in Equation (\ref{target}) can be identified in terms of the TND distribution as
\begin{equation*} \label{identifiability}
    \psi_{mRR} = \frac{\mathbb{E}_{TND}\left[\mathbb{P}(Y = 1 \mid V=v ,  \boldsymbol{C}  = \boldsymbol{c}, H=1 )  \omega_{v}(\boldsymbol{c}) q_{0} \right]}{\mathbb{E}_{TND}\left[\mathbb{P}(Y = 1 \mid V=v_0,  \boldsymbol{C}  = \boldsymbol{c}, H=1 )  \omega_{v_0}(\boldsymbol{c}) q_{0} \right]},
\end{equation*}
where under the control exchangeability assumption that $\{Y = 0, H=1 \} \perp V \mid  \boldsymbol{C}$,  the debiasing weights are identifiable based on the following equation: 
\begin{equation*}
\omega_v(\boldsymbol{c}) :=  \frac{\mathbb{P}(H=1 \mid V=v ,  \boldsymbol{C}  = \boldsymbol{c}  )}{\mathbb{P}(H=1 \mid  \boldsymbol{C}  = \boldsymbol{c} )} = \frac{\mathbb{P}(V=v \mid  \boldsymbol{C}  = \boldsymbol{c} , H=1)}{\mathbb{P}(V=v \mid  \boldsymbol{C}  = \boldsymbol{c} , Y=0, H=1)} =  \frac{ \mathbb{P}(Y = 0\mid  \boldsymbol{C}  = \boldsymbol{c}, H=1 )}{\mathbb{P}(Y = 0 \mid V=v ,  \boldsymbol{C}  = \boldsymbol{c}, H=1 )}.
\end{equation*}
\end{theorem} 
Proofs for the above theorem are provided in Appendix \ref{Appx.A1}. Further, we define convenient notation for the four nuisance functions in the debiasing weights, for$\  v \in \mathcal{V},$
\begin{itemize}
    \item $\pi_v^0(\boldsymbol{c}) := \mathbb{P}_{TND}(V=v \mid \boldsymbol{C}  = \boldsymbol{c}, Y=0),$ propensity score among the controls;
    \item  $\pi_v(\boldsymbol{c}) := \mathbb{P}_{TND}(V=v \mid \boldsymbol{C}  = \boldsymbol{c}),$ propensity score in the overall TND sample;
    \item  $m(\boldsymbol{c}) := \mathbb{P}_{TND}(Y = 1\mid  \boldsymbol{C}  = \boldsymbol{c} ),$ marginal outcome regression functions; and
    \item $\mu_v(\boldsymbol{c}) :=\mathbb{P}_{TND}(Y = 1 \mid V=v ,  \boldsymbol{C}  = \boldsymbol{c} ),$ treatment-specific outcome regression functions.
\end{itemize}
Then, the identified debiasing weights in Theorem \ref{ProP1} can be rewritten as
\begin{align*}
     \omega_v(\boldsymbol{c}) &= \frac{ \mathbb{P}(Y = 0\mid  \boldsymbol{C}  = \boldsymbol{c}, H=1 )}{\mathbb{P}(Y = 0 \mid V=v ,  \boldsymbol{C}  = \boldsymbol{c}, H=1 )} = \frac{1 - m(\boldsymbol{c})}{1 - \mu_v(\boldsymbol{c})} ,\text{and}\ \ 
     \omega_v(\boldsymbol{c}) = \frac{\mathbb{P}(V=v \mid  \boldsymbol{C}  = \boldsymbol{c} , H=1)}{\mathbb{P}(V=v \mid  \boldsymbol{C}  = \boldsymbol{c}, Y=0, H=1 )}= \frac{\pi_v(\boldsymbol{c})}{\pi_v^0(\boldsymbol{c})}.\label{wetlist}
\end{align*}
According to the result of Theorem \ref{ProP1} and canceling out  the constant $q_0$, the estimand $\psi_{mRR}$ can also be identified as $\mathbb{E}_{TND}\left[\mu_{v}(\boldsymbol{c}) \omega_{v}(\boldsymbol{c})  \right]/ \mathbb{E}_{TND}\left[\mu_{v_0}(\boldsymbol{c}) \omega_{v_0}(\boldsymbol{c}) \right].$ Therefore, we define a key statistical estimand as
\begin{equation}\label{tag1}
\psi_{v} := \psi_v(\mathbb{P}_{TND})  = \mathbb{E}_{TND}\left[\mu_v(\boldsymbol{c}) \omega_v(\boldsymbol{c}) \right],
\end{equation}
such that $\psi_{mRR} = \psi_v/\psi_{v_0}$.
Setting  $\hat{\omega}_v(\boldsymbol{c}) = [1- \hat{m}(\boldsymbol{c})]/[1-\hat{\mu}_v(\boldsymbol{c})]$, we propose the following outcome regression-based estimator
\begin{equation} \label{outreg}
    \hat{\psi}^{O}_{mRR} = \frac{1}{n} \sum_{k=1}^{n} \hat{\mu}_v(\boldsymbol{c}_k ) \hat{\omega}_v(\boldsymbol{c}_k) \Biggm/ \frac{1}{n} \sum_{k=1}^{n} \hat{\mu}_{v_0}(\boldsymbol{c}_k) \hat{\omega}_{v_0}(\boldsymbol{c}_k),
\end{equation}
where the superscript $O$ is a shorthand for outcome regression, and the estimated outcome probabilities $\hat{\mu}_v(\boldsymbol{c}):=\hat{\mathbb{P}}_{TND}(Y=1\mid V=v,  \boldsymbol{C}  = \boldsymbol{c})$ can be obtained, for instance, by fitting a logistic regression of $Y$ conditional on $\boldsymbol{C}$, stratifying on or otherwise adjusting for $V=v$ using the data sampled under the TND. The numerator of the weights is based on an outcome regression model also fit using the TND data, but not adjusting for vaccination.  Alternatively, we may set $\hat{\omega}_v(\boldsymbol{c}) = \hat{\pi}_v(\boldsymbol{c})/\hat{\pi}_v^0(\boldsymbol{c})$,
such that the denominator $\hat{\pi}_v^0(\boldsymbol{c})$ is a propensity score estimate from an estimator using only the control data, whereas the numerator  $\hat{\pi}_v(\boldsymbol{c})$ is a propensity score estimate from an estimator using the whole TND-sampled data. Note that the denominators in $\omega_v$ must be positive; therefore, the nuisance functions of $\pi_v^0(\boldsymbol{c})$ and $[1 -\mu_v(\boldsymbol{c})]$ should be bounded away from zero. 

For inference, when parametric models (e.g., generalized linear models) are used for nuisance function estimation, we may use bootstrap or robust (Huber-White) sandwich estimation of the variance for both IPW and weighted outcome regression. For details on the derivation of the latter based on M-estimation theory\cite{stefanski2002calculus, saul2020calculus},  see Appendix \ref{Appx.A1} Section A.2.

\section{Methodology}\label{sec:meth}
In the context of TND sampling, to develop an efficient and doubly robust VE estimator that can integrate machine learning methods, we start by deriving the efficient influence function (EIF) for $\psi_v$ using the Gateaux derivative \citep{hines2022demystifying}. 
We then study the second-order remainder term of the von Mises expansion for $\psi_v$. Next, we provide an estimator (TNDDR) based on the derived EIF, establishing conditions that result in $\sqrt{n}$-consistency, asymptotic normality, and double robustness of the estimator. Lastly, for the efficient estimation of the VE under the TND, we give a double machine learning algorithm to implement the TNDDR estimator, which leverages data splitting and cross-fitting to maintain valid statistical inference for $\psi_{mRR}$ under machine learning to estimate the nuisance functions.
\subsection{Optimality theory for estimation of the estimand}   
The error of regular asymptotically linear estimators can be decomposed into two components: the mean influence function and a remainder term that becomes negligible at root-n rates. We start by deriving the EIF, which characterizes the efficiency bound in a local asymptotic minimax sense. Using the EIF, we can construct estimators with desirable properties \citep{fisher2021visually, kennedy2022semiparametric}. We define the operator $\mathbb{EIF}: \Psi \rightarrow L_2(\mathbb{P}_{TND})$ that maps target parameters (or functionals) $\psi(\mathbb{P}_{TND}): \mathcal{P}_{TND} \rightarrow \mathbb{R}$ to their EIFs, $\varphi(\boldsymbol{Z}) \in L_2(\mathbb{P}_{TND})$  in a nonparametric model. Adopting the specific form of $\omega(\boldsymbol{c}) = [1 - m(\boldsymbol{c})]/[1 - \mu_v(\boldsymbol{c})]$, the EIF for $\psi_v$ was obtained through the computation of the Gateaux derivative, employing techniques outlined in Kennedy (2022)\cite{kennedy2022semiparametric}. See Appendix Sections \ref{Appx.A2} and \ref{Appx.A3} for the proof of the following results.

\begin{lemma} \label{LEMMA1}
$\text{For} \  v \in \mathcal{V}$, the EIF 
for $\psi_v$ defined in Equation (\ref{tag1}) is 
\begin{align}
    \mathbb{EIF}(\psi_{v}) &= \varphi_v(\boldsymbol{Z}, \mathbb{P}_{TND}) \nonumber \\  & = \frac{\mathbb{I}(Y = 1, V=v)}{\pi^0_v(\boldsymbol{C})} - \mu_v(\boldsymbol{C})\left\{ \frac{ \mathbb{I}\left( Y=0, H=1\right) \left[\mathbb{I}(V=v) - \pi^0_{v}(\boldsymbol{C})\right] }{\pi^0_v(\boldsymbol{C})[1 - \mu_v(\boldsymbol{C})]}  \right\}  - \psi_{v}(\mathbb{P}_{TND}) \nonumber \\
    & = \frac{\mathbb{I}(Y = 1, V=v) - odds_{TND}(Y\mid V=v, \boldsymbol{C})\mathbb{I}\left( Y=0, H=1\right) \left[\mathbb{I}(V=v) - \pi^0_{v}(\boldsymbol{C})\right]}{\pi^0_v(\boldsymbol{C})} - \psi_{v}(\mathbb{P}_{TND}), \nonumber
\end{align}
where $odds_{TND}(Y\mid V=v, \boldsymbol{C}):=\mu_v(\boldsymbol{C})/[1-\mu_v(\boldsymbol{C})]$ is the odds under $\mathbb{P}_{TND}$ of $Y=1$ for $V=v$ given $\boldsymbol{C}$.
\end{lemma}

The expression of the EIF of $\psi_v$ indicates that the existence of this EIF in the nonparametric model not only requires that the propensity scores $\pi^0_v(\boldsymbol{C})$ are bounded away from zero  (a common identifiability condition in causal inference), but also that $[1 - \mu_v(\boldsymbol{C})]$ is bounded away from zero so that the odds $odds_{TND}(Y\mid V=v, \boldsymbol{C})$ are finite.

We computed the Gâteaux derivative under the assumption that the covariate variables $\boldsymbol{C}$ are discrete, following Strategy 1 in Kennedy (2022)\cite{kennedy2022semiparametric}. However, the results are valid with both continuous and discrete covariates. Since the derived Gâteaux derivative has finite variance in both cases, we conclude that our target parameter is pathwise differentiable, with the EIF as defined in Lemma \ref{LEMMA1}. Further, we investigated the von Mises-type expansion of $\psi_v$, which functions as a distributional counterpart to the Taylor expansion for real-valued functions. This exploration serves two primary purposes: (1) to validate the results presented in Lemma \ref{LEMMA1}, and (2) to examine the remainder term, as it plays a crucial role in characterizing the error of the EIF-based estimator (see Section \ref{secEst&Inf}). We begin by introducing notation to describe the nuisance functions under an arbitrary distribution $\bar{\mathbb{P}}_{TND}$ potentially different than $\mathbb{P}_{TND}$, and we define: $\bar{m}(\boldsymbol{c}) := \bar{\mathbb{P}}_{TND}(Y = 1 \mid \boldsymbol{C}  = \boldsymbol{c})$, similarly for $\bar{\pi}^0_v(\boldsymbol{c})$ and $\bar{\mu}_v(\boldsymbol{c})$. Then, we have the following lemma about the von Mises expansion of $\psi_v$.

\begin{lemma} \label{lemma2}
$\text{For} \  v \in \mathcal{V}$, the von Mises expansion or distributional Taylor expansion of $\psi_v$ is 
\begin{equation} \label{MiseExp}
    \psi_{v}(\mathbb{P}_{TND})=\psi_{v}(\bar{\mathbb{P}}_{TND})+\int \varphi_{v}(\bar{\mathbb{P}}_{TND}) d(\mathbb{P}_{TND}-\bar{\mathbb{P}}_{TND})+R^v_{2}(\bar{\mathbb{P}}_{TND}, \mathbb{P}_{TND}),
\end{equation} for distributions $\mathbb{P}_{TND}$ and $\bar{\mathbb{P}}_{TND}$, where $\varphi_v(\mathbb{P}_{TND})= \frac{\mathbb{I}(Y = 1, V=v)}{\pi^0_v(\boldsymbol{C})} - \mu_v(\boldsymbol{C})\left\{ \frac{ \mathbb{I}\left( Y=0, H=1\right) \left[\mathbb{I}(V=v) - \pi^0_{v}(\boldsymbol{C})\right] }{\pi^0_v(\boldsymbol{C})[1 - \mu_v(\boldsymbol{C})]}  \right\}  - \psi_{v}$ (derived in Lemma \ref{LEMMA1}),  and $$R^v_{2}(\bar{\mathbb{P}}_{TND}, \mathbb{P}_{TND}) = \int\frac{1}{\bar{\pi}^0_v(\boldsymbol{c})} \left\{ \mathcal{A}_v(\boldsymbol{c}) +  \mathcal{B}_v(\boldsymbol{c}) [1 - \bar{m}(\boldsymbol{c})]  \right\}  d\mathbb{P}_{TND},$$ 
with $\mathcal{A}_v(\boldsymbol{c}) := [\bar{\pi}^0_v(\boldsymbol{c}) - \pi^0_{v}(\boldsymbol{c})][m(\boldsymbol{c}) - \bar{m}(\boldsymbol{c})][odds_{v}(\boldsymbol{c}) - \overline{odds}_v(\boldsymbol{c})]$, and $\mathcal{B}_v(\boldsymbol{c}) := [\bar{\pi}^0_v(\boldsymbol{c}) - \pi^0_{v}(\boldsymbol{c})][odds_{v}(\boldsymbol{c}) - \overline{odds}_v(\boldsymbol{c})]$, in terms of $odds_v(\boldsymbol{c}) =  \mu_v(\boldsymbol{c})/[1 - \mu_v(\boldsymbol{c})]$ and $\overline{odds}_v(\boldsymbol{c}) =  \bar{\mu}_v(\boldsymbol{c})/[1 - \bar{\mu}_v(\boldsymbol{c})].$
\end{lemma}


The remainder term $R^v_{2}(\bar{\mathbb{P}}_{TND}, \mathbb{P}_{TND})$ in Equation (\ref{MiseExp}) is a second-order product of the nuisance function errors, hence the condition of Lemma 2 in Kennedy et al. (2022)\cite{kennedy2021semiparametric} is satisfied, i.e., $\left.\frac{d}{d \epsilon} R^v_{2}\left(\mathbb{P}_{TND}, \mathbb{P}_{{TND}, \epsilon}\right)\right|_{\epsilon=0}=0$ (equivalent to Neyman orthogonality \cite{chernozhukov2018double}), for the parametric submodel $\mathcal{P}_{TND,\epsilon} = \{\mathbb{P}_{TND,\epsilon}: \epsilon \in \mathbb{R}\}\subseteq \mathcal{P}_{TND}$ with $\mathbb{P}_{TND, \epsilon=0}=\mathbb{P}_{TND}$. 
We see that term $\mathcal{A}_v(\boldsymbol{c})$ within $R^v_{2}(\bar{\mathbb{P}}_{TND}, \mathbb{P}_{TND})$ encompasses three nuisance functions: $\pi^0_v(\boldsymbol{c})$, $m(\boldsymbol{c})$, and $\mu_v(\boldsymbol{c})$, while term $\mathcal{B}_v(\boldsymbol{c})$ includes $\pi^0_v(\boldsymbol{c})$ and $\mu_v(\boldsymbol{c})$. The accuracy of estimation of these nuisance functions will determine the second-order errors, as elaborated upon in the following subsection.

\subsection{Efficient estimation and statistical inference of the estimand} \label{secEst&Inf}
For the estimation and inference of $\psi_v$, considering TND sampling distribution, we will employ the following standard notation. For a vector $X_n$, let $X_n = O_{\mathbb{P}_{TND}}(r_n)$ denote that $X_n/r_n$ is bounded in probability, and $X_n = o_{\mathbb{P}_{TND}}(r_n)$ denote that $X_n/r_n$ converges to zero in probability, with respect to the observed data distribution $\mathbb{P}_{TND}$. The squared $L_2(\mathbb{P}_{TND})$ norm of a function $f$ is denoted as $\|f\|^2:=\int[f(\boldsymbol{z})]^2 d \mathbb{P}_{TND}(\boldsymbol{z})$. The empirical measure is denoted as $\mathbb{P}_{TND,n}$, and thus the TND sample averages are written as $\mathbb{P}_{TND, n}(f) = \mathbb{P}_{TND, n}[f(\boldsymbol{Z})]= \frac{1}{n}\sum_{k=1}^{n}f(\boldsymbol{Z}_k)$.

We now present an estimator derived from the EIF of $\psi_v$, and we proceed to explore its properties as outlined in the subsequent theorem and corollaries. 
\begin{theorem} \label{Them1}
Define a one-step estimator for $\psi_v$ as $\hat{\psi}_v:= \mathbb{P}_{TND, n}[\phi_v(\boldsymbol{Z}; \hat{\pi}^{0}_v, \hat{\mu}_v)],$ where 
\begin{align} \label{phi}
    \phi_v(\boldsymbol{Z}; \hat{\pi}^{0}_v, \hat{\mu}_v) =  \frac{\mathbb{I}(Y = 1, V=v)}{\hat{\pi}^0_v(\boldsymbol{c})} - \hat{\mu}_v(\boldsymbol{c}) \frac{ \mathbb{I}\left( Y=0, H=1\right) \left[\mathbb{I}(V=v) - \hat{\pi}^0_{v}(\boldsymbol{c})\right] }{\hat{\pi}^0_v(\boldsymbol{c})[1 - \hat{\mu}_v(\boldsymbol{c})]}. 
\end{align}
Assuming control exchangeability and that the following  assumptions are satisfied: \\
{\bf{A1:}} $\phi_{v}(\boldsymbol{Z}; \hat{\mathbb{P}}_{TND})$ converges in probability to $\phi_{v}(\boldsymbol{Z}; \mathbb{P}_{TND})$ in $L_2(\mathbb{P}_{TND})$ norm, i.e.,
\begin{equation*}
    \left\|\phi_{v}(\boldsymbol{Z}; \hat{\mathbb{P}}_{TND})-\phi_{v}(\boldsymbol{Z}; \mathbb{P}_{TND})\right\|:= \sqrt{\int\left\{\phi_{v}(\boldsymbol{Z}; \hat{\mathbb{P}}_{TND})-\phi_{v}(\boldsymbol{Z}; \mathbb{P}_{TND})\right\}^2 d \mathbb{P}_{TND}}=o_{\mathbb{P}_{TND}}(1);
\end{equation*}
{\bf{A2:}} Nuisance functions ${\pi}^{0}_v, {\mu}_v$ are estimated under sample-splitting and cross-fitting; \\
for some $ \epsilon \in (0, 1)$, {\bf{A3:}} Strong overlap: $\mathbb{P}_{TND}\left(\epsilon<\pi^0_{v}(\boldsymbol{c})\right)=1 \text { and } \mathbb{P}_{TND}\left(\epsilon<\hat{\pi}^0_{v}(\boldsymbol{c})\right)=1$; and
{\bf{A4:}} Boundedness of the conditional outcome probability:  $\mathbb{P}_{TND}\left( \mu_v(\boldsymbol{c}) < 1 - \epsilon\right)=1$ and $\mathbb{P}_{TND}\left( \hat{\mu}_v(\boldsymbol{c}) < 1 - \epsilon\right)=1,$\\
then the estimator $\hat{\psi}_v= \mathbb{P}_{TND,n}[\phi_v(\boldsymbol{Z}; \hat{\pi}^{0}_v, \hat{\mu}_v)]$ satisfies
\begin{align} \label{psikey}
    \hat{\psi}_v - \psi_v  =  \left|R^v_2(\hat{\mathbb{P}}_{TND}, \mathbb{P}_{TND})\right| +  \left(\mathbb{P}_{TND,n} - \mathbb{P}_{TND}\right)\left[\varphi_v(\boldsymbol{Z};\mathbb{P}_{TND})\right] + o_{\mathbb{P}_{TND}}\left(1/\sqrt{n}\right),
\end{align} where
\begin{align}\label{r_2bdd}
\begin{split}
    \left|R^v_2(\hat{\mathbb{P}}_{TND}, \mathbb{P}_{TND})\right|&=O_{\mathbb{P}_{TND}}\left(\left\|\hat{\pi}^{0}_{v}-\pi^{0}_{v}\right\|\left\|\hat{m}- m\right\|\left\|\widehat{odds}_{v}-odds_{v} \right\| +   \left\|\hat{\pi}^{0}_{v}-\pi^{0}_{v}\right\|\left\|\widehat{odds}_{v}-odds_{v}\right\| \right),
\end{split}
\end{align} 
and $\widehat{odds}_v(\boldsymbol{c}) =  \hat{\mu}_v(\boldsymbol{c})/[1 - \hat{\mu}_v(\boldsymbol{c})]$ and $\hat{m}(\boldsymbol{c})$ is implicitly estimated through summing over $\mathbb{I}(Y=0)$ and using an empirical distribution for $\boldsymbol{C}$. 
\end{theorem}
\textbf{Proof:} See Appendix Section \ref{Appx.A4} in supplementary materials.

Note that the third term in the decomposition in Equation (\ref{psikey}) stems from the empirical process term $ ( \mathbb{P}_{TND, n} -\mathbb{P}_{TND})(\phi_v(\boldsymbol{Z};\hat{\mathbb{P}}_{TND}) - \phi_v(\boldsymbol{Z};\mathbb{P}_{TND}))$ and is of order $o_{\mathbb{P}_{TND}}(1/{\sqrt{n}})$, since under A1 and A2, it is asymptotically negligible (Lemma 2 in Kennedy et al. (2020)\cite{kennedy2020sharp}), i.e.,
$$( \mathbb{P}_{TND, n} -\mathbb{P}_{TND})(\phi_v(\boldsymbol{Z};\hat{\mathbb{P}}_{TND}) - \phi_v(\boldsymbol{Z};\mathbb{P}_{TND})) = O_{\mathbb{P}_{TND}}\left(\frac{\left\|\phi_{v}(\boldsymbol{Z}; \hat{\mathbb{P}}_{TND})-\phi_{v}(\boldsymbol{Z}; \mathbb{P}_{TND})\right\|}{\sqrt{n}}\right) = o_{\mathbb{P}_{TND}}\left(\frac{1}{\sqrt{n}}\right).$$

Building on the result in Equation (\ref{r_2bdd}), we further investigate the conditions regarding the convergence rates of the propensity score and conditional expectation of the outcome estimators to analyze the second-order remainder term. The results are presented in the following two corollaries. 
\begin{corollary} \label{cor1}
\textbf{$\sqrt{n}$-consistency and asymptotic normality} \\
Under the assumptions in Theorem \ref{Them1} and the following two conditions: \\
1. $\left\|\hat{\pi}^0_{v}-\pi^0_{v}\right\|=o_{\mathbb{P}_{TND}}\left(n^{-1 / 4}\right);$ 2. $\left\|\hat{\mu}_{v}-\mu_{v}\right\|=O_{\mathbb{P}_{TND}}\left(n^{-1 / 4}\right);$ \\
then, the proposed estimator $\hat{\psi}_v$ is $\sqrt{n}$-consistent and asymptotically normal, and the limiting distribution is $\sqrt{n}\left(\hat{\psi}_{v}-\psi_{v}\right) \leadsto \mathcal{N}\left(0, \operatorname{var}\left(\varphi_v(\boldsymbol{Z}) \right)\right),$ where $\leadsto$ denotes convergence in distribution and $\operatorname{var}\left(\varphi_v(\boldsymbol{Z}) \right) = \mathbb{E}\left(\varphi_v^2(\boldsymbol{Z}) \right)$.
\end{corollary}
First, under the assumptions in Theorem \ref{Them1} and these two conditions, the second-order remainder term $R^v_{2}(\bar{\mathbb{P}}_{TND},\mathbb{P}_{TND})$ is $o_{\mathbb{P}_{TND}}\left(n^{-1 / 2}\right)$. Primarily, we show that $\left\|\widehat{odds}_{v}-odds_{v}\right\|=O_{\mathbb{P}_{TND}}\left(n^{-1 / 4}\right),$ if $\left\|\hat{\mu}_{v}-\mu_{v}\right\|=O_{\mathbb{P}_{TND}}\left(n^{-1 / 4}\right)$ (see Appendix Section \ref{Appx.A5} for proof of this claim). Therefore, if we have $\left\|\hat{\pi}^0_{v}-\pi^0_{v}\right\|=o_{\mathbb{P}_{TND}}\left(n^{-1 / 4}\right)$ and $\left\|\widehat{odds}_{v}-odds_{v}\right\|=O_{\mathbb{P}_{TND}}\left(n^{-1 / 4}\right),$ then the product term is $o_{\mathbb{P}_{TND}}(1/\sqrt{n})$ and asymptotically negligible. Thus, the rate of $\left\|\hat{\pi}^{0}_{v}-\pi^{0}_{v}\right\|\left\|\hat{m}- m\right\|\left\|\widehat{odds}_{v}-odds_{v} \right\|$ is at worst $o_{\mathbb{P}_{TND}}\left(n^{-1 / 2}\right)$ when $\left\|\hat{m}-m\right\|=O_{\mathbb{P}_{TND}}\left(1\right)$ (i.e., if the asymptotic bias of $\hat{m}$ is finite). In fact, within our estimator $m(\boldsymbol{c})$ is implicitly estimated nonparametrically using an empirical mean over $\mathbb{I}(Y=0)$ and so $\left\|\hat{m}-m\right\|$ is $o_{\mathbb{P}_{TND}}(1)$ (i.e., consistent). 
Note that for the product of $\left\|\hat{\pi}^0_{v}-\pi^0_{v}\right\|$ and $\left\|\widehat{odds}_{v}-odds_{v}\right\|$ to be $o_{\mathbb{P}_{TND}}(1/\sqrt{n})$, it is not necessary for $\pi^0_{v}$ and $\mu_{v}$ to be estimated at the same rates; any such combination whose product is $o_{\mathbb{P}_{TND}}(1/\sqrt{n})$ would be sufficient. 

Then, if the second-order remainder term is negligible (i.e., of order $o_{\mathbb{P}_{TND}}(1/{\sqrt{n}})$), the middle term $(\mathbb{P}_{TND,n} - \mathbb{P}_{TND})(\varphi_v(\boldsymbol{Z};\mathbb{P}))$ in Equation (\ref{psikey}), which is the sample average of a fixed function, behaves as a normally distributed random variable with variance $\operatorname{var}\left(\varphi_v(\boldsymbol{Z}) \right)/n$, up to error $o_{\mathbb{P}_{TND}}(1/\sqrt{n})$ according to the classical central limit theorem (CLT). Based on these arguments and  under the conditions in Corollary \ref{cor1}, only the sample average term $(\mathbb{P}_{TND,n} - \mathbb{P}_{TND})(\varphi_v(\boldsymbol{Z};\mathbb{P}_{TND}))$ dominates the decomposition in Equation~(\ref{psikey}) of Theorem \ref{Them1}. Applying the CLT and Slutsky's theorem, we can conclude that
\begin{equation*}
    \sqrt{n}\left(\hat{\psi}_{v}-\psi_{v}\right) = \sqrt{n}\left(\mathbb{P}_{TND,n} - \mathbb{P}_{TND}\right)\left[\varphi_v(\boldsymbol{Z};\mathbb{P}_{TND})\right] + o_{\mathbb{P}_{TND}}\left(1\right)\ \leadsto \mathcal{N}\left(0, \operatorname{var}\left(\varphi_v(\boldsymbol{Z}) \right)\right).
\end{equation*}
As a result, the proposed estimator $\hat{\psi}_v$ is $\sqrt{n}$-consistent and asymptotically normal, and a closed-form expression $\hat{\psi}_v \pm \frac{1}{\sqrt{n}}\Phi^{-1}(1 - \frac{\alpha}{2})\sqrt{\hat{\operatorname{var}}\left(\hat{\varphi}_v(\boldsymbol{Z}) \right)}$ provides asymptotically valid $\alpha-$level confidence intervals for $\psi_v$, where $\Phi(\cdot)$ is the standard Gaussian CDF, and an estimator of the variance of $\hat{\psi}_v$ is
\begin{equation*} \label{varpsiv}
\hat{\operatorname{var}}\left(\hat{\varphi}_v(\boldsymbol{Z}) \right)=      \mathbb {P}_{TND,n}(\hat{\varphi}_v^2) = \mathbb{P}_{TND,n}\left\{  \frac{\mathbb{I}(Y = 1, V=v)}{\hat{\pi}^0_v(\boldsymbol{c})} - \hat{\mu}_v(\boldsymbol{c}) \frac{ \mathbb{I}\left( Y=0, H=1\right) \left[\mathbb{I}(V=v) - \hat{\pi}^0_{v}(\boldsymbol{c})\right] }{\hat{\pi}^0_v(\boldsymbol{c})[1- \hat{\mu}_v(\boldsymbol{c})]}  \right\}^2.
\end{equation*}

As shown by Corollary \ref{cor1} and Theorem \ref{Them1}, the proposed estimator has second-order errors,  bounded by the sum of the product of nuisance estimation errors, leading to root-n rates. That is, we can get a faster rate for our estimator even when we estimate the nuisance functions at slower nonparametric rates. We can thus employ flexible machine learning techniques to nonparametrically estimate the nuisance functions under smoothness or sparsity assumptions, for instance, at $n^{-1/4}$ rates to obtain $n^{-1/2}$ rates for our estimator.

\begin{corollary} \label{cor2} 
\textbf{Double Robustness} Under the assumptions in Theorem \ref{Them1}, the proposed estimator $\hat{\psi}_v$ is doubly robust. That is, $\hat{\psi}_v$ is consistent, i.e., $\left\|\hat{\psi}_v - \psi_v\right\| = o_{\mathbb{P}_{TND}}(1)$, if either the propensity score is consistently estimated, i.e., $\left\|\hat{\pi}^0_{v}-\pi^0_{v}\right\| = o_{\mathbb{P}_{TND}}(1)$ and/or the treatment-specific conditional expectation of the outcome is, i.e., $\left\|\hat{\mu}_{v}-\mu_{v}\right\| =o_{\mathbb{P}_{TND}}(1)$.
\end{corollary}
Appendix Section \ref{Appx.proofdr} provides the proof of this result. Essentially,  the second-order remainder term $|R^v_{2}(\bar{\mathbb{P}}, \mathbb{P})| =o_{\mathbb{P}_{TND}}(1)$ if either  $\left\|\hat{\pi}^0_{v}-\pi^0_{v}\right\| = o_{\mathbb{P}_{TND}}(1)$ and/or $\left\|\hat{\mu}_{v}-\mu_{v}\right\| =o_{\mathbb{P}_{TND}}(1)$, which means that correctly specifying the propensity score and/or outcome models leads to the elimination of the second-order remainder term.

\subsection{Estimation and inference of the estimand mRR and proposed method of VE estimation}
In this subsection we present an efficient double machining learning (DML)\cite{chernozhukov2018double} estimator and variance estimate for the target parameter $\psi_{mRR}$. We apply the functional delta method to obtain the von Mises expansion for $\psi_{mRR} = \text{exp}(\ln(\psi_{v}) - \ln(\psi_{v_0}))$\cite{van2000asymptotic}. We propose a related one-step estimator (denoted as $\hat{\psi}^{eif}_{mRR}$) and explore its properties, including $\sqrt{n}$-consistency, asymptotic normality, and double robustness, in a theorem and a corresponding corollary in Appendix Section \ref{mrrThm}.

\begin{algorithm}[ht] 
\begin{itemize}
    \item
{\textbf{Step 1:}} Randomly split the observed TND data $\boldsymbol{Z}_{1,...n} = (\boldsymbol{Z}_1, ..., \boldsymbol{Z}_n)$ into $J \geq 2$ disjoint, evenly sized folds such that the size of each fold is 
 integer $n/J$. Index the folds by $j = 1, 2,..., J$ and let $Q_{(j)}$ represent the participant indices within the $j^{th}$ fold; then, the whole TND data can be denoted as $(\boldsymbol{Z}_1, ..., \boldsymbol{Z}_n) = (\boldsymbol{Z}_{Q_{(j)}})_{j =1}^{J}$. Denote $Q_{(-j)}$ as the set of participant indices that are not included in $Q_{(j)}$, i.e., $Q_{(-j)} = \{1, 2, ..., n \} \backslash Q_{(j)}$ for $j = 1, 2,..., J$.\\

\item {\textbf{Step 2:}} For each $j = 1,2,..., J$, using the data that excludes the $j$-th fold, i.e., data in $\boldsymbol{Z}_{Q_{(-j)}}$, use machine learning methods (e.g., highly-adaptive lasso or deep neural networks) to estimate the nuisance functions $\pi_{v}^{0}(\cdot)$ and $\mu_{v}^{}(\cdot)$, whose estimators are denoted  $\hat{\pi}_{v}^{0}(\cdot;\boldsymbol{Z}_{Q_{(-j)}})$ and $\hat{\mu}_{v}(\cdot;\boldsymbol{Z}_{Q_{(-j)}})$, respectively.\\

\item {\textbf{Step 3:}} Construct the double machine learning VE estimator i.e., $\widehat{\textit{VE}} = 1 - \hat{\psi}^{eif}_{mRR}= 1 -\hat{\psi}_v/\hat{\psi}_{v_0}$ with the nuisance components estimated in Step 2, where for each $v$ (i.e., $v$ and $v_0 \in \mathcal{V}$), $\hat{\psi}_v =\frac{1}{n} \sum_{j=1}^{J}\sum_{k \in Q_{(j)}} \hat{\phi}_{v}\left[\boldsymbol{Z}_{Q_{(j)}}; \hat{\pi}^{0}_{v}(\cdot;\boldsymbol{Z}_{Q_{(-j)}}), \hat{\mu}_{v}(\cdot;\boldsymbol{Z}_{Q_{(-j)}}) \right], $ with
\begin{equation*} \label{eifest2}
     \hat{\phi}_{v}\left[\boldsymbol{Z}_{Q_{(j)}}; \hat{\pi}^{0}_{v}(\cdot;\boldsymbol{Z}_{Q_{(-j)}}), \hat{\mu}_{v}(\cdot;\boldsymbol{Z}_{Q_{(-j)}}) \right]  = \frac{\mathbb{I}(Y_k = 1, V_k=v)}{\hat{\pi}^{0}_v(\boldsymbol{c}_k;\boldsymbol{Z}_{Q_{(-j)}})} -  \frac{ \hat{\mu}_v(\boldsymbol{c}_k;\boldsymbol{Z}_{Q_{(-j)}}) \mathbb{I}\left( Y_k=0, H_k=1\right) \left[\mathbb{I}(V_k=v) - \hat{\pi}^{0}_{v}(\boldsymbol{c}_k;\boldsymbol{Z}_{Q_{(-j)}})\right] }{\hat{\pi}^{0}_v(\boldsymbol{c}_k;\boldsymbol{Z}_{Q_{(-j)}})[1 - \hat{\mu}_v(\boldsymbol{c}_k;\boldsymbol{Z}_{Q_{(-j)}})]}.
\end{equation*}
An approximate $(1-\alpha) \times 100 \%$ Wald-type confidence interval of VE is
\begin{equation*}
    \mathrm{CI}_n(\textit{VE}):=1 - \left[ \hat{\psi}^{eif}_{mRR} \mp  \Phi^{-1}\left(1-\frac{\alpha}{2}\right)  \sqrt{\hat{\sigma}^2/n}\right],
\end{equation*}
where $\hat{\sigma}^2=\hat{\kappa}^{-2} \frac{1}{nJ} \sum_{j=1}^{J}\sum_{k \in Q_{(j)}} S\left(\boldsymbol{Z}_{Q_{(j)}}; \hat{\psi}^{eif}_{mRR}, \hat{\phi}_{v}, \hat{\phi}_{v_0}\right)^2,$ \text { with the linear score function for} $\psi_{mRR}$
\begin{equation*}
S\left(\boldsymbol{Z}_{Q_{(j)}}; \hat{\psi}^{eif}_{mRR}, \hat{\phi}_{v}, \hat{\phi}_{v_0}\right) = \hat{\psi}_{mRR}\hat{\phi}_{v_0} - \hat{\phi}_{v}, \text{and}\  \ \hat{\kappa}=\frac{1}{nJ} \sum_{j=1}^{J}\sum_{k \in Q_{(j)}}     \hat{\phi}_{v_0}\left[\boldsymbol{Z}_{Q_{(j)}}; \hat{\pi}^{0}_{v_0}(\cdot;\boldsymbol{Z}_{Q_{(-j)}}), \hat{\mu}_{v_0}(\cdot;\boldsymbol{Z}_{Q_{(-j)}}) \right].  
\end{equation*}
\end{itemize}
\caption{TNDDR (TND Doubly Robust) method for VE estimation under TND}\label{AlgTNDDR}
\end{algorithm}

Our DML VE estimation procedure under the TND, referred to as TNDDR (TND Doubly Robust), is summarized in the three-step Algorithm \ref{AlgTNDDR}. The first step involves implementing sample splitting for the cross-fitting. In Step 2, we apply machine learning to estimate the nuisance functions $\pi_{v}^{0}(\cdot)$ and $\mu_{v}^{}(\cdot)$, for instance, using the highly-adaptive lasso (HAL)\citep{benkeser2016highly}. HAL is a supervised learning approach that satisfies the convergence requirements in Corollary \ref{cor1}, provided that the nuisance functions are càdlàg with a finite sectional variation norm, and that belongs to a Donsker class \citep{bibaut2019fast, van2000asymptotic}. 
Alternatively, multivariate adaptive regression splines (MARS) \citep{friedman1991multivariate}, deep neural networks (NN), or random forests may also be employed; however, it is not guaranteed that the latter two methods will achieve the desired convergence rate. 


In Step 3 of the TNDDR algorithm,  we estimate the marginal risk ratio by $\hat{\psi}^{eif}_{mRR} = \hat{\psi}_{v}/\hat{\psi}_{v_0}$, which is a doubly robust and efficient estimator (see Appendix Section \ref{mrrThm}). For inference, we employ an estimator of the asymptotic variance based on the results of DML (see Theorem 3.2. in Chernozhukov et al.\cite{chernozhukov2018double}) as follows. Considering the linear score function for $\psi_{mRR}$ given by $S(\boldsymbol{Z}, \psi_{mRR}) = \psi_{mRR}\phi_{v_0} - \phi_{v}$ where $\phi_{v}$ is defined in Equation (\ref{phi}) for both $v$ and $v_0$, we have
\begin{equation}\label{CIeif}
\hat{\sigma}^2=\hat{\kappa}^{-2} \frac{1}{nJ} \sum_{j=1}^{J}\sum_{k \in Q_{(j)}} S\left(\boldsymbol{Z}_{Q_{(j)}}; \hat{\psi}^{eif}_{mRR}, \hat{\phi}_{v}, \hat{\phi}_{v_0}\right)^2, \text { with } \hat{\kappa}=\frac{1}{nJ} \sum_{j=1}^{J}\sum_{k \in Q_{(j)}}     \hat{\phi}_{v_0}\left[\boldsymbol{Z}_{Q_{(j)}}; \hat{\pi}^{0}_{v_0}(\cdot;\boldsymbol{Z}_{Q_{(-j)}}), \hat{\mu}_{v_0}(\cdot;\boldsymbol{Z}_{Q_{(-j)}}) \right].  
\end{equation}

As demonstrated in Corollary 3.1 of Chernozhukov et al.\cite{chernozhukov2018double}, the confidence interval based on  $\hat{\sigma}^2$ provides uniformly valid confidence bands under regularity conditions. Alternatively, one can use a normal approximation to construct confidence intervals for the natural logarithm of $\psi_{mRR}$, employing the EIF of $\ln(\psi_{v}/\psi_{v_0})$ (see Appendix subsection \ref{App:Alt} for details). Notably, the double machine learning estimator in the TNDDR algorithm solves an aggregated estimating equation with all $J$ folds, and thus is a DML2-type estimator as presented in Chernozhukov et al\cite{chernozhukov2018double}, and an asymptotically equivalent DML1-type estimator, which is an average of $J$ estimators
from all $J$ folds, can also be constructed analogously.

\section{Simulation Studies} \label{sec:sim}
We conducted three simulation studies to evaluate the performance of the proposed estimator.  Study 1 assessed the $\sqrt{n}$-consistency of TNDDR under cross-fitting (sample-splitting) and using different machine learning methods for nonparametric estimation of the nuisance functions. Study 2 investigated the double robustness of the proposed estimator. In both studies, we validated the coverage of the $95\%$ confidence intervals of TNDDR with either machine learning or correctly specified parametric models. These first two studies investigated scenarios where control exchangeability approximately holds, with a low co-infection rate. Study 3 examined the robustness under settings with differing co-infection rates where control exchangeability does not always approximately hold.

\subsection{Basic settings, data generation, and estimators}
We simulated TND studies with confounders and binary vaccination, infection, symptoms due to infection, and hospitalisation. Building on the basic simulation settings in Schnitzer (2022)\cite{schnitzer2022estimands}, and the DAG (a) in Figure \ref{TNDdag}, we first generated a large population with a continuous confounder $C$, and two unmeasured binary variables $U_1$ and $U_2$. Both $U_1$ and $U_2$ affected the probability of contracting the SARS-CoV-2 infection, experiencing COVID-related symptoms, and needing hospitalisation if symptoms were present. $U_1$ also affected the probability of contracting other infections and symptoms due to other infections. Note that disease symptoms for SARS-CoV-2 infection and other infections were generated using distinct models with some common covariates. Only those with disease symptoms were candidates for hospitalisation. Vaccination provided protection against SARS-CoV-2 infection, disease symptoms, and hospitalisation. Specifically, the data generation mechanism for the two studies is as follows. The baseline confounder was generated as a uniform $C \sim U(0.1,3)$. The unmeasured covariates $U_1$ and $U_2$ were independently generated as Bernoulli distribution with a success probability $0.5$. The vaccination $V$ was also generated from a Bernoulli distribution with a success probability $[1 + \text{exp}\{-\lambda_v(\cdot)\}]^{-1}$, where $\lambda_v(\cdot) = 0.25+0.75C - 0.5\log(C) - 1.25\sin(\pi C)$ depended on confounder $C$. Infection with other viruses ($I_1$) was generated from a Bernoulli distribution with a success probability of $\text{expit}(\beta_{I_1}+ 0.35C + 6.5U_1),$ whereas infection with SARS-CoV-2 ($I_2$) was generated from a Bernoulli with a success probability $[1 + \text{exp}\{-\lambda_{COVID}(\cdot)\}]^{-1}$, where $\lambda_{COVID}(\cdot) = \beta_{I_2} + 0.15C + 0.5\exp(C)[1 + 0.15\cos(C)] -\text{log}(3)V  + \beta_{em}V\times C+\text{log}(1.2)U_2(1.5-V)-2U_1$. Note that the coefficient $\beta_{em}$ of the interaction term between $C$ and $V$ affects the vaccination effect modification, which represents a distinct individual effect for each patient based on the individual value of $C$. Parameters $\beta_{I_1}$ and $\beta_{I_2}$ represent baseline infection rates. Setting $\beta_{I_1} = \beta_{I_2} = -11.5$ in Studies 1 and 2 produces low co-infection rates so that the control exchangeability assumption holds approximately. Study 3 varies the co-infection rates to create scenarios which do not satisfy the control exchangeability assumption. Furthermore, disease symptoms due to other viruses ($W_1$) and to SARS-CoV-2 ($W_2$) were also independently generated as Bernoulli, with probabilities $[1 + \text{exp}\{-\lambda_{w_1}(\cdot)\}]^{-1}$ and $[1 + \text{exp}\{-\lambda_{w_2}(\cdot)\}]^{-1}$, respectively. For $W_1$, $\lambda_{w_1}(\cdot)$ was set to $-0.5 + 0.5C -0.5U_1$ and only generated among those with $I_1 = 1$. For $W_2$, $\lambda_{w_2}(\cdot)$ was set to $-3.75 + 2C- \log(2.5)V - U_1 + 0.5U_2(1 - V)$ and only generated among those with $I_2 = 1$. A  disease symptoms variable ($W$) was generated to take the value 1 if $W_1 = 1$ or $W_2 = 1$ and 0 if $W_1 = W_2 = 0$. Then, hospitalisation for symptoms (when $W=1$), in ignorance of the type of infection, was generated from a Bernoulli with a success probability of $[1 + \text{exp}\{-[-1.5 + 0.5C - 0.5U_1]\}]^{-1}$. Finally, according to the design, TND samples were generated by randomly selecting hospitalized individuals. 

In all studies, three different estimators were applied: (1) the IPW estimator \citep{schnitzer2022estimands}, (2) the outcome regression estimator in Equation (\ref{outreg}) based on the outcome probability ratio debiasing weights ($\hat{\psi}^{O}_{mRR}$), and (3) the proposed EIF-based estimator TNDDR with cross-fitting. The nonparametric percentile bootstrap was used to obtain confidence intervals for both the IPW and the outcome regression estimators. For TNDDR, we directly computed confidence intervals based on Equation (\ref{CIeif}), also presented in the TNDDR algorithm. 

\subsection{Assessing the rootn-consistency of TNDDR}
With the goal of assessing the efficiency of TNDDR, in Study 1, we set $\beta_{em} = 0.25$ and generated 500 datasets of size $n = 1000,$ $4000$ and $8000$, where the case percentage in the TND samples was between $58\%$ and $64\%$. The true marginal risk ratio was $\psi_{mRR} = 0.507$.  For all estimators, the nuisance models were estimated using three different machine learning methods: (1) MARS (implemented in the \texttt{R} package \texttt{earth}\cite{milborrow2017earth}), (2) HAL (implemented in package \texttt{hal9001}\cite{hejazi2020hal9001}) and (3) NN (package \texttt{nnet}\cite{ripley2016feed}, and set one hidden layer containing 5 hidden neurons), employing cross-fitting as detailed in algorithm \ref{AlgTNDDR}, with a value of $J = 2$. In addition, to investigate the empirical convergence of the proposed estimators, we estimated the marginal risk ratio with different sample sizes between 300 and 10000, increasing by 200 at each step, with 200 replications for each sample size.

\begin{figure}[ht]\centering
    \begin{subfigure}[ht]{1\textwidth}
        \centering
        \includegraphics[height=3.15in, bb=0 0 700 500]{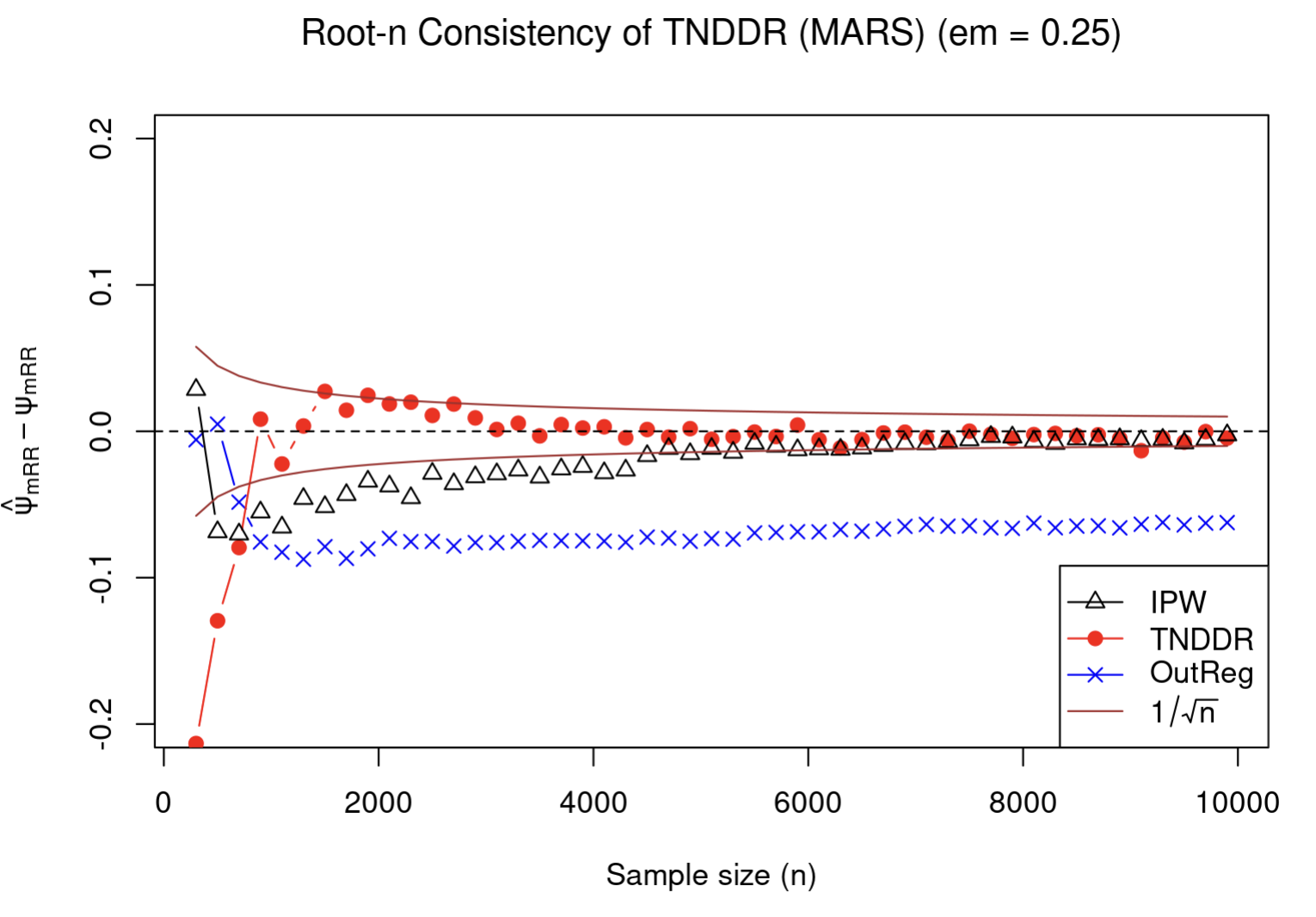}
    \end{subfigure}
        \qquad
    \begin{subfigure}[ht]{1\textwidth}
        \centering
        \includegraphics[height=3.2in, bb=0 0 700 500]{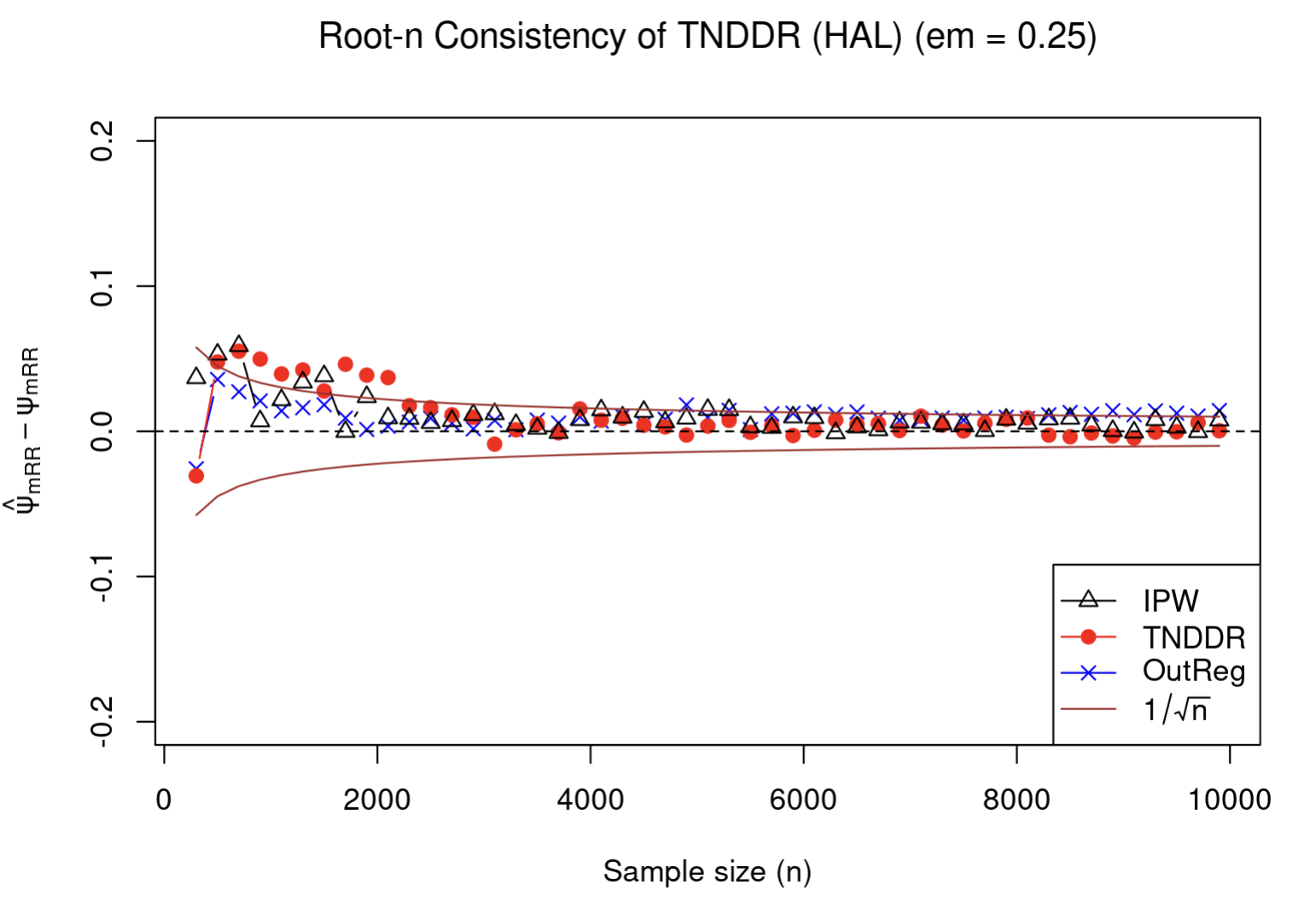}
    \end{subfigure}
    \caption{Simulation results from 200 simulations with varying sample sizes (from 300 to 10000 stepped by 200) under the setting that $\beta_{em} = 0.25,$ and truth is $\psi_{mRR} = 0.507$ with approximate $0.05\%$ the co-infection rate in the TND
sample. For all the estimators, the top figure uses MARS, while the bottom figure uses HAL (see Table \ref{rootNhal} in Appendix \ref{Appx.sim} for the NN results). }
\label{P3study2S2}
\end{figure}

Table \ref{tab:1} presents the performance results of all estimators in terms of bias, Monte Carlo mean squared error (MC MSE), Monte Carlo standard error (MC SE), and the coverage probability of $95\%$ confidence intervals (TNDDR's Wald-style CI based on Equation (\ref{CIeif}), and percentile confidence intervals for the IPW and outcome regression estimators, computed using nonparametric bootstrap). Compared to both the IPW and the outcome regression estimator, regardless of the ML method used, TNDDR estimators always demonstrated lower bias and MC SE, more stable MC MSE and higher coverage probability. The IPW estimator with NN and the outcome regression estimator had relatively larger bias. Contrasting machine learning methods, TNDDR (MARS) exhibited the smallest bias in two sample size scenarios, and TNDDR (HAL) achieved the lowest MC MSE in all scenarios. Furthermore, as anticipated, with these three machine learning methods, TNDDR achieved the desired coverage probability. Figures \ref{P3study2S2} and \ref{rootNhal} illustrate the average error of each estimator across sample sizes (from 300 to 10000 in increments of 200), with lines of $1/\sqrt{n}$ used as benchmarks.  TNDDR converged at parametric rates when implemented with any of the three machine learning methods; however, IPW with NN and the outcome regression method with MARS, both of which are single-robust estimators, did not.

\begin{table}[ht] \centering
    \caption{Results of Simulation Study 1: 500 simulations for sample sizes of $n = 1000,$ $4000$ and $8000$ with case percentages in the TND samples ranging from  $58 \% \sim 64\%$.}
    \begin{threeparttable}[t]
 \begin{tabular}{ccccccc} 
        \hline
       \multirow{3}{*}{$n$} & Scenarios $\&$ Truth & \multicolumn{5}{c}{$\beta_{em} = 0.25$; $\psi_{mRR} = 0.507$; Co-infection (TND): $\sim 0.05\%$}   \\
        \cmidrule{3-7} 
       & Methods & IPW (MARS) &  OutReg (MARS)  & TNDDR (MARS) & TNDDR (HAL) & TNDDR (NN)   \\
        \hline
        \multirow{3}{*}{$1000$} & Median bias   & -0.058 & -0.074  & -0.006 &  0.011 & -0.037   \\
        & MC MSE  & 0.032 & 0.030  & 0.033 & 0.022&  0.062    \\
        & MC SE  & 0.008 & 0.008  & 0.009 & 0.005 &  0.012    \\
        & $\% Cov_{mRR}$ & 94.1 & 96.0  & 97.6  & 96.8 & 98.2   \\
          \hline
        \multirow{3}{*}{$4000$} & Median bias   & -0.023 & 0.076  & 0.006 &  0.008 & 0.015   \\
        & MC MSE  & 0.011 & 0.009  & 0.010 & 0.006&  0.009     \\
        & MC SE  & 0.004 & 0.003  & 0.003 & 0.003&  0.004    \\
        & $\% Cov_{mRR}$ & 89.2 & 61.2  & 96.4  & 96.7 &  96.2   \\
          \hline
        \multirow{3}{*}{$8000$} & Median bias   & -0.016 & -0.067  & -0.005 &  0.003 & 0.011    \\
        & MC MSE  & 0.017 & 0.006  & 0.008 & 0.004&  0.004    \\
        & MC SE  & 0.006 & 0.002  & 0.003 & 0.002&  0.002  \\
        & $\% Cov_{mRR}$ & 87.6 & 46.8  & 96.8 & 97.2 &  92.8  \\
          \hline
    \end{tabular} 
    \begin{tablenotes}
      \item Note: IPW denotes the estimator $\hat{\psi}^{ipw}_{mRR}$, TNDDR corresponds to the proposed TND EIF-based estimator $\hat{\psi}^{eif}_{mRR},$ and OutReg denotes the estimator $\hat{\psi}^{O}_{mRR}$ with the debiasing weights of the outcome probability ratio. All estimators were implemented with cross-fitting. Nuisance functions for the three estimators were estimated using  MARS, HAL, and NN, as indicated in parentheses. See the complementary Table \ref{tab:1comp} in Appendix \ref{Appx.sim} for results of IPW and OutReg with HAL and NN. MC MSE and MC SE stand for Monte Carlo Mean Square Error and Monte Carlo Standard Error, respectively. Co-infection (TND) means the co-infection rate in the TND sample. 
    \end{tablenotes}
    \end{threeparttable}
 \label{tab:1}  
\end{table}

\subsection{Investigating the double robustness of TNDDR}
In Study 2, our objective was to validate the double robustness of TNDDR. For each method, we used generalized linear models (GLMs) with different specifications to estimate the propensity score (PS) and conditional expectation of the outcome functions, with cross-fitting. The investigation included four scenarios to evaluate the performance: (a) both models are correctly specified, meaning they encompass all the terms present in the data-generating process, including both linear and non-linear components. (b) The PS model is correct, but the outcome model is not. (c) The outcome model is correct, but the propensity score model is incorrect, lacking some nonlinear terms. (d) Both the PS and outcome models are incorrect. We note, however, that the ``correct'' models were only approximately correct since we specified models under $\mathbb{P}$ but are modeling the TND-sampled data. 

Table \ref{drtab} displays results from the four scenarios, with sample sizes  1000, 4000, and 8000 (cases percentage between $58 \% - 64\%$), and with true marginal risk ratio $\psi_{mRR} = 0.507$ ($\beta_{em} = 0.25$). As anticipated, TNDDR  demonstrated double robustness as when either the PS model and/or outcome model was correctly specified, its bias was minimal. On the other hand, the IPW and outcome regression with debiasing weights exhibited single robustness, with minimal bias only when the PS model or outcome models were correctly specified for IPW and outcome regression, respectively. The coverage of the TNDDR confidence intervals was near optimal when one or both models were correctly specified. 

\begin{table}[ht]\centering
\setlength{\tabcolsep}{3pt}\caption{ Validation of Double Robustness: Study 2 Results. (True $\psi_{mRR} = 0.507$ ($\beta_{em} = 0.25$), cases percentage: $58 \% - 64\%$).}
    \begin{threeparttable}[t]
    \begin{tabular}{ccccc|ccc|ccc|ccc}
        \hline
       \multicolumn{2}{c}{Scenarios} & \multicolumn{3}{c}{a: Both True} & \multicolumn{3}{c}{b: Only PS True}  & \multicolumn{3}{c}{c: Only Outcome True}  & \multicolumn{3}{c}{d: Both False}\\
        \multicolumn{2}{c}{Methods} &  IPW &  OutReg &  TNDDR&  IPW&  OutReg &  TNDDR&  IPW& OutReg&  TNDDR&  IPW&  OutReg &  TNDDR \\
        \hline
       \multirow{4}{*}{$1000$}& Median  Bias & -0.025 &  -0.021 &  0.017&  -0.023&  -0.045&  0.001&  
       0.166 & -0.022& 0.001 &  0.171 & -0.050 &  0.031\\
     & MC MSE & 0.016&  0.015&  0.014&  0.018 &  0.012&  0.010&  0.052 & 0.020& 0.022&  0.055 &  0.012&  0.026\\
     & MC SE
     & 0.006 &  0.006&  0.006&  0.006 &  0.004&  0.005&  0.007 & 0.006& 0.006&  0.006 &  0.005&  0.006\\
     &$\% Cov$ & 94.6 &  93.7 &  97.3&  92.8 &  88.8 &  97.8&  64.6 & 92.1 & 93.7 &  65.4 &  89.7 &  90.8\\
     \hline 
    \multirow{4}{*}{$4000$}&Median   Bias & -0.005 &  -0.004 &  0.007&  -0.009&  -0.036&  -0.001&  0.171 & -0.004& -0.002&  0.163 & -0.040 &  0.011\\
    & MC MSE 
    & 0.004 &  0.004&  0.005&  0.005 &  0.008&  0.005&  0.050 & 0.006& 0.005&  0.033 &  0.008&  0.006\\
    & MC SE
     & 0.003 &  0.003&  0.004&  0.003 &  0.003&  0.002&  0.002 & 0.004& 0.003&  0.003 &  0.003&  0.003\\
    &$\% Cov$ & 92.8 &  93.2 &  97.6 &  91.4 &  83.2 &  95.2&  15.8 & 90.4 & 92.8 &  17.6 &  80.4 &  87.4\\
    \hline   
    \multirow{4}{*}{$8000$}&Median   Bias & -0.006 &  -0.004 &  -0.002&  -0.007&  -0.032&  -0.004&  0.176 & -0.006 & -0.004&  0.176 & -0.030 &  0.025\\
    & MC MSE 
    & 0.002 &  0.002&  0.001&  0.003 &  0.004&  0.003&  0.034 & 0.003& 0.003&  0.035 &  0.003&  0.004\\
    & MC SE
     & 0.002 &  0.002&  0.002&  0.002 &  0.002&  0.002&  0.003 & 0.002& 0.002&  0.003 &  0.002&  0.002\\
    &$\% Cov$ & 91.6 &  90.8 &  98.2&  90.6 &  78.2 &  97.2&  13.4 & 90.5 & 93 &  15.6 &  77.0 &  82.8\\
    \hline   
    \end{tabular}
     \begin{tablenotes}
      \item  Note: MC MSE stands for Monte Carlo Mean Square Error and MC SE stands for Monte Carlo Standard Error. Generalized linear models (logistic regression) are used to estimate both the PS and outcome regression functions, and in this study, TNDDR refers specifically to TNDDR-GLM, with cross-fitting. 
    \end{tablenotes}
        \end{threeparttable}    \label{drtab}
\end{table}

\subsection{Exploring the impact of co-infection rates on TNDDR, violating control exchangeability}
Study 3 aimed to assess how deviations from the control exchangeability assumption  due to high co-infection rates impact the performance of each estimator. In each scenario, we report: (1) the co-infection rate within the TND samples and (2) the co-infection rate within the general population. 
In addition, in the data-generating process, we implemented a control exchangeability testing procedure: in the population dataset generated for the TND samples, we regressed the indicator of being a control on \( V \) and covariates to test whether \( V \) is conditionally independent of the control status. We report a non-significant (negative) test as ``Ctrl. exch. approx. holds'' (indicating that the control exchangeability assumption approximately holds) or a significant (positive) test as ``Ctrl. exch. does not hold''.

\begin{table}[ht] \centering
    \caption{Simulation results in different settings with different co-infection rates: 500 simulations of sample size $n = 8000$.}
    \begin{threeparttable}[t]
 \begin{tabular}{ccccccccc} 
        \hline
      & \multirow{3}{*}{Scenarios $\&$ Truth} & &\multicolumn{5}{c}{}   \\
      & & Methods & IPW &  OutReg  & TNDDR & TNDDR  & TNDDR  & TNDDR  \\
      & & Metrics & (MARS) & (MARS) & (MARS)& (HAL) & (NN) & (GLM)\\
        \hline
            \multirow{4}{*}{(I)} & \multirow{1}{*}{ Co-infec (TND): $\sim 0.05\%$} & Median bias   & -0.016 & -0.067 & -0.005  & 0.003  & 0.011   & -0.002  \\
      &  \multirow{1}{*}{ Co-infec (Pop): $\sim 0.0001\%$} &  MC MSE  & 0.017 & 0.006  & 0.008 & 0.004 &  0.004   & 0.001   \\
      & Ctrl. exch. approx. holds  & MC SE  & 0.006 & 0.002  & 0.003 & 0.002&  0.002  & 0.002\\
     &   $\psi_{mRR} = 0.507$ & $\% Cov_{mRR}$ & 87.6 & 46.8  & 96.8  & 97.2 &  90.8  & 98.2 \\
          \hline
          \multirow{3}{*}{(II)} &   \multirow{1}{*}{ Co-infec (TND): $\sim 4.53\%$}  & Median bias   & -0.062 & -0.152 & -0.051  & -0.046  & -0.048 & -0.038    \\
     &  \multirow{1}{*}{ Co-infec (Pop): $\sim 1.21\%$} & MC MSE  & 0.009 & 0.024  & 0.008 & 0.005&  0.006 &0.005    \\
    &    Ctrl. exch. does not hold & MC SE  & 0.003 & 0.001  & 0.002 & 0.002&  0.002 & 0.002   \\
     &   $\psi_{mRR} = 0.621$ & $\% Cov_{mRR}$ & 67.8 & 1.67  & 94.8  & 93.5 &  86.2 & 95.6   \\
          \hline
          \multirow{3}{*}{(III)} &   \multirow{1}{*}{ Co-infec (TND): $\sim 38.60\%$}  & Median bias   & -0.132 & -0.222 & -0.121  & -0.194 & -0.202 & -0.285   \\
    &   \multirow{1}{*}{ Co-infec (Pop): $\sim 22.10\%$} & MC MSE  & 0.023 & 0.050 & 0.020 & 0.045 &  0.072 & 0.085     \\
   &   Ctrl. exch. does not hold  & MC SE  & 0.003 & 0.001  & 0.003 & 0.005&  0.008& 0.007   \\
     &   $\psi_{mRR} = 0.690$ & $\% Cov_{mRR}$ & 22.8 & 0.25  & 65.4  & 63.7 &  65.2 & 64.8   \\
          \hline
    \end{tabular} 
    \begin{tablenotes}
      \item Note:  Co-infec (TND) and Co-infec (Pop) refer to the co-infection rates in the TND sample and the general population, respectively.  Ctrl. exch. refers to the control exchangeability assumption which was tested in the population dataset; a significant (positive) test result was reported as Ctrl. exch. does not hold; a non-significant (negative) test was reported as Ctrl. exch. approx. holds. IPW denotes the estimator of $\hat{\psi}^{ipw}_{mRR}$, TNDDR corresponds to the proposed TND EIF-based estimator of $\hat{\psi}^{eif}_{mRR}$ with cross-fitting, and OutReg denotes the estimator $\hat{\psi}^{O}_{mRR}$ with the debiasing weights of the outcome probability ratio. Nuisance functions for the three estimators were estimated using MARS, HAL, NN and correctly specified GLM, as indicated in parentheses. See the complementary Table \ref{tab:3co-infcom} in Appendix \ref{Appx.sim} for results of IPW and OutReg with HAL and NN. MC MSE and MC SE stand for Monte Carlo Mean Square Error and Monte Carlo Standard Error, respectively.
    \end{tablenotes}
    \end{threeparttable}
 \label{tab:3co-inf}  
\end{table}

Table (\ref{tab:3co-inf}) presents results from all estimators across three scenarios with varying co-infection rates. Scenario (I) involved rare co-infection, where the control exchangeability assumption approximately held. In contrast, Scenarios (II) and (III) had higher co-infection rates, and the control exchangeability assumption did not hold. Specifically, the moderately severe Scenario (II) had around $5\%$ infection overlap in the TND samples, and Scenario (III) exhibited a very high co-infection rate, with around $39\%$ in the TND samples. In both Scenario (II) and Scenario (III), all estimators showed some degree of bias. The best-performing estimator, which is based on TNDDR, exhibited relative biases of approximately $6\%$ in Scenario (II) and $17\%$ in Scenario (III), with coverage probabilities around $95\%$ for Scenario (II) and $65\%$ for Scenario (III). In the moderately severe Scenario (II), TNDDR estimators with HAL and correctly specified GLM models showed relatively stronger performance, whereas the outcome regression using MARS performed poorly. Under the most severe Scenario (III), all methods showed significant performance declines; however, TNDDR estimators using HAL and MARS demonstrated comparatively better results, while the outcome regression with a correctly specified GLM model performed the worst.

\section{Real Data Analysis} \label{sec:real}
\subsection{Setting}
We applied TNDDR to a TND study carried out in Québec, Canada by the \emph{Institut national de santé publique du Québec} (INSPQ). This study covered the timeframes characterized by the Omicron BA.5 dominance period, specifically from July 3 to November 5, 2022. The study data were harmonized from various sources, including provincial laboratories, immunization records, hospitalization records, and chronic disease surveillance. In this study, we considered the subpopulation of older adults (aged 60 years or older). The study design selected those exhibiting symptoms associated with COVID-19 who had undergone testing in the emergency room of an acute-care hospital. Cases ($Y=1$) corresponded to those patients who tested positive and were hospitalized due to their COVID-19 symptoms within 14 days from testing, while controls ($Y=0$) were those patients who tested negative. 

\subsection{Exposure,  covariates, and analytical strategy}
Regarding the exposure denoted as $V$, if an individual received a booster mRNA vaccine dose (3rd, 4th, or 5th dose) within a span of less than 6 months but with at least 7 days before undergoing testing, then $V$ was assigned a value of 1. This is in comparison to individuals who had been vaccinated but had their final dose administered 6 months or more prior to the testing date, for whom $V$ is set to 0. Data from individuals with other vaccine statuses were not included. 

We adjusted for covariates $\boldsymbol{C}$ defined as (1) age groups in four categories $60-69, 70-79, 80-89, 90+$, (2) sex (Female/Male), (3) presence of multimorbidity (Yes/No), which refers to having a minimum of two underlying health conditions that elevate the susceptibility to severe COVID-19 \citep{carazo2023effectiveness}; (4) timeframe of observation, spanning from epidemiological weeks 27-44 of 2022, with eighteen categories in all. Table \ref{INSPQtb1} and Figure \ref{figEpiWeek} in Appendix Section~\ref{Appx.realdata} provide an overview of the distribution of these covariates, as well as the occurrence of symptomatic SARS-CoV-2 infection within the groups of subjects categorized as $V=1$ and $V=0$.  As mentioned in the introduction, a convincing negative control, such as prior influenza vaccination, is unavailable in these data, thus precluding VE estimation using the negative control approach of Li et al. (2023)\cite{li2023double}. Thus, the statistical interpretation of our VE estimate relies on control exchangeability and the causal interpretation further relies on the no unmeasured confounders assumption.
\begin{table}[ht] \centering
    \caption{Estimated $\psi_{mRR}$ along with $95\%$ confidence intervals (CI), E-values for point estimates and for the confidence interval limit obtained from IPW estimator, outcome regression estimator (OutReg), TNDDR estimator, and logistic regression, using generalized linear models (GLM) and ML (i.e., MARS) to compute the nuisance functions.}
    \begin{threeparttable}[t]
\begin{tabular}{c|ccc|ccc}
\hline
Methods  & \multicolumn{3}{l|}{\ \ \ \ \ \ \ \ \ \ \ \ \ \ \ \ \ \ \ \ \ \ \ GLM} & \multicolumn{3}{l}{\ \ \ \ \ \ \ \ \ \ \ \ \ \ \ \ \ \ \ \ \ ML (MARS)}                          \\ 
Results & \multicolumn{1}{l|}{\ \ \ \ \ \ \ Est. $\&\ $ CI} & \multicolumn{1}{l|}{$\mathbf{E}_{est}$} & $\mathbf{E}_{CI}$ & \multicolumn{1}{l|}{\ \ \ \ \ \ \ \ Est. $\&\ $ CI} & \multicolumn{1}{l|}{$\mathbf{E}_{est}$} & $\mathbf{E}_{CI}$ \\ \hline
 IPW & \multicolumn{1}{l|}{0.655 (0.610, 0.703)} & \multicolumn{1}{l|}{2.42} & 2.20 & \multicolumn{1}{l|}{0.659 (0.614, 0.707)} & \multicolumn{1}{l|}{2.40} & 2.18 \\ \hline
OutReg  & \multicolumn{1}{l|}{ 0.655 (0.609, 0.719)} & \multicolumn{1}{l|}{2.42} & 2.13 & \multicolumn{1}{l|}{0.669 (0.619, 0.793)} & \multicolumn{1}{l|}{2.35} & 1.83 \\ \hline
TNDDR & \multicolumn{1}{l|}{0.650 (0.602, 0.703)} & \multicolumn{1}{l|}{2.45} & 2.20 & \multicolumn{1}{l|}{0.646 (0.599, 0.696)} & \multicolumn{1}{l|}{2.47} & 2.23 \\ \hline
Logistic regression  & \multicolumn{1}{l|}{0.655 (0.630, 0.680)} & \multicolumn{1}{l|}{2.42} & 2.30 & $-$ & $-$ & $-$\\ \hline
\end{tabular}
    \begin{tablenotes}
      \item Note: Est. denotes the estimate and CI stands for confidence interval. $\mathbf{E}_{est}$ and $\mathbf{E}_{CI}$ represent the E-values for the point estimates and the $95\%$ confidence interval (evaluated for the upper limit of the CI closest to the null).
    \end{tablenotes}
    \end{threeparttable}
    \label{tbrealdat2}  
\end{table}

\subsection{Results}
There were a total of 3,101 cases and 61,088 controls in this TND sample, and we observed imbalances between the exposure groups where people with a booster dose within 6 months were overall older and more often had multimorbidity. Table \ref{tbrealdat2} presents the estimates of $\psi_{mRR}$ with their corresponding $95\%$ confidence intervals 
from four distinct estimators, including a logistic regression with covariate main terms. In our implementations of IPW, the outcome regression approach, and TNDDR, we utilized both generalized linear models (GLM) and machine learning-based estimation to compute the nuisance functions. For the machine learning-based estimation, since we were not able to install the \texttt{HAL} package on internal servers housing the INSPQ data, we opted for MARS.  In our study, we observed that obtaining nuisance function estimates using MARS required approximately 3 seconds. 
Moreover, VE estimates were similar regardless of which estimator was used and whether machine learning methods were used or not. Notably, there was no increase in the variance for methods that target the marginal risk ratio (IPW, OutR, TNDDR) when using machine learning instead of parametric methods. While the parametric logistic regression yielded a shorter confidence interval, it  targets a different estimand, namely a conditional risk ratio under the assumed model, making direct comparisons infeasible.

In addition, due to potential unmeasured confounders, we investigated E-values, which is a sensitivity analysis technique that quantifies the evidence of causation in observational studies in the presence of unmeasured confounding \cite{ding2016sensitivity, vanderweele2017sensitivity}. Table \ref{tbrealdat2} presents E-values for point estimates ($\mathbf{E}_{est}$), which represent the magnitude of association between an unmeasured confounder and both the exposure and outcome, needed to shift the estimate to the null (i.e., risk ratio of 1). The table also presents E-values for the upper confidence limit ($\mathbf{E}_{CI}$) which represents the magnitude of these associations needed to shift the confidence interval to include the null.  To interpret these E-values \citep{mathur2018web,vanderweele2017sensitivity, haneuse2019using}, for instance, consider the estimated RR of 0.646 from TNDDR with MARS; an unmeasured confounder associated with both the compound outcome and vaccination by an RR of 2.47 each, beyond the measured confounders, could explain away the point estimate, whereas weaker confounding could not. Similarly, an unmeasured confounder associated with both the compound outcome and vaccination by an RR of 2.23-fold each, beyond the measured confounders, could shift the confidence interval to include the null, but weaker confounding could not achieve this. More details and an extended E-value analysis -- that looks at the strength of confounding needed to shift the estimate or confidence interval to different hypothetical true mRR -- are presented in the Appendix Section~\ref{Appx.realdata}.  Expert knowledge suggests that unmeasured confounders, such as socioeconomic status or frailty, would be less important confounders than age. The odds ratio for age was 1.48 in the outcome model and 1.45 in the propensity score (vaccination) model. Based on this and the additional E-values in the Appendix, it is improbable that the true mRR exceeds 0.8 or falls below 0.5 after accounting for unmeasured confounding. 

To conclude, for the composite outcome of interest $Y=1$, namely infection with SARS-CoV-2, symptom development, and subsequent hospitalization resulting from those symptoms, our TNDDR results suggest that the booster mRNA vaccine had protective effects during the period of Omicron BA.5 dominance with VE (i.e., $1 - \psi_{mRR}$) estimates of $35\% \ 
(95\%\ \text{CI}: 30\%, 40\%)$ with E-values equal to 2.5 for the point estimate and 2.2 for the confidence interval.


\section{Conclusion and Discussion}\label{sec:conc}
In the context of monitoring COVID-19 vaccine effectiveness, this study focused on the estimation of VE under the TND, with the goal of constructing and justifying data-adaptive methods that alleviate reliance on strict statistical assumptions. We developed a double machine learning approach centred around deriving the efficient influence function to create a doubly robust and locally efficient VE estimator. We theoretically and empirically justified the integration of machine learning in the estimation of the nuisance functions. 

Other evaluations of the theoretical foundation of TND have focused on the effectiveness in the healthcare-seeking subpopulation as the target parameter of interest (e.g., Smith et al., 2017\cite{smith2017tnd}; Johnson et al., 2019\cite{johnson2019tnd}). In contrast, our approach incorporates the restriction of medically-attended cases and controls into the definition of the outcome -- hospitalized (or medically-attended) symptomatic SARS-CoV-2 infection -- with the entire population as the target for inference. 
Our approach to identifiability of the statistical risk ratio parameter assumes control exchangeability, which given our assumed DAG holds with an additional ``no overlap'' condition between target and other infections. However, even under ``rare overlap'' and moderately severe overlap scenarios, our TNDDR demonstrated robust performance in simulations.  
 Our interpretation of the statistical parameter as causal relies on an additional conditional exchangeability assumption. Given that our outcome includes viral infection and hospitalization, this assumption would be violated by behavioural and structural characteristics such as healthcare-seeking, including the level of access to healthcare resources. While the TND has the potential to limit bias from differential healthcare-seeking behavior by including medical attendance as a component of the outcome,  complete elimination of bias is not guaranteed \citep{sullivan2016theoretical,lewnard2021theoretical}. 


The analysis of the Québec administrative data demonstrated that the proposed method is stable in a real-world scenario, highlighting its potential to support ongoing monitoring of VE. The VE estimates from our double machine learning TNDDR approach were around $35\%$, indicating moderate protection among older individuals who received the recent booster shot compared to those who did not. We utilized E-values to assess the sensitivity of our results to unmeasured confounders, representing the minimum strength of association on the RR scale that an unmeasured confounder would require with both vaccination and the outcome to shift the confidence interval to include the null or different hypothetical true mRR, allowing us to more confidently exclude certain levels of the potential mRR. Thus, our causal interpretation suggests that, during this Omicron period, recent booster shots increased protection, thereby reducing the likelihood of hospitalized symptomatic SARS-CoV-2 infection. In addition, our analysis is specific to the prevailing SARS-CoV-2 variant, particularly focusing on Omicron BA.5 as the predominant strain at the time. Generalizing these results to newer subvariants is inappropriate, as VE for rapidly mutating variants requires continuous monitoring; however, our research contributes to improving the accuracy of online estimation and enabling timely surveillance efforts. 

 In our application, both treatment groups received the primary COVID-19 vaccination series, so our results may be less susceptible to confounding by healthcare-seeking behavior \citep{shi2023current}.  Residual bias can be addressed through further sensitivity analysis \citep{mccandless2017comparison} and bias quantification\cite{greenland2003quantifying}. If both a negative control exposure and outcome are available, and in the setting that there is no effect modification by unmeasured confounders, double negative control methods can be used \citep{li2023double}. Past influenza vaccination could serve as an appropriate negative control exposure \citep{shi2020multiply}; however, in our analysis of data from Qu\'ebec administrative records, access to this information is not provided by the current data usage mandate.
Another limitation in the application is that recent booster vaccines are set as the treatment group. The causal consistency assumption implies that all individuals classified in this group are receiving essentially the same intervention. However, this assumption could be violated due to different histories of vaccination type or timing.
Another causal assumption that may be violated is the no interference assumption, which implies that one patient’s infection, illness, and hospitalization under their vaccination status do not depend on another patient’s vaccination status. This assumption is violated by ``herd immunity'', where members of a population are indirectly protected from infection and disease due to others' vaccination status. The evaluation of vaccine effectiveness under the TND in the presence of interference; bias quantification under violations of conditional exchangeability and control exchangeability; and an extension to Targeted Maximum Likelihood Estimation (TMLE)\cite{van2011targeted} to improve finite-sample performance are good topics for future research. 

\newpage
\renewcommand\refname{Bibliography}
\bibliographystyle{plainnat}
\bibliography{reference}

\newpage
\begin{center}
{\large\bf SUPPLEMENTARY MATERIAL FOR\\Efficient and doubly robust estimation of COVID-19 vaccine effectiveness under the test-negative design}
\end{center}
\appendix
\numberwithin{equation}{section}
\addcontentsline{toc}{section}{Appendices}
\begin{description}
\item[Description:] This supplementary material includes the proofs of the theoretical results and additional simulation studies. All the code for conducting simulation studies and analyzing real data is accessible in the online repository \url{https://github.com/CONGJIANG/TNDDR}.
\end{description}

\section{Proof of Theorem 1 and Empirical Sandwich Variance Estimators:}\label{Appx.A1}
This section contains two subsections: (A.1) the proof of Theorem \ref{ProP1}, and (A.2) the derivation of sandwich variance estimators for the IPW and proposed outcome regression estimators of VE.

\subsection{Proof of Theorem 1:}\label{SecA.1}
This subsection presents the proof of Theorem \ref{ProP1}. The proof consists of three parts: first, we show that
\begin{equation*}
\omega_v(\boldsymbol{c}) := \frac{\mathbb{P}(H=1 \mid V=v ,  \boldsymbol{C}  = \boldsymbol{c}  )}{\mathbb{P}(H=1 \mid  \boldsymbol{C}  = \boldsymbol{c} )}.
\end{equation*} Then, we show that
\begin{equation*}
\frac{\mathbb{P}(H=1 \mid V=v ,  \boldsymbol{C}  = \boldsymbol{c}  )}{\mathbb{P}(H=1 \mid  \boldsymbol{C}  = \boldsymbol{c} )} = \frac{\mathbb{P}(V=v \mid  \boldsymbol{C}  = \boldsymbol{c} , H=1)}{\mathbb{P}(V=v \mid  \boldsymbol{C}  = \boldsymbol{c} )}.
\end{equation*}
Finally, we demonstrate that
\begin{equation*}
\frac{\mathbb{P}(H=1 \mid V=v ,  \boldsymbol{C}  = \boldsymbol{c}  )}{\mathbb{P}(H=1 \mid  \boldsymbol{C}  = \boldsymbol{c} )} = \frac{ \mathbb{P}(Y = 0\mid  \boldsymbol{C}  = \boldsymbol{c}, H=1 )}{\mathbb{P}(Y = 0 \mid V=v ,  \boldsymbol{C}  = \boldsymbol{c}, H=1 )} = \frac{ 1- \mathbb{P}(Y = 1\mid  \boldsymbol{C}  = \boldsymbol{c}, H=1 )}{ 1 - \mathbb{P}(Y = 1 \mid V=v ,  \boldsymbol{C}  = \boldsymbol{c}, H=1 )}.
\end{equation*}
Firstly, we have
\begin{align*}
    & \mathbb{E}_{TND}\left \{\mathbb{P}(Y =1 \mid V=v,  \boldsymbol{C}  = \boldsymbol{c} , H=1)  \omega_v(\boldsymbol{c})  q_0\right\}\\
    =&  \int_{ \boldsymbol{\mathscr{C}} } \mathbb{P}(Y =1 \mid V=v,  \boldsymbol{C}  = \boldsymbol{c} , H=1)  \frac{\mathbb{P}(H=1 \mid V=v ,  \boldsymbol{C}  = \boldsymbol{c}  )}{\mathbb{P}(H=1 \mid  \boldsymbol{C}  = \boldsymbol{c} )}  q_0  p_{ \boldsymbol{C}  }(\boldsymbol{c}  \mid H=1) d\boldsymbol{c}   \\
    =& \int_{ \boldsymbol{\mathscr{C}} } \frac{\mathbb{P}(Y =1, H=1\mid V=v,  \boldsymbol{C}  = \boldsymbol{c} ) }{\mathbb{P}(H=1\mid V=v,  \boldsymbol{C}  = \boldsymbol{c} ) } \frac{\mathbb{P}(H=1 \mid V=v ,  \boldsymbol{C}  = \boldsymbol{c}  )}{\mathbb{P}(H=1 \mid  \boldsymbol{C}  = \boldsymbol{c} )} \mathbb{P}(H=1 \mid  \boldsymbol{C}  = \boldsymbol{c} )  p_{ \boldsymbol{C}  }(\boldsymbol{c}  ) d\boldsymbol{c}   \\
    =& \int_{ \boldsymbol{\mathscr{C}} } \mathbb{P}(Y =1, H=1\mid V=v,  \boldsymbol{C}  = \boldsymbol{c} )   p_{ \boldsymbol{C}  }(\boldsymbol{c}  ) d\boldsymbol{c}    \\
    =& \int_{ \boldsymbol{\mathscr{C}} } \mathbb{P}(Y =1 \mid V=v,  \boldsymbol{C}  = \boldsymbol{c} )   p_{ \boldsymbol{C}  }(\boldsymbol{c}  ) d\boldsymbol{c}   \\
    =& \mathbb{E}\{\mathbb{P}(Y=1 \mid V=v,  \boldsymbol{C}  = \boldsymbol{c} )\},
\end{align*}
where the first equality follows by plugging in the definition of $\omega_v(\boldsymbol{c})$, the second equality is applying the conditional Bayes rule on $p_{ \boldsymbol{C}  }(\boldsymbol{c} \mid H=1),$ i.e., $p_{ \boldsymbol{C}  }(\boldsymbol{c} \mid H=1) = \mathbb{P}(H=1\mid \boldsymbol{C} = \boldsymbol{c})p_{ \boldsymbol{C}  }(\boldsymbol{c})/\mathbb{P}(H=1),$ and the forth equation follows by the fact that $\{ Y = 1\} = \{ Y = 1, H=1\}$ since $\{ Y = 1\} \subset \{ H=1\}.$

Secondly, building on the conditional Bayes rule, we have
\begin{align*}
    \mathbb{P}(H=1 \mid V=v ,  \boldsymbol{C}  = \boldsymbol{c}  ) 
    =  \frac{\mathbb{P}(V=v \mid H=1,  \boldsymbol{C}  = \boldsymbol{c} ) \mathbb{P}(H=1 \mid  \boldsymbol{C}  = \boldsymbol{c} )}{\mathbb{P}(V=v \mid   \boldsymbol{C}  = \boldsymbol{c} )},
\end{align*}
thus, we have $$\frac{\mathbb{P}(H=1 \mid V=v ,  \boldsymbol{C}  = \boldsymbol{c}  )}{\mathbb{P}(H=1 \mid  \boldsymbol{C}  = \boldsymbol{c} )} = \frac{\mathbb{P}(V=v \mid  \boldsymbol{C}  = \boldsymbol{c} , H=1)}{\mathbb{P}(V=v \mid  \boldsymbol{C}  = \boldsymbol{c} )}.$$
Further, under the assumption that being hospitalized for symptoms of another infection is independent of vaccination conditional on covariates, i.e., $\{Y = 0, H=1 \} \perp V \mid  \boldsymbol{C},$ (i.e., control exchangeability assumption) we then have 
\begin{align*}
 \mathbb{P}(V=v \mid \boldsymbol{C}  = \boldsymbol{c}, Y = 0, H=1) = \frac{\mathbb{P}(Y = 0, H=1 \mid V=v,   \boldsymbol{C}  = \boldsymbol{c} ) \mathbb{P}(V=v \mid  \boldsymbol{C})}{\mathbb{P}(Y = 0, H=1 \mid   \boldsymbol{C}  = \boldsymbol{c} )} = \mathbb{P}(V=v \mid  \boldsymbol{C}  = \boldsymbol{c} ).
\end{align*}

Thus, the debiasing weights can be written as 
\begin{equation} \label{debiasw}
    \omega_v(\boldsymbol{c}) = \frac{\mathbb{P}(V=v \mid  \boldsymbol{C}  = \boldsymbol{c} , H=1)}{\mathbb{P}(V=v \mid  \boldsymbol{C}  = \boldsymbol{c}, Y = 0, H=1 )},
\end{equation}
and it is the ratio of propensity scores which are estimated by the whole TND sample data (numerator) and the only controls (denominator). \\
Lastly, using the identity that $$   \mathbb{P}(V=v \mid  \boldsymbol{C}  = \boldsymbol{c}, H=1 ) 
    =  \frac{ \mathbb{P}(H=1 \mid V=v ,  \boldsymbol{C}  = \boldsymbol{c}  ) \mathbb{P}(V=v \mid   \boldsymbol{C}  = \boldsymbol{c} )  }{ \mathbb{P}(H=1 \mid  \boldsymbol{C}  = \boldsymbol{c} )},$$
    $$ \mathbb{P}(V=v \mid  \boldsymbol{C}  = \boldsymbol{c}, Y=0, H=1 ) 
    =  \frac{ \mathbb{P}(Y=0, H=1 \mid V=v ,  \boldsymbol{C}  = \boldsymbol{c}  ) \mathbb{P}(V=v \mid   \boldsymbol{C}  = \boldsymbol{c} )  }{ \mathbb{P}(Y =0, H=1 \mid  \boldsymbol{C}  = \boldsymbol{c} )};$$
we can write Equation (\ref{debiasw}) as
\begin{align*}
        \omega_v(\boldsymbol{c}) & = \frac{\mathbb{P}(V=v \mid  \boldsymbol{C}  = \boldsymbol{c} , H=1)}{\mathbb{P}(V=v \mid  \boldsymbol{C}  = \boldsymbol{c}, Y = 0, H=1 )} = \frac{\mathbb{P}(H=1 \mid V=v ,  \boldsymbol{C}  = \boldsymbol{c}  )}{\mathbb{P}(H=1 \mid  \boldsymbol{C}  = \boldsymbol{c} )}\Biggm/ \frac{\mathbb{P}(Y = 0, H=1 \mid V=v ,  \boldsymbol{C}  = \boldsymbol{c}  )}{\mathbb{P}(Y = 0, H=1 \mid  \boldsymbol{C}  = \boldsymbol{c} )} \\ &= \frac{\mathbb{P}(Y = 0, H=1 \mid  \boldsymbol{C}  = \boldsymbol{c} ) }{\mathbb{P}(H=1 \mid  \boldsymbol{C}  = \boldsymbol{c} )}*\frac{\mathbb{P}(H=1 \mid V=v ,  \boldsymbol{C}  = \boldsymbol{c}  ) }{\mathbb{P}(Y = 0, H=1 \mid V=v ,  \boldsymbol{C}  = \boldsymbol{c}  ) }. 
\end{align*}

The first component is $$\frac{\mathbb{P}(Y = 0, H=1 \mid  \boldsymbol{C}  = \boldsymbol{c} ) }{\mathbb{P}(H=1 \mid  \boldsymbol{C}  = \boldsymbol{c} )} = \mathbb{P}(Y = 0 \mid  \boldsymbol{C}  = \boldsymbol{c}, H=1 ).$$
The second component is $$\frac{\mathbb{P}(H=1 \mid V=v ,  \boldsymbol{C}  = \boldsymbol{c}  ) }{\mathbb{P}(Y = 0, H=1 \mid V=v ,  \boldsymbol{C}  = \boldsymbol{c}  ) } = \frac{1}{\mathbb{P}(Y = 0 \mid V=v ,  \boldsymbol{C}  = \boldsymbol{c}, H=1 )}.$$
To sum up, the debiasing weights can also be identified as
\begin{equation} \label{wetform2}
     \omega_v(\boldsymbol{c}) = \frac{ \mathbb{P}(Y = 0\mid  \boldsymbol{C}  = \boldsymbol{c}, H=1 )}{\mathbb{P}(Y = 0 \mid V=v ,  \boldsymbol{C}  = \boldsymbol{c}, H=1 )} = \frac{ 1- \mathbb{P}(Y = 1\mid  \boldsymbol{C}  = \boldsymbol{c}, H=1 )}{ 1 - \mathbb{P}(Y = 1 \mid V=v ,  \boldsymbol{C}  = \boldsymbol{c}, H=1 )}.
\end{equation}
Equations (\ref{debiasw}) and (\ref{wetform2}) represent two distinct forms of debiasing weights, with Equation (\ref{debiasw}) representing the propensity score ratio and Equation (\ref{wetform2}) representing the outcome probability ratio. They are mathematically identical but rely on distinct models.

Finally, we have two remarks regarding the IPW estimator as follows.
\textit{Remark 1.} Regarding identifiability of the IPW estimator, under the control exchangeability and no co-infection assumption, i.e., $\{Y = 0, H=1 \} \perp V \mid  \boldsymbol{C}$, the propensity score model based on simple random sampling can be fit using only the control data of the TND samples, i.e., $\mathbb{P}(V=v\mid \boldsymbol{C})=\mathbb{P}_{TND}(V=v\mid Y = 0, \boldsymbol{C})$. Because $\{Y = 0, H=1 \} \perp V \mid  \boldsymbol{C}$, we have $\mathbb{P}(Y=0, H=1 \mid V=v, \boldsymbol{C}) = \mathbb{P}(Y=0, H=1 \mid \boldsymbol{C}),$ and thus
\begin{equation*}
\mathbb{P}_{TND}(V=v \mid Y=0, \boldsymbol{C})  =\mathbb{P}(V=v \mid Y=0, S= 1, \boldsymbol{C}) 
=\frac{\mathbb{P}(Y=0, H=1 \mid V=v, \boldsymbol{C}) \mathbb{P}(V=v \mid \boldsymbol{C})}{\mathbb{P}(Y=0, H=1 \mid \boldsymbol{C})} =  \mathbb{P}(V=v \mid \boldsymbol{C}). 
\end{equation*}

\textit{Remark 2.} The debiasing weights adjust for bias arising from outcome-dependent sampling. By applying these weights, the TND estimators (including both IPW and outcome regression-based estimators) can be viewed as standard estimators incorporating the debiasing weights. As given in Theorem \ref{ProP1}, our outcome regression-based estimator $\hat{\psi}^{O}_{mRR}$ takes a weighted mean of the  standard conditional outcome probabilities $\mu_v(\boldsymbol{c})$. 
Similarly, the IPW estimator of $\psi_v$ proposed by \cite{schnitzer2022estimands} can be represented as, for $v = \mathcal{V}$, $$\frac{1}{n}\sum_k^n \frac{Y_k \mathbb{I}(V_k = v)}{\hat{\pi}^0_v(\boldsymbol{c}_k)} =  \frac{1}{n}\sum_k^n  \frac{ \hat{\pi}_v(\boldsymbol{c}_k ) }{ \hat{\pi}^0_v(\boldsymbol{c}_k )} \frac{Y_k \mathbb{I}(V_k = v)}{\hat{\pi}_v(\boldsymbol{c}_k)}=  \frac{1}{n}\sum_k^n  \hat{\omega}_v(\boldsymbol{c}_k) \frac{Y_k \mathbb{I}(V_k = v)}{\hat{\pi}_v(\boldsymbol{c}_k)}.$$ 

\subsection{Empirical Sandwich Variance Estimators}\label{SecA.2}
In this subsection, we derive the sandwich variance estimators for the IPW and outcome regression estimators of VE, and explain how the empirical sandwich variance estimator is obtained using M-estimation theory \cite{stefanski2002calculus}. 

Under parametric models, we can formulate the outcome regression-based estimator using a weighted estimating equation for logistic regression, and thus construct a sandwich estimator for the variance. Recall that we have
\begin{itemize}
    \item  $m(\boldsymbol{c}) := \mathbb{P}_{TND}(Y = 1\mid  \boldsymbol{C}  = \boldsymbol{c} ),$ marginal outcome regression functions; and
    \item $\mu_v(\boldsymbol{c}) :=\mathbb{P}_{TND}(Y = 1 \mid V=v ,  \boldsymbol{C}  = \boldsymbol{c} ),$ treatment-specific outcome regression functions.
\end{itemize}
and the debiasing weight that $\hat{\omega}_v(\boldsymbol{c}) = [1- \hat{m}(\boldsymbol{c})]/[1-\hat{\mu}_v(\boldsymbol{c})]$, the outcome regression-based estimator is solved by
\begin{equation*}
    \frac{1}{n} \sum_{k=1}^{n} \hat{\mu}_v(\boldsymbol{c}_k ) \hat{\omega}_v(\boldsymbol{c}_k) - \psi_v = 0. 
\end{equation*}

The estimating equations of the outcome regression-based estimator based on weighted logistic regression: for $\boldsymbol{\theta} = (\beta, \gamma, \psi_{v}, \psi_{v_0}, \psi^{O}_{mRR}),$
\begin{equation}\label{mEST}
\sum_{i=1}^n M\left(O_i ; \hat{\boldsymbol{\theta}}\right) = \sum_{i=1}^{n}\left(
\begin{gathered}
\left(Y_i-\operatorname{expit}\left(\left(V_i, \boldsymbol{C}_i\right)^{\top} \hat{\beta}\right)\right) \left(V_i, \boldsymbol{C}_i\right) \\
\left(Y_i-\operatorname{expit}\left(\boldsymbol{C}_i^{\top} \hat{\gamma}\right)\right) \boldsymbol{C}_i \\
q_0\operatorname{expit}\left(\left(v_i, \boldsymbol{C}_i\right)^{\top} \hat{\beta}\right)\frac{1 - \operatorname{expit}\left(\boldsymbol{C}_i^{\top} \hat{\gamma}\right)}{1 - \operatorname{expit}\left(\left(v_i, \boldsymbol{C}_i\right)^{\top} \hat{\beta}\right)}-\hat{\psi}_v \\
q_0\operatorname{expit}\left(\left(v_{0i}, \boldsymbol{C}_i\right)^{\top} \hat{\beta}\right)\frac{1 - \operatorname{expit}\left(\boldsymbol{C}_i^{\top} \hat{\gamma}\right)}{1 - \operatorname{expit}\left(\left(v_{0i}, \boldsymbol{C}_i\right)^{\top} \hat{\beta}\right)}-\hat{\psi}_{v_0} \\
\left(\hat{\psi}_v/\hat{\psi}_{v_0}\right)-\hat{\psi}^{O}_{mRR}
\end{gathered}
\right) = \boldsymbol{0},
\end{equation}
where, for $i=1,...,n,$ $O_i$ are independent observations; then the M-estimator is the solution for the target $\psi_{mRR}$ in the above equations, i.e., $\hat{\psi}_{mRR}$ in Equation (\ref{mEST}). 

Building on the M-theory and estimating equations (\ref{mEST}), the asymptotic sandwich variance estimator provides a robust estimate of the variance of the estimator $\hat{\psi}_{mRR}$, which is given by: 
\[
V(\boldsymbol{\theta}) = B(\boldsymbol{\theta})^{-1} F(\boldsymbol{\theta}) B(\boldsymbol{\theta})^{-1{\top}},
\]
where \( B(\boldsymbol{\theta}) \) represents the expected negative derivative of the estimating equation (\ref{mEST}), and \( F(\boldsymbol{\theta}) \) is the variance of the estimating equations. The empirical counterpart, used in finite samples, is:
\[
V_n(O_i; \hat{\boldsymbol{\theta}}) = B_n(O_i; \hat{\boldsymbol{\theta}})^{-1} F_n(O_i; \hat{\boldsymbol{\theta}}) B_n(O_i; \hat{\boldsymbol{\theta}})^{-1\top},
\]
where \( B_n(O_i; \hat{\boldsymbol{\theta}}) \) is the sample average of the negative derivative of the estimating function (i.e., $n^{-1} \sum_{i=1}^n-\partial M\left(O_i ; \hat{\boldsymbol{\theta}}\right)/\partial \boldsymbol{\theta}^\top$), while \( F_n(O_i; \hat{\boldsymbol{\theta}}) \) is the sample variance of the estimating equations (i.e., $n^{-1} \sum_{i=1}^n M\left(O_i ; \hat{\boldsymbol{\theta}}\right) M\left(O_i ; \hat{\boldsymbol{\theta}}\right)^\top$). Note that, when the correct parametric family is assumed, we then have $B(\boldsymbol{\theta}) = F(\boldsymbol{\theta})$ \cite{stefanski2002calculus}.

Regarding computational M-estimation, the equations (\ref{mEST}) can be applied using the \texttt{geex}\cite{saul2020calculus} package to obtain the sandwich estimators for the variance of the weighted outcome regression-based estimator, employing generalized linear models.

Similarly, for the IPW estimator, we omit the details but note that the corresponding empirical sandwich variance estimators can be derived using the following estimation equations.
For $\boldsymbol{\theta} = (\alpha, \psi_{1}, \psi_{0}, \psi^{ipw}_{mRR}),$
\begin{equation}
\sum_{i=1}^n M\left(O_i ; \hat{\boldsymbol{\theta}}\right) = \sum_{i=1}^{n}\left(
\begin{gathered}
\mathbb{I}(Y_i = 0)\left(V_i-\operatorname{expit}\left(\boldsymbol{C}_i^{\top} \hat{\alpha}\right)\right) \boldsymbol{C}_i \\
q_0\frac{Y_i V_i}{\operatorname{expit}\left(\boldsymbol{C}_i^{\top} \hat{\alpha}\right)}-\hat{\psi}_1 \\
q_0\frac{Y_i\left(1-V_i\right)}{1-\operatorname{expit}\left(\boldsymbol{C}_i^{\top} \hat{\alpha}\right)}-\hat{\psi}_{0} \\
\left(\hat{\psi}_1/\hat{\psi}_{0}\right)-\hat{\psi}^{ipw}_{mRR}
\end{gathered}
\right) = \boldsymbol{0}.
\end{equation}

Recall that for the propensity score among the controls $\pi_v^0(\boldsymbol{c}) := \mathbb{P}_{TND}(V=v \mid \boldsymbol{C}  = \boldsymbol{c}, Y=0),$ we have the IPW estimator that 
\begin{align*}
   \hat{\psi}^{ipw}_{mRR} =  \left.\frac{1}{n}\sum_{i=1}^n \frac{Y_k\mathbb{I}(V_k=1)}{\hat{\pi}_1^0(\boldsymbol{c})} \middle/ \frac{1}{n}\sum_{i=1}^n\frac{Y_k\mathbb{I}(V_k=0)}{\hat{\pi}_0^0(\boldsymbol{c})}\right.. 
\end{align*}

\section{Proof of Lemma \ref{LEMMA1}:}\label{Appx.A2}
In this section, we will derive the efficient influence function of $\psi_v$. We will begin by deriving the efficient influence function for $\mathbb{P}_{TND}(Y = 1 \mid X = x)$ using the Gateaux derivative or ``point mass contamination'' \citep{hines2022demystifying}. Note that we employ the variables $Y$ and $X$ as two arbitrary variables here, and the $Y$ variable we introduce here may differ from the $Y$ variable referenced elsewhere in this manuscript.  It enables us to establish a general expression for EIFs under the TND. Subsequently, we utilize these expressions as building blocks for the computation of the specific EIFs that are of interest \citep{kennedy2022semiparametric}.

Let $Z = (X, Y)$ and $\delta_z = \mathbb{I}(Z = z)$ be the the Dirac measure at $Z = z$, where $z = (x, 1)$. For the submodel $\mathbb{P}_{TND, \epsilon}(Z=z)=(1-\epsilon) \mathbb{P}_{TND}(Z=z)+\epsilon \mathbb{I}(Z = \tilde{z})$, we have
\begin{equation*}
\mathbb{P}_{TND,\epsilon}(Y=1 \mid X=x)=\frac{\mathbb{P}_{TND,\epsilon}(Z=z)}{\mathbb{P}_{TND,\epsilon}(X=x)}=\frac{(1-\epsilon) \mathbb{P}_{TND}(Z=z)+\epsilon \mathbb{I}\left(z=\tilde{z}\right)}{(1-\epsilon) \mathbb{P}_{TND}(X=x)+\epsilon \mathbb{I}\left(x=\tilde{x}\right)}. 
\end{equation*}
Therefore, the Gateaux derivative is 
\begin{align*}
&\left.\frac{d}{d \epsilon} \psi\left\{(1-\epsilon) \mathbb{P}_{TND}(z)+\epsilon \delta_{\tilde{z}}\right\}\right|_{\epsilon=0}\\  =&\left.\frac{d}{d \epsilon} \left[ \frac{(1-\epsilon) \mathbb{P}_{TND}(Z=z)+\epsilon \mathbb{I}\left(z=\tilde{z}\right)}{(1-\epsilon) \mathbb{P}_{TND}(X=x)+\epsilon \mathbb{I}\left(x=\tilde{x}\right)} \right]\right|_{\epsilon=0}   \\
 =& \frac{\left\{\mathbb{I}\left(z=\tilde{z}\right)-\mathbb{P}_{TND}(Z=z)\right\} \mathbb{P}_{TND}(X=x)-\left\{\mathbb{I}\left(x=\tilde{x}\right)-\mathbb{P}_{TND}(X=x)\right\} \mathbb{P}_{TND}(Z=z)}{\mathbb{P}_{TND}(X=x)^2}\\
 =& \frac{\mathbb{I}\left(z=\tilde{z}\right)-\mathbb{P}_{TND}(Z=z)}{\mathbb{P}_{TND}(X=x)}-\frac{\mathbb{I}\left(x=\tilde{x}\right)-\mathbb{P}(X=x)}{\mathbb{P}_{TND}(X=x)} \mathbb{P}_{TND}(Y=1 \mid X=x) \\
 =& \frac{\mathbb{I}\left(z=\tilde{z}\right)-\mathbb{I}\left(x=\tilde{x}\right) \mathbb{P}_{TND}(Y=1 \mid X=x)}{\mathbb{P}_{TND}(X=x)}=\varphi\left(\tilde{z} ; \mathbb{P}_{TND}\right)
\end{align*}

Therefore, the EIF of $\mathbb{P}_{TND}(Y = 1 \mid X = x)$ is 
\begin{equation} \label{EIFmodel}
     \frac{\mathbb{I}\left(Y = 1, X = x\right)-\mathbb{I}\left(X = x\right) \mathbb{P}_{TND}(Y=1 \mid X=x)}{\mathbb{P}_{TND}(X=x)}. 
\end{equation}
The consequence of this computation is that EIFs for $\pi_v(\boldsymbol{c}) := \mathbb{P}_{TND}(V=v \mid \boldsymbol{C}  = \boldsymbol{c})$, $\pi^0_v(\boldsymbol{c}) := \mathbb{P}_{TND}(V=v \mid \boldsymbol{C}  = \boldsymbol{c}, Y = 0, H=1)$, and $\mu_v(\boldsymbol{c}) := \mathbb{P}_{TND}(Y = 1 \mid V=v, \boldsymbol{C}  = \boldsymbol{c})$ are as follows. 
\begin{equation} \label{2propen}
     \mathbb{EIF}(\pi_v(\boldsymbol{c})) = \frac{\mathbb{I}\left(V=v, \boldsymbol{C}  = \boldsymbol{c}\right)-\mathbb{I}\left(\boldsymbol{C}  = \boldsymbol{c}\right) \mathbb{P}_{TND}(V=v \mid \boldsymbol{C}  = \boldsymbol{c})}{\mathbb{P}_{TND}(\boldsymbol{C}  = \boldsymbol{c})},
\end{equation}

\begin{equation}\label{2propen0}
     \mathbb{EIF}(\pi^0_v(\boldsymbol{c})) = \frac{\mathbb{I}\left(V=v, \boldsymbol{C}  = \boldsymbol{c}, Y=0, H=1\right)-\mathbb{I}\left(\boldsymbol{C}  = \boldsymbol{c}, Y=0, H=1\right) \mathbb{P}_{TND}(V=v \mid \boldsymbol{C}  = \boldsymbol{c}, Y=0, H=1)}{\mathbb{P}_{TND}(\boldsymbol{C}  = \boldsymbol{c}, Y=0, H=1)},
\end{equation}

\begin{equation} \label{2eifout}
     \mathbb{EIF}(\mu_v(\boldsymbol{c}))= \frac{\mathbb{I}\left(Y = 1,V=v, \boldsymbol{C}  = \boldsymbol{c}\right)-\mathbb{I}\left(V=v,\boldsymbol{C}  = \boldsymbol{c}\right) \mathbb{P}_{TND}(Y=1 \mid V=v, \boldsymbol{C}  = \boldsymbol{c})}{\mathbb{P}_{TND}(V=v,\boldsymbol{C}  = \boldsymbol{c})}. 
\end{equation}

Building on Equations (\ref{2propen}) and (\ref{2propen0}), we have that the EIF for $\omega_v(\boldsymbol{c}) = \pi_v(\boldsymbol{c}) / \pi^0_v(\boldsymbol{c})$ is
\begin{equation} \label{2eifwet}
     \mathbb{EIF}(\omega_v(\boldsymbol{c})) = \frac{\pi^0_v(\boldsymbol{c}) \mathbb{EIF}(\pi_v(\boldsymbol{c})) - \pi_v(\boldsymbol{c})\mathbb{EIF}(\pi^0_v(\boldsymbol{c}) )}{[\pi^0_v(\boldsymbol{c})]^2} = \frac{ \mathbb{EIF}(\pi_v(\boldsymbol{c}))}{\pi^0_v(\boldsymbol{c})} - \frac{\pi_v(\boldsymbol{c})\mathbb{EIF}(\pi^0_v(\boldsymbol{c}) )}{[\pi^0_v(\boldsymbol{c})]^2}. 
\end{equation}
Building on Theorem \ref{ProP1} and Equation (\ref{tag1}) we have the estimand of interest, that is,
\begin{equation*}
\psi_{v}(\mathbb{P}_{TND}) = \mathbb{E}_{TND}\left \{\mathbb{P}_{TND}(Y=1 \mid V=v,  \boldsymbol{C}  = \boldsymbol{c})  \omega_v(\boldsymbol{c})  \right\}.
\end{equation*}
Then, building on Example 2 in \cite{kennedy2022semiparametric}, pretending the data are discrete, we have the EIF for $\psi_{v}$, that is, 
\begin{align} \label{2EIFmost}
    \mathbb{EIF}(\psi_{v}) = \mathbb{EIF}&\left[\sum_{ \boldsymbol{c} } \mu_v(\boldsymbol{c})\omega_v(\boldsymbol{c})\mathbb{P}_{TND}(\boldsymbol{C}  = \boldsymbol{c})\right] \nonumber \\
    = \sum_{ \boldsymbol{c} }[ &\underbrace{\mathbb{EIF}(\mu_v(\boldsymbol{c}))\omega_v(\boldsymbol{c})\mathbb{P}_{TND}(\boldsymbol{C}  = \boldsymbol{c})}_{A} \\ \nonumber &+ \underbrace{\mu_v(\boldsymbol{c})\mathbb{EIF}(\omega_v(\boldsymbol{c}))\mathbb{P}_{TND}(\boldsymbol{C}  = \boldsymbol{c})}_{B} + \underbrace{\mu_v(\boldsymbol{c})\omega_v(\boldsymbol{c})\mathbb{EIF}(\mathbb{P}_{TND}(\boldsymbol{C}  = \boldsymbol{c}))}_{C}],
\end{align}
where the third equality follows by Trick 2a in \cite{kennedy2022semiparametric}. Equation \ref{2EIFmost} primarily consists of three terms. Next, we will calculate each of these terms as outlined below.

For term A, plugging in Equation (\ref{2eifout}), we have
\begin{align*}
\begin{split}
    &\sum_{ \boldsymbol{c} }\mathbb{EIF}(\mu_v(\boldsymbol{c}))\omega_v(\boldsymbol{c})\mathbb{P}_{TND}(\boldsymbol{C}  = \boldsymbol{c}) \\ =& \sum_{ \boldsymbol{c} } \frac{\mathbb{I}\left(Y = 1,V=v, \boldsymbol{C}  = \boldsymbol{c}\right)-\mathbb{I}\left(V=v,\boldsymbol{C}  = \boldsymbol{c}\right) \mu_v(\boldsymbol{c})}{\mathbb{P}_{TND}(V=v,\boldsymbol{C}  = \boldsymbol{c})} \omega_v(\boldsymbol{c})\mathbb{P}_{TND}(\boldsymbol{C}  = \boldsymbol{c})  \\
    =& \sum_{ \boldsymbol{c} } \frac{\mathbb{I}\left(Y = 1,V=v, \boldsymbol{C}  = \boldsymbol{c}\right)-\mathbb{I}\left(V=v,\boldsymbol{C}  = \boldsymbol{c}\right) \mu_v(\boldsymbol{c})}{\mathbb{P}_{TND}(V=v \mid \boldsymbol{C}  = \boldsymbol{c})} \omega_v(\boldsymbol{c}) \\
    =& \sum_{ \boldsymbol{c} } \frac{\mathbb{I}\left(Y = 1,V=v, \boldsymbol{C}  = \boldsymbol{c}\right)-\mathbb{I}\left(V=v,\boldsymbol{C}  = \boldsymbol{c}\right) \mu_v(\boldsymbol{c})}{\mathbb{P}_{TND}(V=v \mid \boldsymbol{C}  = \boldsymbol{c})} \frac{\pi_v(\boldsymbol{c})}{\pi^0_v(\boldsymbol{c})} \\ 
       =& \sum_{ \boldsymbol{c} } \frac{\mathbb{I}\left(Y = 1,V=v, \boldsymbol{C}  = \boldsymbol{c}\right)-\mathbb{I}\left(V=v,\boldsymbol{C}  = \boldsymbol{c}\right) \mu_v(\boldsymbol{c})}{\pi^0_v(\boldsymbol{c})}\\
    =&\frac{\mathbb{I}(Y = 1, V=v) - \mathbb{I}(V=v)\mu_v(\boldsymbol{c})}{\pi^{0}_v(\boldsymbol{c})}
\end{split}
\end{align*}
For term B, to compute $\sum_{ \boldsymbol{c} }\mu_v(\boldsymbol{c})\mathbb{EIF}(\omega_v(\boldsymbol{c}))\mathbb{P}_{TND}(\boldsymbol{C}  = \boldsymbol{c})$, plugging in Equation (\ref{2eifwet}), we first have
\begin{align*}
    & \sum_{ \boldsymbol{c} }\mu_v(\boldsymbol{c}) \frac{ \mathbb{EIF}(\pi_v(\boldsymbol{c}))}{\pi^0_v(\boldsymbol{c})} \mathbb{P}_{TND}(\boldsymbol{C}  = \boldsymbol{c})\\ =& \sum_{ \boldsymbol{c} }\mu_v(\boldsymbol{c}) \frac{ \mathbb{I}\left(V=v, \boldsymbol{C}  = \boldsymbol{c}\right)-\mathbb{I}\left(\boldsymbol{C}  = \boldsymbol{c}\right) \mathbb{P}_{TND}(V=v \mid \boldsymbol{C}  = \boldsymbol{c})}{\pi^0_v(\boldsymbol{c})\mathbb{P}_{TND}(\boldsymbol{C}  = \boldsymbol{c})} \mathbb{P}_{TND}(\boldsymbol{C}  = \boldsymbol{c}) \\ 
    =& \frac{ \mathbb{I}(V=v)\mu_v(\boldsymbol{c}) - \mu_v(\boldsymbol{c})\pi_v(\boldsymbol{c})}{\pi^{0}_v(\boldsymbol{c})}    \\
    =& \frac{ \mathbb{I}(V=v)\mu_v(\boldsymbol{c}) }{\pi^{0}_v(\boldsymbol{c})}   - \mu_v(\boldsymbol{c})\omega_v(\boldsymbol{c}) ,
\end{align*}
then, we have
\begin{align*}
& -\sum_{ \boldsymbol{c} }\mu_v(\boldsymbol{c})\frac{\pi_v(\boldsymbol{c})\mathbb{EIF}[\pi^0_v(\boldsymbol{c})] }{[\pi^0_v(\boldsymbol{c})]^2}\mathbb{P}_{TND}(\boldsymbol{C}  = \boldsymbol{c}) \\ =& -\sum_{ \boldsymbol{c} }\mu_v(\boldsymbol{c})\frac{\pi_v(\boldsymbol{c})\left\{ \mathbb{I}\left(V=v, \boldsymbol{C}  = \boldsymbol{c}, Y=0, H=1\right)-\mathbb{I}\left(\boldsymbol{C}  = \boldsymbol{c}, Y=0, H=1\right) \pi^0_{v}(\boldsymbol{c})\right\} }{[\pi^0_v(\boldsymbol{c})]^2\mathbb{P}_{TND}(\boldsymbol{C}  = \boldsymbol{c}, Y=0, H=1)}\mathbb{P}_{TND}(\boldsymbol{C}  = \boldsymbol{c}) \\
=& -\sum_{ \boldsymbol{c} }\mu_v(\boldsymbol{c})\frac{\pi_v(\boldsymbol{c})\left\{ \mathbb{I}\left(V=v, \boldsymbol{C}  = \boldsymbol{c}, Y=0, H=1\right)-\mathbb{I}\left(\boldsymbol{C}  = \boldsymbol{c}, Y=0, H=1\right) \pi^0_{v}(\boldsymbol{c})\right\} }{[\pi^0_v(\boldsymbol{c})]^2\mathbb{P}_{TND}(Y=0, H=1 \mid \boldsymbol{C}  = \boldsymbol{c})} \\ 
=& -\mu_v(\boldsymbol{c})\frac{\pi_v(\boldsymbol{c})\left\{ \mathbb{I}\left(V=v, Y=0, H=1\right)-\mathbb{I}\left( Y=0, H=1\right) \pi^0_{v}(\boldsymbol{c})\right\} }{[\pi^0_v(\boldsymbol{c})]^2\mathbb{P}_{TND}(Y=0 \mid \boldsymbol{C}  = \boldsymbol{c})}.
\end{align*}
Finally, for term C, note that $\mathbb{EIF}(\mathbb{P}_{TND}(\boldsymbol{C}  = \boldsymbol{c})) = \mathbb{I}(\boldsymbol{C}  = \boldsymbol{c}) - \mathbb{P}_{TND}(\boldsymbol{C}  = \boldsymbol{c})$, we have
\begin{align*}
    \sum_{ \boldsymbol{c} }\mu_v(\boldsymbol{c})\omega_v(\boldsymbol{c})\mathbb{EIF}(\mathbb{P}_{TND}(\boldsymbol{C}  = \boldsymbol{c})) &= \sum_{ \boldsymbol{c} }\mu_v(\boldsymbol{c})\omega_v(\boldsymbol{c})[\mathbb{I}(\boldsymbol{C}  = \boldsymbol{c}) - \mathbb{P}_{TND}(\boldsymbol{C}  = \boldsymbol{c})] \\ &= \mu_v(\boldsymbol{c})\omega_v(\boldsymbol{c}) - \psi_{v}(\mathbb{P}_{TND}).
\end{align*}
To conclude, by combining these elementary results for terms A, B, and C, based on Equation (\ref{2EIFmost}), the EIF regarding $\psi_{v}$ can be expressed as follows:
\begin{align} \label{2EIFmost2}
    \mathbb{EIF}(\psi_{v}) &= \left[ \frac{\mathbb{I}(Y = 1, V=v)}{\pi^{0}_v(\boldsymbol{c})} - \mu_v(\boldsymbol{c})\frac{\pi_v(\boldsymbol{c})\left\{ \mathbb{I}\left(V=v, Y=0, H=1\right)-\mathbb{I}\left( Y=0, H=1\right) \pi^0_{v}(\boldsymbol{c})\right\} }{[\pi^0_v(\boldsymbol{c})]^2\mathbb{P}_{TND}(Y=0 \mid \boldsymbol{C}  = \boldsymbol{c})} \right] - \psi_{v}(\mathbb{P}_{TND}) \nonumber \\
    & =\left[ \frac{\mathbb{I}(Y = 1, V=v)}{\pi^0_v(\boldsymbol{c})} - \omega_v(\boldsymbol{c})\mu_v(\boldsymbol{c})\left\{ \frac{ \mathbb{I}\left( Y=0, H=1\right) \left[\mathbb{I}(V=v) - \pi^0_{v}(\boldsymbol{c})\right] }{\pi^0_v(\boldsymbol{c})[1 - m(\boldsymbol{c})]}  \right\}\right] - \psi_{v}(\mathbb{P}_{TND}) \nonumber \\
    &= \frac{\mathbb{I}(Y = 1, V=v)}{\pi^0_v(\boldsymbol{C})} - \mu_v(\boldsymbol{C})\left\{ \frac{ \mathbb{I}\left( Y=0, H=1\right) \left[\mathbb{I}(V=v) - \pi^0_{v}(\boldsymbol{C})\right] }{\pi^0_v(\boldsymbol{C})[1 - \mu_v(\boldsymbol{C})]}  \right\}  - \psi_{v}(\mathbb{P}_{TND}) \nonumber \\
    & = \frac{\mathbb{I}(Y = 1, V=v) - odds_{TND}(Y \mid V=v,\boldsymbol{C})\mathbb{I}\left( Y=0, H=1\right) \left[\mathbb{I}(V=v) - \pi^0_{v}(\boldsymbol{C})\right]}{\pi^0_v(\boldsymbol{C})} - \psi_{v}(\mathbb{P}_{TND}), 
\end{align}
where the specific form of $\omega(\boldsymbol{c}) = [1 - m(\boldsymbol{c})]/[1 - \mu_v(\boldsymbol{c})]$ leads to the third equality, and $odds_{TND}(Y \mid V=v,\boldsymbol{C})$ is the odds under $\mathbb{P}_{TND}$ of $Y$ given $V=v$ and $\boldsymbol{C}$.

\section{Proof of Lemma 2:}\label{Appx.A3}
In this section, we present the proof of Lemma \ref{lemma2}.\\
\textbf{Proof}:
  The EIF of $\psi_v$, $\varphi_v(\boldsymbol{Z}; \mathbb{P}_{TND})= \frac{\mathbb{I}(Y = 1, V=v)}{\pi^0_v(\boldsymbol{c})} - \mu_v(\boldsymbol{c})\left\{ \frac{ \mathbb{I}\left( Y=0, H=1\right) \left[\mathbb{I}(V=v) - \pi^0_{v}(\boldsymbol{c})\right] }{\pi^0_v(\boldsymbol{c})[1 - \mu_v(\boldsymbol{c})]}  \right\}  - \psi_{v}(\mathbb{P}_{TND}),$ gives
  \begin{align*}
  &R^v_{2}(\bar{\mathbb{P}}_{TND}, \mathbb{P}_{TND})\\ =& \psi_{v}(\mathbb{P}_{TND}) - \psi_{v}(\bar{\mathbb{P}}_{TND}) - \int \varphi_{v}(\bar{\mathbb{P}}_{TND}) d(\mathbb{P}_{TND}-\bar{\mathbb{P}}_{TND}) \\
  =& \psi_{v}(\mathbb{P}_{TND}) - \psi_{v}(\bar{\mathbb{P}}_{TND}) -  \int \left( \frac{\mathbb{I}(Y = 1, V=v)}{\bar{\pi}^0_v(\boldsymbol{c})} - \bar{\mu}_v(\boldsymbol{c}) \frac{\mathbb{I}\left( Y=0, H=1\right) \left[\mathbb{I}(V=v) - \bar{\pi}^0_{v}(\boldsymbol{c})\right] }{\bar{\pi}^0_v(\boldsymbol{c})[1 - \bar{\mu}_v(\boldsymbol{c})]}- \psi_{v}(\bar{\mathbb{P}}_{TND})\right) d\mathbb{P}_{TND} \\ 
  =& \psi_{v}(\mathbb{P}_{TND})  -  \int \left( \frac{\mathbb{I}(Y = 1, V=v)}{\bar{\pi}^0_v(\boldsymbol{c})} - \bar{\mu}_v(\boldsymbol{c}) \frac{ \mathbb{I}\left( Y=0, H=1, V=v\right) - \mathbb{I}\left( Y=0, H=1\right)\bar{\pi}^0_{v}(\boldsymbol{c})  }{\bar{\pi}^0_v(\boldsymbol{c})[1 - \bar{\mu}_v(\boldsymbol{c})]}\right) d\mathbb{P}_{TND} \\
  =& \psi_{v}(\mathbb{P}_{TND})  -  \int \left( \frac{\mathbb{P}_{TND}(Y = 1, V=v \mid \boldsymbol{c})}{\bar{\pi}^0_v(\boldsymbol{c})} \right. \\ &\left. \ \ \ \ \ \ \ \ \ \ \ \ \ \ \ \ \ \ \ \ \ \ \ \ \ \ \ \ \ - \bar{\mu}_v(\boldsymbol{c}) \frac{ \mathbb{P}_{TND}\left( Y=0, H=1, V=v \mid \boldsymbol{c} \right) - \mathbb{P}_{TND}\left( Y=0, H=1 \mid \boldsymbol{c} \right) \bar{\pi}^0_{v}(\boldsymbol{c}) }{\bar{\pi}^0_v(\boldsymbol{c})[1 - \bar{\mu}_v(\boldsymbol{c})]}\right) d\mathbb{P}_{TND}  \\ 
  =& \int \mu_v(\boldsymbol{c})\omega_v(\boldsymbol{c}) d\mathbb{P}_{TND}  -  \int \left( \frac{\mu_v(\boldsymbol{c})\omega_v(\boldsymbol{c})\pi^0_v(\boldsymbol{c})}{\bar{\pi}^0_v(\boldsymbol{c})}  - \bar{\mu}_v(\boldsymbol{c}) \frac{ \pi^0_{v}(\boldsymbol{c}) [1 - m(\boldsymbol{c})] - [1 - m(\boldsymbol{c})] \bar{\pi}^0_{v}(\boldsymbol{c}) }{\bar{\pi}^0_v(\boldsymbol{c})[1 - \bar{\mu}_v(\boldsymbol{c})]}\right) d\mathbb{P}_{TND}  \\  
     =& \int   \frac{\mu_v(\boldsymbol{c})\omega_v(\boldsymbol{c})\bar{\pi}^0_v(\boldsymbol{c}) 
 -\mu_v(\boldsymbol{c})\omega_v(\boldsymbol{c})\pi^0_v(\boldsymbol{c})}{\bar{\pi}^0_v(\boldsymbol{c})}  + \bar{\mu}_v(\boldsymbol{c}) \frac{ \pi^0_{v}(\boldsymbol{c}) [1 - m(\boldsymbol{c})] - [1 - m(\boldsymbol{c})] \bar{\pi}^0_{v}(\boldsymbol{c}) }{\bar{\pi}^0_v(\boldsymbol{c})[1 - \bar{\mu}_v(\boldsymbol{c})]}d\mathbb{P}_{TND}  \\ 
      =& \int   \frac{\mu_v(\boldsymbol{c})[1-m(\boldsymbol{c})][\bar{\pi}^0_v(\boldsymbol{c}) 
 -\pi^0_v(\boldsymbol{c})]}{\bar{\pi}^0_v(\boldsymbol{c})[1-\mu(\boldsymbol{c})]}  - \bar{\mu}_v(\boldsymbol{c}) \frac{  [1 - m(\boldsymbol{c})] [\bar{\pi}^0_{v}(\boldsymbol{c}) - \pi^0_{v}(\boldsymbol{c}) ] }{\bar{\pi}^0_v(\boldsymbol{c})[1 - \bar{\mu}_v(\boldsymbol{c})]}d\mathbb{P}_{TND}\\
 =& \int \frac{[\bar{\pi}^0_v(\boldsymbol{c}) 
 -\pi^0_v(\boldsymbol{c})]}{\bar{\pi}^0_v(\boldsymbol{c})}\left[ \underbrace{ \frac{\mu_v(\boldsymbol{c})[1-m(\boldsymbol{c})]}{[1-\mu(\boldsymbol{c})]}}_{(I)}  - \underbrace{\bar{\mu}_v(\boldsymbol{c}) \frac{  [1 - m(\boldsymbol{c})]  }{[1 - \bar{\mu}_v(\boldsymbol{c})]}}_{(II)}\right]d\mathbb{P}_{TND}
  \end{align*}
  where the second equality is based on the fact that $\int \varphi_{v}(\bar{\mathbb{P}}_{TND}) d \bar{\mathbb{P}}_{TND}=0$ and then plugging in $ \varphi_{v}(\bar{\mathbb{P}}_{TND})$, the third equality cancels the term $\psi_{v}(\bar{\mathbb{P}}_{TND})$, and the fourth equality applies iterated expectation. The fifth equality follows by the definition of $\psi_{v}(\mathbb{P}_{TND})$ and the facts that $\mathbb{P}_{TND}(Y = 1, V=v \mid \boldsymbol{c}) = \mathbb{P}_{TND}(Y = 1\mid V=v,\boldsymbol{c})\mathbb{P}_{TND}(V=v\mid \boldsymbol{c})=\mu_v(\boldsymbol{c})\pi_v(\boldsymbol{c}) = \mu_v(\boldsymbol{c})\omega_v(\boldsymbol{c})\pi^0_v(\boldsymbol{c}) ,$ $\mathbb{P}_{TND}\left( Y=0, H=1 \mid \boldsymbol{c} \right) = [1 - m(\boldsymbol{c})],$ by previous definitions, and 
  \begin{align*}
      & \mathbb{P}_{TND}\left( Y=0, H=1, V=v \mid \boldsymbol{c} \right) \\ =& \mathbb{P}_{TND}\left(  V=v \mid Y=0, H=1,\boldsymbol{c} \right)\mathbb{P}_{TND}\left( Y=0, H=1 \mid \boldsymbol{c} \right) = \pi^0_{v}(\boldsymbol{c}) [1 - m(\boldsymbol{c})],
  \end{align*}
the sixth equality aggregates the first two terms, and the seventh equality uses $\omega_v(\boldsymbol{c})=[1 - m(\boldsymbol{c})]/[1 - \mu_v(\boldsymbol{c})]$. The last equality extracts the common factor $[\bar{\pi}^0_v(\boldsymbol{c}) - \pi^0_{v}(\boldsymbol{c}]/\bar{\pi}^0_v(\boldsymbol{c})$.
For the terms (I) and (II), we have that
  \begin{align*}  
  (I) - (II) & =  [1 - m(\boldsymbol{c})]\left(\frac{\mu_v(\boldsymbol{c})}{1 - \mu_v(\boldsymbol{c})} -  \frac{\bar{\mu}_v(\boldsymbol{c})}{1 - \bar{\mu}_v(\boldsymbol{c})}\right) \\ &= \left[\left([1 - m(\boldsymbol{c})] - [1 - \bar{m}(\boldsymbol{c})]\right)\left(\frac{\mu_v(\boldsymbol{c})}{1 - \mu_v(\boldsymbol{c})} -  \frac{\bar{\mu}_v(\boldsymbol{c})}{1 - \bar{\mu}_v(\boldsymbol{c})}\right) \right. \\ &\left. \ \ \ \ \ \ \ \ \ \ \ \ \ \ \ \ \ \ \ \ \ \ \ \ \ + [1 - \bar{m}(\boldsymbol{c})] \left(\frac{\mu_v(\boldsymbol{c})}{1 - \mu_v(\boldsymbol{c})} -  \frac{\bar{\mu}_v(\boldsymbol{c})}{1 - \bar{\mu}_v(\boldsymbol{c})}\right)\right]
  \\ &= \left[\left( m(\boldsymbol{c}) - \bar{m}(\boldsymbol{c})\right)\left(odds_v(\boldsymbol{c}) -  \overline{odds}_v(\boldsymbol{c})\right)  + [1 - \bar{m}(\boldsymbol{c})] \left(odds_v(\boldsymbol{c}) -  \overline{odds}_v(\boldsymbol{c})\right)\right]
  \end{align*}
  where $odds_v(\boldsymbol{c}) =  \mu_v(\boldsymbol{c})/[1 - \mu_v(\boldsymbol{c})],$ and $\overline{odds}_v(\boldsymbol{c}) =  \bar{\mu}_v(\boldsymbol{c})/[1 - \bar{\mu}_v(\boldsymbol{c})].$
    
    Therefore, building on the above two results, for $R^v_{2}(\bar{\mathbb{P}}_{TND}, \mathbb{P}_{TND})$, we further have
  \begin{align*}
  \begin{split}
  R^v_{2}(\bar{\mathbb{P}}_{TND}, \mathbb{P}_{TND})& =   \int \frac{[\bar{\pi}^0_v(\boldsymbol{c}) 
 -\pi^0_v(\boldsymbol{c})]}{\bar{\pi}^0_v(\boldsymbol{c})}\left[ \underbrace{ \frac{\mu_v(\boldsymbol{c})[1-m(\boldsymbol{c})]}{[1-\mu(\boldsymbol{c})]}}_{(I)}  - \underbrace{\bar{\mu}_v(\boldsymbol{c}) \frac{  [1 - m(\boldsymbol{c})]  }{[1 - \bar{\mu}_v(\boldsymbol{c})]}}_{(II)}\right]d\mathbb{P}_{TND},\\
  &= \int\frac{1}{\bar{\pi}^0_v(\boldsymbol{c})} \left( \mathcal{A}_v(\boldsymbol{c}) +  \mathcal{B}_v(\boldsymbol{c}) [1 - \bar{m}(\boldsymbol{c})]  \right)  d\mathbb{P}_{TND}.
  \end{split}
  \end{align*}
  where $\mathcal{A}_v(\boldsymbol{c}) := [\bar{\pi}^0_v(\boldsymbol{c}) - \pi^0_{v}(\boldsymbol{c})][m(\boldsymbol{c}) - \bar{m}(\boldsymbol{c})][odds_{v}(\boldsymbol{c}) - \overline{odds}_v(\boldsymbol{c})]$, and $\mathcal{B}_v(\boldsymbol{c}) := [\bar{\pi}^0_v(\boldsymbol{c}) - \pi^0_{v}(\boldsymbol{c})][odds_{v}(\boldsymbol{c}) - \overline{odds}_v(\boldsymbol{c})]$.

\section{Proof of Theorem 2:}\label{Appx.A4}
In this section, we provide the proof of Theorem \ref{Them1}.\\
\textbf{Proof:}
By definition of our proposed estimator, we have the important decomposition
\begin{align*}
    &\psi_v(\mathbb{P}_{TND}) - \mathbb{P}_{TND, n}[\phi_v(\boldsymbol{Z}; \hat{\pi}^{0}_v, \hat{\mu}_v )] \\ =& \psi_v(\mathbb{P}_{TND})  + (\mathbb{P}_{TND}- \mathbb{P}_{TND, n})(\phi_v(\boldsymbol{Z};\hat{\mathbb{P}}_{TND})) -  \mathbb{P}_{TND}(\phi_v(\boldsymbol{Z};\hat{\mathbb{P}}_{TND}))\\
    =& \psi_v(\mathbb{P}_{TND}) - \mathbb{P}_{TND}(\phi_v(\boldsymbol{Z};\hat{\mathbb{P}}_{TND})) + (\mathbb{P}_{TND}- \mathbb{P}_{TND, n})(\phi_v(\boldsymbol{Z};\hat{\mathbb{P}}_{TND})) \\ &\ \ \ \ \ \ \ \ \ \ \ \ \ \ \ - (\mathbb{P}_{TND}- \mathbb{P}_{TND, n})(\phi_v(\boldsymbol{Z};\mathbb{P}_{TND})) + (\mathbb{P}_{TND}- \mathbb{P}_{TND, n})(\phi_v(\boldsymbol{Z};\mathbb{P}_{TND}))\\
    =&  \underbrace{\psi_v(\mathbb{P}_{TND}) - \mathbb{P}_{TND}(\phi_v(\boldsymbol{Z};\hat{\mathbb{P}}_{TND}))}_{(1)} \\ &\ \ \ \ \ \ \ \ \ \ \ \ \ \ \ \ + \underbrace{(\mathbb{P}_{TND}- \mathbb{P}_{TND, n})(\phi_v(\boldsymbol{Z};\hat{\mathbb{P}}_{TND}) - \phi_v(\boldsymbol{Z};\mathbb{P}_{TND})) }_{(2)}+ \underbrace{(\mathbb{P}_{TND}- \mathbb{P}_{TND, n})(\phi_v(\boldsymbol{Z};\mathbb{P}_{TND}))}_{(3)},
\end{align*}
where the first equation follows by adding and subtracting $\mathbb{P}_{TND}(\phi_v(\boldsymbol{Z};\hat{\mathbb{P}}_{TND}))$, and the second equation is obtained by subtracting and adding $(\mathbb{P}_{TND}- \mathbb{P}_{TND, n})(\phi_v(\boldsymbol{Z};\mathbb{P}_{TND}))$.
For term (1), we have that
\begin{align*}
    &\psi_v(\mathbb{P}_{TND}) - \mathbb{P}_{TND}(\phi_v(\boldsymbol{Z};\hat{\mathbb{P}}_{TND}))\\ = &\psi_v(\mathbb{P}_{TND}) - \mathbb{P}_{TND}(\varphi_v(\hat{\mathbb{P}}_{TND}) + \psi_v (\hat{\mathbb{P}}_{TND})) \\=& \psi_v(\mathbb{P}_{TND}) - \psi_v (\hat{\mathbb{P}}_{TND}) - \int \varphi_v(\boldsymbol{Z}; \hat{\mathbb{P}}_{TND}) d\mathbb{P}_{TND} = R^v_{2}(\hat{\mathbb{P}}_{TND}, \mathbb{P}_{TND}),
\end{align*}
with the fact that $\int \varphi_{v}(\hat{\mathbb{P}}_{TND}) d \hat{\mathbb{P}}_{TND}=0$. 
Applying the results of Lemma \ref{lemma2} and the conditions given in Theorem \ref{Them1} to the above expression and using Cauchy-Schwarz inequality under TND distribution ($\mathbb{P}_{TND}(f g) \leq\|f\| \ \|g\|$), we have that, with a constant $\mathcal{C}$,
\begin{align*}
\begin{split}
    \psi_v(\mathbb{P}_{TND}) - \mathbb{P}_{TND}(\phi_v(\boldsymbol{Z};\hat{\mathbb{P}}_{TND})) & = R^v_{2}(\hat{\mathbb{P}}_{TND}, \mathbb{P}_{TND}) \\  &\leq \mathcal{C}\left(\left\|\hat{\pi}^{0}_{v}-\pi^{0}_{v}\right\|\left\|\hat{m}- m\right\|\left\|\widehat{odds}_{v}-odds_{v} \right\| +   \left\|\hat{\pi}^{0}_{v}-\pi^{0}_{v}\right\|\left\|\widehat{odds}_{v}-odds_{v}\right\| \right).
\end{split}
\end{align*}
That is, we have that (up to a multiplicative constant) $|\psi_v(\mathbb{P}_{TND}) - \mathbb{P}_{TND}(\phi_v(\boldsymbol{Z};\hat{\mathbb{P}}_{TND}))|$ or $|R^v_{2}(\hat{\mathbb{P}}_{TND}, \mathbb{P}_{TND})|$ is bounded above by $\left\|\hat{\pi}^{0}_{v}-\pi^{0}_{v}\right\|\left\|\hat{m}- m\right\|\left\|\widehat{odds}_{v}-odds_{v} \right\| +   \left\|\hat{\pi}^{0}_{v}-\pi^{0}_{v}\right\|\left\|\widehat{odds}_{v}-odds_{v}\right\|.$
For term (2), since $\mathbb{P}_{TND, n}$ is the empirical measure on an independent sample from $\hat{\mathbb{P}}_{TND}$, we can apply Lemma 2 of \cite{kennedy2020sharp} with the assumption A1 that $\left\|\phi_{v}(Z ; \hat{\mathbb{P}}_{TND})-\phi_{v}(Z ; \mathbb{P}_{TND})\right\|=o_{\mathbb{P}_{TND}}(1)$:
\begin{align*}
    (\mathbb{P}_{TND}- \mathbb{P}_{TND, n})(\phi_v(\boldsymbol{Z};\hat{\mathbb{P}}_{TND}) - \phi_v(\boldsymbol{Z};\mathbb{P}_{TND})) = O_{\mathbb{P}_{TND}}\left(\frac{\left\|\phi_{v}(Z ; \hat{\mathbb{P}}_{TND})-\phi_{v}(Z ; \mathbb{P}_{TND})\right\|}{\sqrt{n}}\right) = o_{\mathbb{P}_{TND}}\left(\frac{1}{\sqrt{n}}\right).
\end{align*}
For term (3), we have $(\mathbb{P}_{TND}- \mathbb{P}_{TND, n})(\phi_v(\boldsymbol{Z};\mathbb{P}_{TND})) =  (\mathbb{P}_{TND}- \mathbb{P}_{TND, n})(\phi_v(\boldsymbol{Z};\mathbb{P}_{TND}) - \psi_v(\mathbb{P}_{TND})) = (\mathbb{P}_{TND}- \mathbb{P}_{TND, n})(\varphi_v(\boldsymbol{Z};\mathbb{P}_{TND})),$ because $(\mathbb{P}_{TND}- \mathbb{P}_{TND, n})(\psi_v) = 0.$ 

Thus, the following results are obtained when combining (1) and (2), plus term (3):\begin{align*}
    & \psi_v(\mathbb{P}_{TND}) - \mathbb{P}_{TND, n}[\phi_v(\boldsymbol{Z}; \hat{\pi}^{0}_v, \hat{\mu}_v )] \\  =& O_{\mathbb{P}_{TND}}\left(\left\|\hat{\pi}^{0}_{v}-\pi^{0}_{v}\right\|\left\|\hat{m}- m\right\|\left\|\widehat{odds}_{v}-odds_{v} \right\| +   \left\|\hat{\pi}^{0}_{v}-\pi^{0}_{v}\right\|\left\|\widehat{odds}_{v}-odds_{v}\right\| \right) \\ &\ \ \ \ \ \ \ \ \ \ \ \ \ \ \ \ \ \ \   +  (\mathbb{P}_{TND}- \mathbb{P}_{TND, n})(\varphi_v(\boldsymbol{Z};\mathbb{P}_{TND})) + o_{\mathbb{P}_{TND}}\left(\frac{1}{\sqrt{n}}\right),
\end{align*}
where $\widehat{odds}_v(\boldsymbol{c}) =  \hat{\mu}_v(\boldsymbol{c})/[1 - \hat{\mu}_v(\boldsymbol{c})],$ and $odds_v(\boldsymbol{c}) =  \mu_v(\boldsymbol{c})/[1 - \mu_v(\boldsymbol{c})].$

\section{Proof of the claim in Corollary 1:}\label{Appx.A5}
Herein, we provide the proof substantiating the claim made in Corollary \ref{cor1}.\\
\textbf{Claim 1:}\label{clim1} If $\left\|\hat{\mu}_{v}-\mu_{v}\right\|=O_{\mathbb{P}_{TND}}\left(n^{-1 / 4}\right),$ then $\left\|\widehat{odds}_{v}-odds_{v}\right\|=O_{\mathbb{P}_{TND}}\left(n^{-1 / 4}\right),$ where $odds_v(\boldsymbol{c}) =  \mu_v(\boldsymbol{c})/[1 - \mu_v(\boldsymbol{c})],$ and $\widehat{odds}_v(\boldsymbol{c}) =  \hat{\mu}_v(\boldsymbol{c})/[1 - \hat{\mu}_v(\boldsymbol{c})].$\\
\textbf{Proof:}
For $\left\|\hat{\mu}_{v}-\mu_{v}\right\|=O_{\mathbb{P}_{TND}}\left(n^{-1 / 4}\right)$, we have  $\left\|(1 - \hat{\mu}_{v})-(1-\mu_{v})\right\|=O_{\mathbb{P}_{TND}}\left(n^{-1 / 4}\right),$ where $L_2(\mathbb{P}_{TND})$ norm is defined as 
\begin{equation*}
    \left\|\hat{\mu}_{v}-\mu_{v}\right\| := \sqrt{\int(\hat{\mu}_{v}-\mu_{v})^2 d \mathbb{P}_{TND}} = \sqrt{\mathbb{P}_{TND}\{ (\hat{\mu}_{v}-\mu_{v})^2 \} }.
\end{equation*}
First, we have that if $\left\|\hat{\mu}_{v}-\mu_{v}\right\|=O_{\mathbb{P}_{TND}}\left(n^{-1 / 4}\right),$ and $0 < \epsilon < \mu_v < 1 - \epsilon$ and $0 < \epsilon < \hat{\mu}_v < 1 - \epsilon$, then $\left\|\hat{\mu}_{v}\right\|=O_{\mathbb{P}_{TND}}\left(1\right),$ and $\left\| \frac{1}{\hat{\mu}_{v}}\right\|=O_{\mathbb{P}_{TND}}\left(1\right).$ Similarly, we have $\left\|(1 - \hat{\mu}_{v})-(1-\mu_{v})\right\|=O_{\mathbb{P}_{TND}}\left(n^{-1 / 4}\right),$ and $\left\|1 - \hat{\mu}_{v}\right\|=O_{\mathbb{P}_{TND}}\left(1\right),$ and $\left\| \frac{1}{1 - \hat{\mu}_{v}}\right\|=O_{\mathbb{P}_{TND}}\left(1\right).$ Then, we have 
\begin{align*}
    \left\|odds_{v} - \widehat{odds}_{v}\right\| & = \left\|\frac{\mu_v}{1-\mu_v} - \frac{\hat{\mu}_v}{1-\hat{\mu}_v}\right\| \\
    & = \left\|\frac{\mu_v(1-\hat{\mu}_v) - \hat{\mu}_v(1-\mu_v)}{(1-\mu_v)(1-\hat{\mu}_v)} \right\|   \\
    & = \left\|\frac{\mu_v(1-\hat{\mu}_v) - \mu_v(1-\mu_v) + \mu_v(1-\mu_v)- \hat{\mu}_v(1-\mu_v)}{(1-\mu_v)(1-\hat{\mu}_v)} \right\| \\
    & = \left\|\frac{\mu_v[(1-\hat{\mu}_v) - (1-\mu_v)] + (1-\mu_v)(\mu_v- \hat{\mu}_v)}{(1-\mu_v)(1-\hat{\mu}_v)} \right\| \\
    &\leq  \left\|\frac{\mu_v[(1-\hat{\mu}_v) - (1-\mu_v)] }{(1-\mu_v)(1-\hat{\mu}_v)} \right\| +  \left\|\frac{ (1-\mu_v)(\mu_v- \hat{\mu}_v)}{(1-\mu_v)(1-\hat{\mu}_v)} \right\| \\
    &<  \left\|\frac{(1 - \epsilon)[(1-\hat{\mu}_v) - (1-\mu_v)] }{\epsilon(1-\hat{\mu}_v)} \right\| +  \left\|\frac{ \mu_v- \hat{\mu}_v}{1-\hat{\mu}_v} \right\|\\
    &\leq  \left\|\frac{(1 - \epsilon)[(1-\hat{\mu}_v) - (1-\mu_v)] }{\epsilon} \right\| \left\|\frac{1}{1-\hat{\mu}_v} \right\| +  \left\| \mu_v- \hat{\mu}_v \right\| \left\|\frac{1}{1-\hat{\mu}_v} \right\|\\
    &= \frac{(1 - \epsilon) }{\epsilon}\left\|(1 - \hat{\mu}_{v})-(1-\mu_{v})\right\| \left\|\frac{1}{1-\hat{\mu}_v} \right\|  +  \left\| \mu_v- \hat{\mu}_v \right\| \left\|\frac{1}{1-\hat{\mu}_v} \right\| \\
    &= O_{\mathbb{P}_{TND}}\left(n^{-1 / 4}\right) O_{\mathbb{P}_{TND}}\left(1\right) + O_{\mathbb{P}_{TND}}\left(n^{-1 / 4}\right)O_{\mathbb{P}_{TND}}\left(1\right) \\
    &= O_{\mathbb{P}_{TND}}\left(n^{-1 / 4}\right).
\end{align*}
where the inequality in the fifth line uses the triangle inequality, the inequality in the sixth line follows by $0 < \epsilon < \mu_v < 1 - \epsilon$, and the inequality in the seventh line follows by the Cauchy-Schwarz inequality under TND distribution ($\mathbb{P}_{TND}(f g) \leq\|f\| \ \|g\|$, where $g = 1/(1 - \hat{\mu}_v)$).\\

\section{Proof of Corollary 2: Double Robustness}\label{Appx.proofdr}
In this section, we show the double robustness property of $\hat{\psi}_v$ (i.e., Corollary \ref{cor2}). According to conditions and results of Theorem \ref{Them1}, the estimator proposed $\hat{\psi}_v$ is consistent with the rate of convergence given by 
\begin{align*}
1/\sqrt{n} + \left\|\hat{\pi}^{0}_{v}-\pi^{0}_{v}\right\|\left\|\hat{m}- m\right\|\left\|\widehat{odds}_{v}-odds_{v} \right\| +   \left\|\hat{\pi}^{0}_{v}-\pi^{0}_{v}\right\|\left\|\widehat{odds}_{v}-odds_{v}\right\|,
\end{align*}
where the second and the third parts in the above equation are from
\begin{align*}
|R^v_{2}(\bar{\mathbb{P}}_{TND}, \mathbb{P}_{TND})| = O_{\mathbb{P}_{TND}}  \left(\left\|\hat{\pi}^{0}_{v}-\pi^{0}_{v}\right\|\left\|\hat{m}- m\right\|\left\|\widehat{odds}_{v}-odds_{v} \right\| +   \left\|\hat{\pi}^{0}_{v}-\pi^{0}_{v}\right\|\left\|\widehat{odds}_{v}-odds_{v}\right\| \right).
\end{align*}
That is, $|R^v_{2}(\bar{\mathbb{P}}_{TND}, \mathbb{P}_{TND})| \leq \left\|\hat{\pi}^{0}_{v}-\pi^{0}_{v}\right\| \left( \left\|\hat{m}- m\right\|\left\|\widehat{odds}_{v}-odds_{v} \right\| + \left\|\widehat{odds}_{v}-odds_{v} \right\|\right).$
If $\left\|\hat{\pi}^0_{v}-\pi^0_{v}\right\| = o_{\mathbb{P}_{TND}}(1)$, then $|R^v_{2}(\bar{\mathbb{P}}_{TND}, \mathbb{P}_{TND})| =o_{\mathbb{P}_{TND}}(1)$. \\
On the other hand, consider $odds_v(\boldsymbol{c})  = \mu(\boldsymbol{c})/[1-\mu_v(\boldsymbol{c})].$ If we have $\left\|\hat{\mu}_{v}-\mu_{v}\right\| =o_{\mathbb{P}_{TND}}(1)$, then based on the proof of the claim in Corollary \ref{cor1}, 
\begin{align*}
    0 \leq \left\|odds_{v} - \widehat{odds}_{v}\right\|
    \leq \frac{(1 - \epsilon) }{\epsilon}\left\|(1 - \hat{\mu}_{v})-(1-\mu_{v})\right\| \left\|\frac{1}{1-\hat{\mu}_v} \right\|  +  \left\| \mu_v- \hat{\mu}_v \right\| \left\|\frac{1}{1-\hat{\mu}_v} \right\| = o_{\mathbb{P}_{TND}}(1),
\end{align*}
we have $\left\|odds_{v} - \widehat{odds}_{v}\right\| =o_{\mathbb{P}_{TND}}(1)$; thus, $|R^v_{2}(\bar{\mathbb{P}}_{TND}, \mathbb{P}_{TND})| =o_{\mathbb{P}_{TND}}(1)$.

Therefore, if either the propensity score model is consistently estimated, i.e., $\left\|\hat{\pi}^0_{v}-\pi^0_{v}\right\| = o_{\mathbb{P}_{TND}}(1)$ and/or the outcome models are, i.e., $\left\|\hat{\mu}_{v}-\mu_{v}\right\| =o_{\mathbb{P}_{TND}}(1)$, $|R^v_{2}(\bar{\mathbb{P}}_{TND}, \mathbb{P}_{TND})| =o_{\mathbb{P}_{TND}}(1)$ and $\hat{\psi}_v$ is consistent.

\section{Efficiency Theory and Estimation of mRR}\label{mrrThm}
This section presents the efficiency theory about $\psi_{mRR}$,  building upon the foundations established for $\psi_{v}$. It encompasses a theorem regarding the von Mises expansion for $\psi_{mRR}$ and a one-step estimator. Additionally, it explores one corollary that delves into the estimator's properties, including $\sqrt{n}$-consistency, asymptotic normality, and double robustness.
\begin{theorem}\label{mrrThmandEst}
    For the target marginal risk ratio $\psi_{mRR}  = \psi_{v}/\psi_{v_0}$, where $v$ and $v_0$ in $\mathcal{V}$, the von Mises expansion is
    \begin{equation}\label{eqforln1}
\psi_{mRR}(\mathbb{P}_{TND})=\psi_{mRR}(\bar{\mathbb{P}}_{TND})+\psi_{mRR}(\bar{\mathbb{P}}_{TND}) \int \left[ \frac{\varphi_{v}(\bar{\mathbb{P}}_{TND}) }{\psi_{v}(\bar{\mathbb{P}}_{TND})} - \frac{\varphi_{v_0}(\bar{\mathbb{P}}_{TND}) }{\psi_{v_0}(\bar{\mathbb{P}}_{TND})} \right]  d(\mathbb{P}_{TND}-\bar{\mathbb{P}}_{TND})+R_{2}(\bar{\mathbb{P}}_{TND}, \mathbb{P}_{TND})
\end{equation}
where $R_{2}(\bar{\mathbb{P}}_{TND}, \mathbb{P}_{TND}):= \psi_{mRR}(\bar{\mathbb{P}}_{TND})\left\{ [\psi_{v}(\bar{\mathbb{P}}_{TND})]^{-1}R^{v}_{2}(\bar{\mathbb{P}}_{TND}, \mathbb{P}_{TND}) - [\psi_{v_0}(\bar{\mathbb{P}}_{TND})]^{-1}R^{v_0}_{2}(\bar{\mathbb{P}}_{TND}, \mathbb{P}_{TND})\right\},$ each term in the $R_{2}(\bar{\mathbb{P}}_{TND}, \mathbb{P}_{TND})$ is a second-order nuisance function error, for $v  \in \mathcal{V}$, $\varphi_v(Z)=  \mathbb{EIF}(\psi_v)$ is given in Lemma \ref{LEMMA1}, and $R^v_{2}(\bar{\mathbb{P}}_{TND}, \mathbb{P}_{TND})$ is given in Lemma \ref{lemma2}.
By plugging in $\hat{\psi}_v = \mathbb{P}_{TND, n}[\phi_v(\boldsymbol{Z}; \hat{\pi}^{0}_v, \hat{\mu}_v)]$, the one-step estimator of $\psi_{mRR}$, which can be viewed as an extension of Newton's methods used to emulate maximum likelihood estimators in parametric models (\cite{bickel1993efficient,pfanzagl1982contributions}) is given by 
    \begin{equation}\label{eifestor}
        \hat{\psi}^{eif}_{mRR} = \frac{\mathbb{P}_{TND, n}[\phi_v(\boldsymbol{Z}; \hat{\pi}^{0}_v, \hat{\mu}_v )]}{\mathbb{P}_{TND, n}[\phi_{v_0}(\boldsymbol{Z}; \hat{\pi}^{0}_{v_0}, \hat{\mu}_{v_0})]},
    \end{equation}
where for $v  \in \mathcal{V}$, $\hat{\psi}_v:= \mathbb{P}_{TND, n}[\phi_v(\boldsymbol{Z}; \hat{\pi}^{0}_v, \hat{\mu}_v )]$ is given in Theorem \ref{Them1}.
\end{theorem}
\textbf{Proof:} See subsection \ref{Appx.proofThm2}.

According to the von Mises expansion of $\psi_{mRR}$ in Equation (\ref{eqforln1}), by Lemma 2 of \cite{kennedy2021semiparametric}, our target $\psi_{mRR}(\mathbb{P}_{TND})$ is pathwise differentiable with efficient influence function $\mathbb{EIF}(\psi_{mRR}) = \left[ \varphi_{v}(\boldsymbol{Z}) /\psi_{v}(\boldsymbol{Z}) - \varphi_{v_0}(\boldsymbol{Z})/\psi_{v_0}(\boldsymbol{Z}) \right]\psi,$ where $\varphi_v(Z)=  \mathbb{EIF}(\psi_v)$ is given in Lemma \ref{LEMMA1} for $v  \in \mathcal{V}$. It's worth mentioning that \cite{kennedy2022semiparametric}'s Strategy 2 also offers a way to derive the EIF of $\psi_{mRR}$ through some tricks. Example 6 in \cite{kennedy2022semiparametric} presents a similar one-step estimator for the target parameter that is in a ratio form. In addition, we observe that the estimator $\hat{\psi}^{eif}_{mRR}$ in Equation (\ref{eifestor}) can be derived as the quotient of individual one-step estimators for both the numerator and the denominator.

\begin{corollary}
    The estimator $\hat{\psi}^{eif}_{mRR}$ is $\sqrt{n}$-consistent and asymptotically normal under the assumptions in Theorem \ref{Them1} and in Corollary \ref{cor1} for $v  \in \mathcal{V}$. The limiting distribution is given as $\sqrt{n}\left(\hat{\psi}^{eif}_{mRR}-\psi_{mRR}\right) \leadsto \mathcal{N}\left(0, \operatorname{var}\left(\mathbb{EIF}(\psi_{mRR}) \right)\right),$ where $\operatorname{var}\left(\mathbb{EIF}(\psi_{mRR}) \right) = \psi^2_{mRR}\operatorname{var}\left[\mathbb{EIF}(ln(\psi_{mRR}) )\right]$. Moreover, $\hat{\psi}^{eif}_{mRR}$ is doubly robust under the assumptions in Corollary \ref{cor2} for $v  \in \mathcal{V}$.
\end{corollary}

\subsection{Proof of Theorem \ref{mrrThmandEst}:}\label{Appx.proofThm2}
Firstly, we investigate the von Mises expansion for the  marginal risk ratio $\psi_{mRR}= \psi_{v}/\psi_{v_0} = \text{exp}[ln(\psi_{v}) - ln(\psi_{v_0})]$ by broadly examining the expansion of $\eta := g(\psi_v)$ for any twice continuously differentiable functions $g: \mathbb{R} \rightarrow \mathbb{R}$. Subsequently, we consider setting $g(x) = \ln(x)$ and then $g(x) = \exp(x)$ to derive the results. Finally, we show the one-step estimator of $\psi$.

\textbf{Proof:} Define $\eta:= g(\psi_v)$ for any $g: \mathbb{R} \rightarrow \mathbb{R}$ functions whose first two derivatives ${ g^{\prime }(x),g^{\prime \prime }(x)}$ exist and are continuous over the domain of the function, then we apply Taylor's Theorem, for a $\eta^*$ such that $g^{-1}(\eta^*) \in [g^{-1}(\eta) , g^{-1}(\bar{\eta})],$
\begin{align*}
    g[g^{-1}(\eta)] = g[g^{-1}(\bar{\eta})] + g^{\prime }(g^{-1}(\bar{\eta}))[g^{-1}(\eta) - g^{-1}(\bar{\eta})] + \frac{1}{2}g^{\prime \prime}(g^{-1}(\eta^*))[g^{-1}(\eta) - g^{-1}(\bar{\eta})]^{2}.
\end{align*}
That is,
\begin{align} \label{eq17}
    \eta = \bar{\eta} + g^{\prime }(\bar{\psi}_v)(\psi_v - \bar{\psi}_v) + \frac{1}{2}g^{\prime \prime}(g^{-1}(\eta^*))[\psi_v - \bar{\psi}_v]^{2}.
\end{align}
We further employ Lemma \ref{lemma2} to acquire the expression corresponding to the first-order term $(\psi_v - \bar{\psi}_v)$ and substitute it back into Equation (\ref{eq17}) yields
\begin{align*}
    g[\psi_{v}(\mathbb{P}_{TND})] - g[\psi_{v}(\bar{\mathbb{P}}_{TND})]  = g^{\prime }[\psi_{v}(\bar{\mathbb{P}}_{TND})] \left[\int \varphi_{v}(\bar{\mathbb{P}}_{TND}) d(\mathbb{P}_{TND}-\bar{\mathbb{P}}_{TND})+R^v_{2}(\bar{\mathbb{P}}_{TND}, \mathbb{P}_{TND})\right],
\end{align*}
where each term in the $R^v_{2}(\bar{\mathbb{P}}_{TND}, \mathbb{P}_{TND})$ is a second-order nuisance function. 
Thus, for $g(x) = ln(x)$, we have \begin{align} \label{eqforln}
    ln[\psi_{v}(\mathbb{P}_{TND})] - ln[\psi_{v}(\bar{\mathbb{P}}_{TND})]  = [\psi_{v}(\bar{\mathbb{P}}_{TND})]^{-1}\left[\int \varphi_{v}(\bar{\mathbb{P}}_{TND}) d(\mathbb{P}_{TND}-\bar{\mathbb{P}}_{TND})+R^v_{2}(\bar{\mathbb{P}}_{TND}, \mathbb{P}_{TND})\right].
\end{align}
Further, for the target parameter mRR, $\psi_{mRR}= \psi_{v}/\psi_{v_0} = \text{exp}[ln(\psi_{v}) - ln(\psi_{v_0})]$, we apply Taylor’s
Theorem for a $\psi^*_{mRR}$
such that $ln(\psi^*_{mRR})$ lies between $ ln[\psi_{mRR}(\mathbb{P}_{TND})]$ and $ln[\psi_{mRR}(\bar{\mathbb{P}}_{TND})]$:
\begin{align}\label{eqforln2}
&\psi_{mRR}(\mathbb{P}_{TND})-\psi_{mRR}(\bar{\mathbb{P}}_{TND}) \nonumber \\ & =\psi_{mRR}(\bar{\mathbb{P}}_{TND})\left[ln (\psi_{mRR}(\mathbb{P}_{TND}))- ln (\psi_{mRR}(\bar{\mathbb{P}}_{TND}))\right]+\frac{1}{2}\left[ln(\psi_{mRR}(\mathbb{P}_{TND}))-ln (\psi_{mRR}(\bar{\mathbb{P}}_{TND}))\right]^2 \psi^*_{mRR}
\end{align}
For the first order expression $\left[ln (\psi_{mRR}(\mathbb{P}_{TND}))- ln (\psi_{mRR}(\bar{\mathbb{P}}_{TND}))\right]$ we apply the results of Equation (\ref{eqforln}) to obtain
\begin{align*}
    &ln (\psi_{mRR}(\mathbb{P}_{TND}))- ln (\psi_{mRR}(\bar{\mathbb{P}}_{TND}))\\ =& [\psi_{v}(\bar{\mathbb{P}}_{TND})]^{-1} \int \varphi_{v}(\bar{\mathbb{P}}_{TND}) d(\mathbb{P}_{TND}-\bar{\mathbb{P}}_{TND})\\ &\ \ \ \ \ \ \ \ \ \ \ \ \ \ \ \  - [\psi_{v_0}(\bar{\mathbb{P}}_{TND})]^{-1}\int \varphi_{v_0}(\bar{\mathbb{P}}_{TND}) d(\mathbb{P}_{TND}-\bar{\mathbb{P}}_{TND})+ R_{2}(\bar{\mathbb{P}}_{TND}, \mathbb{P}_{TND}),
\end{align*}
where each term in $$R_{2}(\bar{\mathbb{P}}_{TND}, \mathbb{P}_{TND}):= \psi_{mRR}(\bar{\mathbb{P}}_{TND})\left\{ [\psi_{v}(\bar{\mathbb{P}}_{TND})]^{-1}R^{v}_{2}(\bar{\mathbb{P}}_{TND}, \mathbb{P}_{TND}) - [\psi_{v_0}(\bar{\mathbb{P}}_{TND})]^{-1}R^{v_0}_{2}(\bar{\mathbb{P}}_{TND}, \mathbb{P}_{TND})\right\}$$ is a second-order nuisance function.
Substituting back into Equation (\ref{eqforln2}), we have
\begin{equation*}\label{eqforln3}
\psi_{mRR}(\mathbb{P}_{TND})=\psi_{mRR}(\bar{\mathbb{P}}_{TND})+\psi_{mRR}(\bar{\mathbb{P}}_{TND}) \int \left[ \frac{\varphi_{v}(\bar{\mathbb{P}}_{TND}) }{\psi_{v}(\bar{\mathbb{P}}_{TND})} - \frac{\varphi_{v_0}(\bar{\mathbb{P}}_{TND}) }{\psi_{v_0}(\bar{\mathbb{P}}_{TND})} \right]  d(\mathbb{P}_{TND}-\bar{\mathbb{P}}_{TND})+R_{2}(\bar{\mathbb{P}}_{TND}, \mathbb{P}_{TND})
\end{equation*}
Therefore, the efficient influence function of $\psi$ is 
\begin{align}\label{eifratio}
        \varphi(\boldsymbol{Z} ; \mathbb{P}_{TND}) := \mathbb{EIF}(\psi_{mRR}) = \mathbb{EIF}(\psi_{v}/\psi_{v_0}) = \left[ \frac{\varphi_{v}(\boldsymbol{Z}) }{\psi_{v}(\boldsymbol{Z})} - \frac{\varphi_{v_0}(\boldsymbol{Z}) }{\psi_{v_0}(\boldsymbol{Z})} \right]\psi_{mRR},
\end{align}
where for $v  \in \mathcal{V}$, $$\varphi_v(\boldsymbol{Z}; \mathbb{P}_{TND})=  \mathbb{EIF}(\psi_v)= \frac{\mathbb{I}(Y = 1, V=v)}{\pi^0_v(\boldsymbol{c})} - \mu_v(\boldsymbol{c})\left\{ \frac{ \mathbb{I}\left( Y=0, H=1\right) \left[\mathbb{I}(V=v) - \pi^0_{v}(\boldsymbol{c})\right] }{\pi^0_v(\boldsymbol{c})[1 -\mu_v(\boldsymbol{c})]}  \right\}  - \psi_{v}.$$

For the one-step estimator of $\psi_{mRR}$, Equation (\ref{eifratio}) can also be written as   
\begin{equation*} 
   \varphi(\boldsymbol{Z} ; \mathbb{P}_{TND}) := \mathbb{EIF}(\psi_{mRR}) = \frac{\mathbb{EIF}\left(\psi_{v}\right)}{\psi_{v_0}} - \left(\frac{\psi_{v}}{\psi_{v_0}}\right)\frac{\mathbb{EIF}(\psi_{v_0})}{\psi_{v_0}}. 
\end{equation*}
Recall the definition that $\hat{\psi}_v:= \mathbb{P}_{TND, n}[\phi_v(\boldsymbol{Z}; \hat{\pi}^{0}_v, \hat{\mu}_v )],$ where 
\begin{align*} 
    \phi_v(\boldsymbol{Z}; \hat{\pi}^{0}_v, \hat{\mu}_v ) =  \frac{\mathbb{I}(Y = 1, V=v)}{\hat{\pi}^0_v(\boldsymbol{c})} - \hat{\mu}_v(\boldsymbol{c}) \frac{ \mathbb{I}\left( Y=0, H=1\right) \left[\mathbb{I}(V=v) - \hat{\pi}^0_{v}(\boldsymbol{c})\right] }{\hat{\pi}^0_v(\boldsymbol{c})[1-\hat{\mu}_v(\boldsymbol{c})]},
\end{align*} then the one-step estimator is given by 
\begin{align*}
    \hat{\psi}_{mRR} & =\psi_{mRR}(\hat{\mathbb{P}}_{TND})+\mathbb{P}_{TND, n}\{\varphi(\boldsymbol{Z} ; \hat{\mathbb{P}}_{TND})\} \\
    &= \frac{\psi_{v}(\hat{\mathbb{P}}_{TND})}{\psi_{v_0}(\hat{\mathbb{P}}_{TND})} + \frac{1}{\mathbb{P}_{TND, n}[\phi_{v_0}(\boldsymbol{Z}; \hat{\pi}^{0}_{v_0}, \hat{\mu}_{v_0} )]}\left(\mathbb{P}_{TND, n}[\phi_v(\boldsymbol{Z}; \hat{\pi}^{0}_v, \hat{\mu}_v )] - \frac{\psi_{v}(\hat{\mathbb{P}}_{TND})}{\psi_{v_0}(\hat{\mathbb{P}}_{TND})} {\mathbb{P}_{TND, n}[\phi_0(\boldsymbol{Z}; \hat{\pi}^{0}_{v_0}, \hat{\mu}_{v_0} )]} \right)\\
    &= \frac{\mathbb{P}_{TND, n}[\phi_v(\boldsymbol{Z}; \hat{\pi}^{0}_v, \hat{\mu}_v )]}{\mathbb{P}_{TND, n}[\phi_0(\boldsymbol{Z}; \hat{\pi}^{0}_{v_0}, \hat{\mu}_{v_0} )]}
\end{align*}
the second equation is derived by substituting the definitions of $\psi_{mRR}$, $\varphi(\boldsymbol{Z}; \mathbb{P}_{TND})$ (as shown in equation \ref{eifratio}) and plugging in $\psi_{v_0}(\hat{\mathbb{P}}_{TND}) = \hat{\psi}_{v_0}= \mathbb{P}_{TND, n}[\phi_v(\boldsymbol{Z}; \hat{\pi}^{0}_{v_0}, \hat{\mu}_{v_0} )]$, and the third equation follows from cancelling term $\psi_{v}(\hat{\mathbb{P}}_{TND})/\psi_{v_0}(\hat{\mathbb{P}}_{TND})$.

\subsection{Alternative confidence interval of TNDDR}\label{App:Alt}
For the confidence interval of TNDDR, we can also employ a logarithmic transformation to improve the accuracy of the normal approximation \citep{agresti2012categorical}. To obtain the asymptotic variance of the estimator of the natural logarithm of $\psi_{mRR}$, using the EIF of $\ln(\psi_{mRR}):=\ln(\psi_{v}/\psi_{v_0}),$
\begin{align*}
\begin{split}
    \mathbb{EIF}(\ln(\psi_{v}/\psi_{v_0}) ) & = \mathbb{EIF}\left[\ln(\psi_{v}) - \ln(\psi_{v_0})\right]  = \frac{\mathbb{EIF}(\psi_{v})}{\psi_{v}} - \frac{\mathbb{EIF}(\psi_{v_0})}{\psi_{v_0}} = \frac{{\varphi}_v(\boldsymbol{Z}) }{\psi_{v}} - \frac{{\varphi}_{v_0}(\boldsymbol{Z}) }{\psi_{v_0}},
\end{split}
\end{align*}
we can estimate the variance of $\ln(\hat{\psi}_{v}/\hat{\psi}_{v_0})$, denoted as
$\hat{\operatorname{var}}(\ln \hat{\psi}^{eif}_{mRR}) =\hat{\operatorname{var}}\left(\mathbb{EIF}(\ln(\hat{\psi}_{v}/\hat{\psi}_{v_0}) ) \right),$ by using the estimates of $\hat{\psi}_{v}$, $\hat{\varphi}_v(\boldsymbol{Z})$ and those of $v_0$.  Further, we use the normal approximation to produce the symmetric $\alpha$-level confidence intervals for $\ln(\psi_{mRR}):$
\begin{equation*} 
    \ln(\psi_{mRR}) = \ln\left(\frac{\psi_{v}}{\psi_{v_0}} \right)\in\left(\ln(\hat{\psi}^{eif}_{mRR}) \pm \frac{1}{\sqrt{n}} \Phi^{-1}\left(1-\frac{\alpha}{2}\right)  \sqrt{ \hat{\operatorname{var}}(\ln \hat{\psi}^{eif}_{mRR})}\right).
\end{equation*}
Lastly, we exponentiate the upper and lower confidence limits of $\ln(\psi_{mRR})$ to derive the asymmetric confidence interval for the mRR (i.e., $\psi_{mRR}$). An approximate $(1-\alpha) \times 100 \%$ confidence interval is 
\begin{equation*}
    \mathrm{CI}_n:=1 - \left[ \exp\left(\ln(\hat{\psi}^{eif}_{mRR}) \mp \frac{1}{\sqrt{n}} \Phi^{-1}\left(1-\frac{\alpha}{2}\right)  \sqrt{ \hat{\operatorname{var}}(\ln \hat{\psi}^{eif}_{mRR})}\right)\right].
\end{equation*}

\section{Complementary results of Studies 1 and 3}\label{Appx.sim}
\begin{table}[ht] \centering
    \caption{Complementary results for Table \ref{tab:1}: Results of Simulation Study 1, 500 simulations for sample sizes of $n = 1000,$ $4000$ and $8000$ with case percentages in the TND samples ranging from  $42 \% \sim 48\%$.}
    \begin{threeparttable}[t]
 \begin{tabular}{cccccc} 
        \hline
       \multirow{3}{*}{$n$} & Scenarios $\&$ Truth & \multicolumn{4}{c}{$\beta_{em} = 0.25$; $\psi_{mRR} = 0.507$; Co-infection (TND): $\sim 0.05\%$}   \\
        \cmidrule{3-6} 
       & Methods & IPW (HAL) &  OutReg (HAL)  & IPW (NN) & OutReg (NN)  \\
        \hline
        \multirow{3}{*}{$1000$} & Median bias   & 0.012 & 0.002  & 0.077 &  -0.024  \\
        & MC MSE  & 0.024 & 0.020  & 0.055 & 0.082    \\
        & MC SE  & 0.005 & 0.004  & 0.010 & 0.012    \\
        & $\% Cov_{mRR}$ & $-$ & $-$  & 97.4  & 93.6   \\
          \hline
        \multirow{3}{*}{$4000$} & Median bias   & 0.003 & 0.003  & 0.059 &  -0.008   \\
        & MC MSE  & 0.007 & 0.005  & 0.017 & 0.007    \\
        & MC SE  & 0.003 & 0.002  & 0.005 & 0.004  \\
        & $\% Cov_{mRR}$ & $-$ & $-$  & 93.2  & 91.8   \\
          \hline
        \multirow{3}{*}{$8000$} & Median bias   & -0.003 & 0.002  & 0.085 &  -0.002    \\
        & MC MSE  & 0.005 & 0.004  & 0.017 & 0.004    \\
        & MC SE  & 0.002 & 0.002  & 0.004 & 0.003  \\
        & $\% Cov_{mRR}$ & $-$ & $-$  & 80.8  & 82.8    \\
          \hline
    \end{tabular} 
    \begin{tablenotes}
      \item Note: IPW denotes the estimator of $\hat{\psi}^{ipw}_{mRR}$, TNDDR corresponds to the proposed TND EIF-based estimator of $\hat{\psi}^{eif}_{mRR}$ with corss-fitting, and OutReg denotes the estimator $\hat{\psi}^{O}_{mRR}$ with the debiasing weights of the outcome probability ratio. Nuisance functions for IPW and OutReg were estimated using HAL, and NN, as indicated in parentheses. MC MSE and MC SE stand for Monte Carlo Mean Square Error and Monte Carlo Standard Error, respectively. Co-infection (TND) means the co-infection rate in the TND sample. 
    \end{tablenotes}
    \end{threeparttable}
 \label{tab:1comp}  
\end{table}

\begin{table}[ht] \centering
    \caption{Complementary results for Table \ref{tab:3co-inf}: simulation results in different settings with different co-infection rates: 500 simulations for the sample size of $n = 8000$.}
    \begin{threeparttable}[t]
 \begin{tabular}{cccccccc} 
        \hline
       \multirow{3}{*}{Scenarios $\&$ Truth} & & \multicolumn{5}{c}{}   \\
       & Methods & IPW &  OutReg  & IPW &  OutReg  & IPW &  OutReg  \\
       & Metrics & (HAL) & (HAL) & (NN)& (NN) & (GLM) & (GLM)\\
        \hline
        \multirow{1}{*}{ Co-infec (TND): $\sim 0.05\%$} & Median bias   & -0.003 & 0.002 & 0.085 & -0.002  & -0.006   & -0.004  \\
        \multirow{1}{*}{ Co-infec (Pop): $\sim 0.0001\%$} &  MC MSE  & 0.005 & 0.004  & 0.017 & 0.004 &  0.002   & 0.002  \\
       Ctrl. exch. approx. holds  & MC SE  & 0.002 & 0.002  & 0.004 & 0.003 &  0.002  & 0.002\\
        $\psi_{mRR} = 0.507$ & $\% Cov_{mRR}$ & $-$ & $-$  & 80.8  & 82.8 &  91.6  & 90.8 \\
          \hline
        \multirow{1}{*}{ Co-infec (TND): $\sim 4.53\%$}  & Median bias   & -0.056 & -0.059 & 0.015  & -0.058  & -0.046 & -0.149   \\
       \multirow{1}{*}{ Co-infec (Pop): $\sim 1.21\%$} & MC MSE  & 0.006 & 0.005  & 0.008 & 0.006&  0.004  &0.023   \\
        Ctrl. exch. does not hold & MC SE  & 0.002 & 0.002  & 0.004 & 0.002&  0.002 & 0.001  \\
        $\psi_{mRR} = 0.621$ & $\% Cov_{mRR}$ & $-$ & $-$  & 97.2  & 73.8 &  82.2 & 2.56   \\
          \hline
        \multirow{1}{*}{ Co-infec (TND): $\sim 38.60\%$}  & Median bias   & -0.193 & -0.196 & -0.160  & -0.239 & -0.237 & -0.322   \\
       \multirow{1}{*}{ Co-infec (Pop): $\sim 22.10\%$} & MC MSE  & 0.060 & 0.045 & 0.045 & 0.070 &  0.059 & 0.104     \\
      Ctrl. exch. does not hold  & MC SE  & 0.006 & 0.003  & 0.006 & 0.006 &  0.003& 0.002   \\
        $\psi_{mRR} = 0.690$ & $\% Cov_{mRR}$ & $-$ & $-$  & 61.4  & 46.6 &  8.3 &  0   \\
          \hline
    \end{tabular} 
    \begin{tablenotes}
      \item Note:  Co-infec (TND) and Co-infec (Pop) refer to the co-infection rates in the TND sample and the general population, respectively. Ctrl. exch. refers to the control exchangeability assumption which was tested in the population dataset; a significant (positive) test result was reported as Ctrl. exch. does not hold; a non-significant (negative) test was reported as Ctrl. exch. approx. holds. IPW denotes the estimator of $\hat{\psi}^{ipw}_{mRR}$,  and OutReg denotes the estimator $\hat{\psi}^{O}_{mRR}$ with the debiasing weights of the outcome probability ratio. Nuisance functions for IPW and OutReg were estimated using HAL, NN and correctly specified GLM. MC MSE and MC SE stand for Monte Carlo Mean Square Error and Monte Carlo Standard Error, respectively.
    \end{tablenotes}
    \end{threeparttable}
 \label{tab:3co-infcom}  
\end{table}

\begin{figure}[ht]
    \centering    \includegraphics[height=3.15in, bb=0 0 700 500]{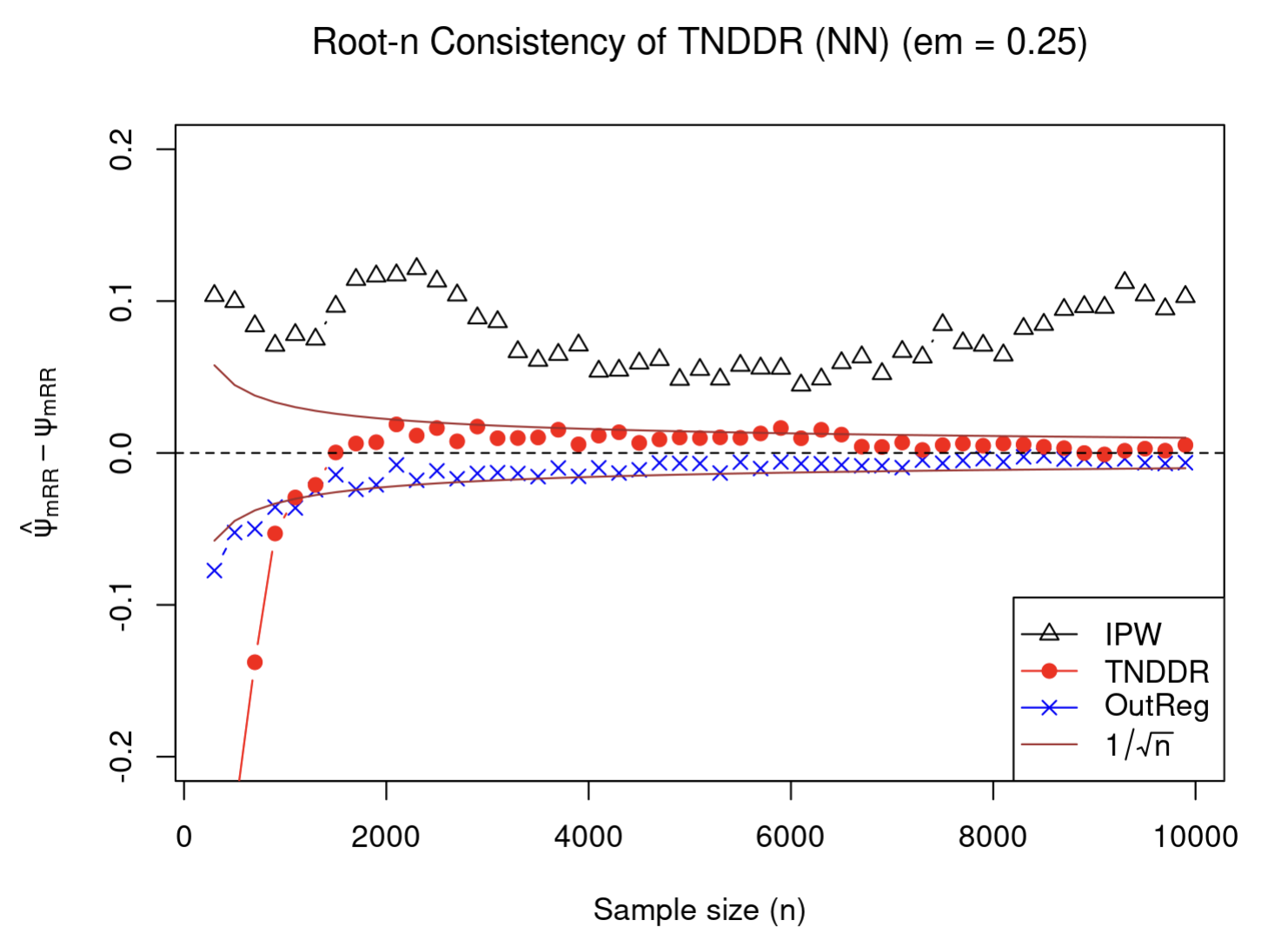}
    \caption{Simulation results of TNDDR (NN) from 200 simulations with varying sample sizes under the setting $\beta_{em}= 0.25\ (\psi_{mRR}= 0.507)$. }
    \label{rootNhal}
\end{figure}

This subsection presents Tables \ref{tab:1comp} and \ref{tab:3co-infcom} complement Tables \ref{tab:1} (from Study 1) and \ref{tab:3co-inf} (from Study 3), providing results for the IPW and outcome regression estimators. 

In Table \ref{tab:1comp}, the confidence intervals were not produced for the IPW (HAL) and outcome regression (HAL) due to the computationally intensive nature of bootstrapping using HAL. Figure \ref{tab:1comp} depicts the average bias of each estimator across sample sizes ranging from 300 to 10,000, with increments of 200, using $1/\sqrt{n}$ lines as benchmarks. As the sample size increases, the IPW (HAL), outcome regression (HAL), and TNDDR (HAL) estimators all can converge at parametric rates, as indicated by the bias approaching zero within the $1/\sqrt{n}$ rate.

\section{Additional information about the real data analysis:}\label{Appx.realdata}
\begin{figure}[ht]
    \centering    \includegraphics[height=3.15in, bb=0 0 700 500]{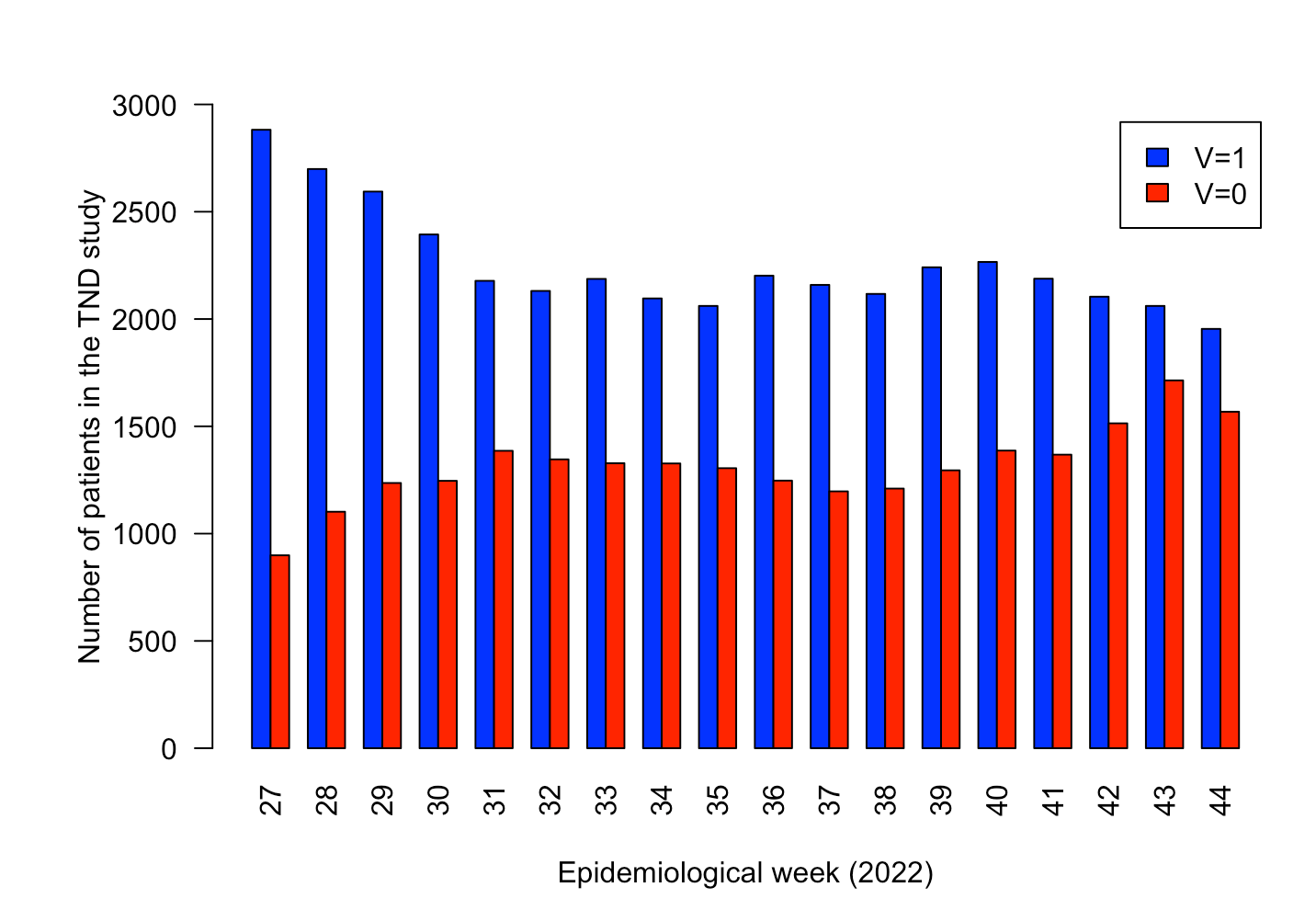}
    \caption{Bar chart depicting the patient counts in the TND study for $V = 1$ and $V = 0$, sorted by the epidemiological observation period from weeks 27 to 44 in 2022.}
    \label{figEpiWeek}
\end{figure}

This section includes an extra table (Table \ref{INSPQtb1}) detailing the summary of descriptive statistics derived from a TND study conducted by INSPQ. Because there are multiple categories within the epidemiological observation timeframe (spanning from weeks 27 to 44 in 2022, totaling 18 categories), we have included a side-by-side bar graph in Figure \ref{figEpiWeek}.

\begin{table}[ht]
\setlength\extrarowheight{0.3pt}
\centering
\caption{Summary of the descriptive statistics from a TND study conducted by the Institut national de santé publique du Québec, Canada, on individuals exhibiting symptoms and undergoing hospitalization, from July 3 to November 5, 2022. MM: multimorbidity.}
\begin{tabular}{lccc}
\hline
&\multicolumn{2}{c}{Vaccination} & Total\\
\cline{2-3}
Variables & Unexposed ($V=0$): 23675 ($36.9\%$) & Exposed ($V=1$): 
40514 ($63.1\%$) & 64189\\
\hline
Age \hspace{13pt} 60-69 & $8542\ (36.1\%)$ & $8615\ (21.3\%)$ & $
17157$ \\
\hspace{35pt} 70-79 & $7811\ (33.0\%)$ & $13817\ (34.1\%)$ & $21628$\\
\hspace{36pt} 80-89 & $5614\ (23.7\%)$ & $
13164\ (32.5\%)$ & $18778$\\
\hspace{37pt} 90+ & $1708\ (7.2\%)$ & $
4918\ (12.1\%)$ & $
6626$\\
\hline
Sex\hspace{15pt} Male & $11510\ (48.6\%)$ & $18607\ (45.9\%)$ &  $30117$\\
\hspace{33pt} Female & $12165\ (51.4\%)$ & $
21907\ (54.1\%)$ & $34072$\\
\hline
MM \hspace{7pt} Yes & $16084\ (67.9\%)$ & $30307\ (74.8\%)$ & $46391$ \\
\hspace{32pt} No & $7591\ (32.1\%)$ & $10207\ (25.2\%)$ & $17798$\\
\hline
Outcomes\hspace{15pt} Cases & $1323\ (5.6\%)$ & $
1778\ (4.4\%)$ & $3101$ \\
\hspace{55pt} Controls & $22352\ (94.4\%)$ & $
38736\ (95.6\%)$ & $61088$\\
\hline 
\end{tabular}
\label{INSPQtb1}
\end{table}

As emphasized in the main text, a causal interpretation of the results of our analysis relies on the assumption of no unmeasured confounding, which necessitates controlling for all confounding factors of the treatment and outcome. However, this assumption cannot be verified with the observed data. Therefore, in our analysis, we report E-values, introduced by VanderWeele and Ding (2017) \cite{vanderweele2017sensitivity} based on the theory of Ding and VanderWeele (2016) \cite{ding2016sensitivity}, which is one sensitivity analysis technique that quantifies the evidence of causation in observational studies in the presence of unmeasured confounding. The E-value measures how strongly an unmeasured confounder must be associated with both the treatment and the outcome to nullify the observed treatment-outcome association. It is expressed on a ratio scale. Larger E-values indicate that a more substantial confounder would be needed to negate the results. The minimum E-value is 1, implying that no unmeasured confounding is required to nullify the association.

For any given risk ratio (RR) reported in a study, two E-values are typically calculated: one for the observed outcome and one for the $95\%$ confidence interval (CI) of the outcome, specifically at the limit closest to the null. For our RR less than 1, the E-value for the observed outcome is calculated as:
\begin{equation*}
\text{E-value} = \frac{1}{RR} + \sqrt{\frac{1}{RR} \times \left(\frac{1}{RR} - 1\right)},
\end{equation*}

and the E-value for the $95\%$ CI is calculated as, for the lower limit of the $95\%$ CI (denoted as $UL$):

\begin{equation*}
\text{E-value of the } 95\% \text{ CI} =  
\begin{cases} 
1 & \text{if } UL \geq 1, \\
\frac{1}{UL} + \sqrt{\frac{1}{UL} \left(\frac{1}{UL} - 1\right)} & \text{if } UL < 1. 
\end{cases}
\end{equation*}
An online E-value calculator (https://www.evalue-calculator.com/evalue/)  and R Package (\texttt{EValue}) for computing E-values are available\cite{mathur2018web}.

In addition, for E-value for a hypothetical true value (e.g., non-null E-values), we
want to calculate the strength of confounding needed to shift the observed RR to the hypothetical true value, \( RR_{hyp} \), rather than the null value of 1. For a hypothetical true RR value below 1, the formula would be adjusted as follows:
\[
\text{E-value} = \frac{RR_{hyp}}{RR} + \sqrt{\frac{RR_{hyp}}{RR} \times \left(\frac{RR_{hyp}}{RR} - 1\right)},
\]
and for the lower limit of the $95\%$ CI:
\begin{equation*}
\text{E-value of the } 95\% \text{ CI for hypothetical true value} =  
\frac{RR_{hyp}}{UL} + \sqrt{\frac{RR_{hyp}}{UL} \left(\frac{RR_{hyp}}{UL} - 1\right)} 
\end{equation*}

\begin{table}[ht]\centering
    \caption{TNDDR $\psi_{mRR}$ estimates and E-values for point estimates and $95\%$ confidence intervals upper limit via GLM and MARS nuisance function estimations with different hypothetical true causal effects values. ($\mathbf{E}_{est}$ and $\mathbf{E}_{CI}$ represent the E-values for the point estimates and the upper confidence interval limit.) }
\begin{tabular}{c|cc|cc}\hline
Method (TNDDR)  & \multicolumn{2}{l|}{GLM: 
0.650 (0.602, 0.703)} & \multicolumn{2}{l}{ML (MARS):0.646 (0.599, 0.696)}                          \\ 
Hypothetical true $\psi_{mRR}$& $\mathbf{E}_{est}$ & $\mathbf{E}_{CI}$ & $\mathbf{E}_{est}$ & $\mathbf{E}_{CI}$ \\ \hline
0.9   & 2.11 & 1.88  & 2.13 & 1.91 \\ \hline
0.8   & 1.76 & 1.53  & 1.78 & 1.56 \\ \hline
0.7   & 1.36 & 1.00  & 1.38 & 1.08 \\ \hline
0.6   & 1.38 & 1.06  & 1.36 & 1.00 \\ \hline
0.5   & 1.92 & 1.70  & 1.91 & 1.69 \\ \hline
0.4   & 2.63 & 2.38  & 2.61 & 2.36 \\ \hline
0.3   & 3.76 & 3.43  & 3.73 & 3.41 \\ \hline
0.2   & 5.95 & 5.47  & 5.91 & 5.44 \\ \hline
0.1   & 12.48 & 11.52  & 12.40 & 11.46 \\ \hline
\end{tabular}    \label{tbrealdatEval}  
\end{table}

Table \ref{tbrealdatEval} presents E-values for point estimates ($\mathbf{E}_{est}$) and E-values for the upper confidence limit  ($\mathbf{E}_{CI}$) derived from sensitivity analyses addressing unmeasured confounding in the study of vaccination and hospitalized symptomatic SARS-CoV-2 infection, with different hypothesized true causal effects. To interpret these E-values \citep{mathur2018web,vanderweele2017sensitivity}, for instance, given the estimated $\psi_{mRR}$ of $0.646$ from TNDDR with MARS, an unmeasured confounder associated with both the compound outcome and vaccination by an RR of 3.73 each, beyond the measured confounders, could shift the observed association to the hypothetical values of the true $\psi_{mRR} = 0.3$, whereas weaker confounding could not. Similarly, an unmeasured confounder associated with both the compound outcome and vaccination by an RR of 3.41 each, beyond the measured confounders, could shift the confidence interval to include the hypothetical value of $\psi_{mRR} = 0.3$, but weaker confounding could not achieve this. This analysis allows us to more confidently exclude certain levels of the potential mRR.

\end{document}